\documentclass[aps,prb,superscriptaddress,amssym,floatfix,notitlepage,eqsecnum]{revtex4-1}
\usepackage{amsfonts}
\usepackage{amsmath}
\usepackage{amssymb}
\usepackage{graphicx}

\DeclareMathOperator{\tr}{tr}

\begin{document}

\title{Conductance of closed and open long Aharonov-Bohm-Kondo rings}
\author{Zheng Shi}
\affiliation{Department of Physics and Astronomy, University of British Columbia,
Vancouver, BC V6T 1Z1, Canada}
\author{Yashar Komijani}
\affiliation{Department of Physics and Astronomy, Rutgers University, Piscataway, New
Jersey, 08854, USA}
\date{\today }

\begin{abstract}
We calculate the finite temperature linear DC conductance of a generic
single-impurity Anderson model containing an arbitrary number of Fermi
liquid leads, and apply the formalism to closed and open long
Aharonov-Bohm-Kondo (ABK) rings. We show that, as with the short ABK ring,
there is a contribution to the conductance from the connected 4-point
Green's function of the conduction electrons. At sufficiently low
temperatures this contribution can be eliminated, and the conductance can be
expressed as a linear function of the T-matrix of the screening channel. For
closed rings we show that at temperatures high compared to the Kondo
temperature, the conductance behaves differently for temperatures above and
below $v_{F}/L$ where $v_{F}$ is the Fermi velocity and $L$ is the
circumference of the ring. For open rings, when the ring arms have both a
small transmission and a small reflection, we show from the microscopic
model that the ring behaves like a two-path interferometer, and that the
Kondo temperature is unaffected by details of the ring. Our findings confirm
that ABK rings are potentially useful in the detection of the size of the
Kondo screening cloud, the $\pi/2$ scattering phase shift from the Kondo
singlet, and the suppression of Aharonov-Bohm oscillations due to inelastic
scattering.
\end{abstract}

\maketitle

\section{Introduction\label{sec:intro}}

The Kondo problem, one of the most influential problems in condensed matter
physics, emerges from a deceptively unembellished model: a localized
impurity spin coupled with a Fermi sea of conduction electrons.\cite%
{ProgTheorPhys.32.37,hewson1997kondo} Perturbation theory in the coupling
constant is plagued by infrared divergence, but after much theoretical
endeavor\cite{JPhysC.3.2346,RevModPhys.47.773,JLowTempPhys.17.31} it has
been recognized that the model has a relatively simple low energy behavior.
For the single-channel spin-$1/2$ model, at temperatures well below the
Kondo temperature $T_{K}$, the impurity spin is \textquotedblleft
screened\textquotedblright\ by the conduction electrons, forming a local
singlet state. The spatial extent of this singlet state, commonly termed the
\textquotedblleft Kondo screening cloud\textquotedblright , is expected to
be $L_{K}=v_{F}/T_{K}$ where $v_{F}$ is the Fermi velocity. The remaining
conduction electrons are well described by a Fermi liquid theory at zero
temperature, and acquire a phase shift ($\pi /2$ in the presence of
particle-hole symmetry) upon being elastically scattered by the Kondo
singlet. Moreover, at finite temperatures, scattering by the Kondo impurity
can have both elastic and inelastic contributions\cite%
{PhysRevLett.93.107204,*PhysRevB.75.235112}, and it has been suggested that
the inelastic scattering can be the origin of decoherence in mesoscopic
structure, as measured for example by weak localization.\cite%
{PhysRevLett.93.107204,PhysRevLett.96.226601,PhysRevLett.89.206804}

Advances in semiconductor technology have made it possible to imitate the
impurity spin with a quantum dot (QD):\cite%
{Nature.391.156,Science.281.540,PhysRevLett.83.804,Science.289.2105,JPhysCondensMatter.16.R513,*condmat0501007}
when the ground state number of electrons in the QD is odd, the QD hosts a
nonzero spin at temperatures lower than the charging energy. This
possibility has triggered renewed experimental and theoretical interest in
mesoscopic manifestations of Kondo physics in QD devices, such as the
observation of the length scale $L_{K}$, the $\pi /2$ phase shift, and also
decoherence effects of inelastic Kondo scattering.

Many mesoscopic configurations have been proposed in order to observe $L_{K}$%
. These include QD-terminated finite quantum wires,\cite{PhysRevB.77.125327}
and also various geometries with an embedded QD, including finite quantum
wires,\cite%
{PhysRevLett.89.206602,*PhysRevB.68.115304,PhysRevLett.90.216801,PhysRevLett.110.246603}
small metallic grains/larger QDs,\cite%
{PhysRevLett.82.2143,PhysRevB.66.115303,EurophysLett.71.973,PhysRevB.73.205325,PhysRevLett.96.176802,*EurophysLett.97.17006,*PhysRevB.85.155455}
and in particular, closed long Aharonov-Bohm (AB) rings with\cite%
{PhysRevB.72.245313,PhysRevB.83.165310} and without\cite%
{PhysRevLett.86.2854,*PhysRevB.64.085308} external electrodes. (A closed
ring conserves the electric current and there is no leakage current.)
Another motivation for quantum rings is that they may be used to answer the
question of whether or not the inelastic scattering from the Kondo QD can
cause decoherence by suppressing the amplitude of AB oscillations. A common
feature of all these configurations is that they introduce at least one
additional mesoscopic length scale $L$. When the bare Kondo coupling
strength is adjusted so that $L_{K}$ crosses the scale $L$, the dependence
of observables on other control parameters changes qualitatively. In the
closed long AB ring with an embedded QD [also known as the
Aharonov-Bohm-Kondo (ABK) ring], for instance, $L$ is the circumference of
the ring: it is known that both $L_{K}$ itself and the conductance through
the ring can have drastically different AB phase dependences for $L_{K}\gg L$
and $L_{K}\lesssim L$.\cite{PhysRevB.72.245313,PhysRevB.83.165310} In the
\textquotedblleft large Kondo cloud\textquotedblright\ regime $L_{K}\gg L$,
corresponding to a relatively small bare Kondo coupling, the Kondo cloud
\textquotedblleft leaks\textquotedblright\ out of the ring and the size of
the cloud becomes strongly influenced by the ring size and other mesoscopic
details of the system. For a given bare Kondo coupling, $L_{K}$ can be
extremely sensitive to the AB phase at certain values of Fermi energy,
varying by many orders of magnitude. This sensitivity is completely lost in
the opposite \textquotedblleft small Kondo cloud\textquotedblright\ regime $%
L_{K}\lesssim L$, where the bare Kondo coupling is relatively large.

The conductance calculation of ABK rings, however, involves an additional
layer of complication\cite{PhysRevB.88.245104} that went neglected in a
number of early works. In mesoscopic Kondo problems with Fermi liquid
electrodes, it is usually convenient to work with the scattering states and
rotate to the basis of the so-called screening and non-screening channels:
the screening channel $\psi $ is coupled to the QD and therefore has a
nonzero T-matrix, while the non-screening channel $\phi $ is described by a
decoupled non-interacting theory.\cite%
{JETPLett.47.452,PhysRevLett.76.114,PhysRevB.64.045328,PhysRevB.82.165426,PhysRevB.83.165310}
A careful evaluation by Kubo formula at finite temperatures reveals that,
unlike a QD directly coupled to external leads, the interaction effects on
the linear DC conductance of short ABK rings are generally \emph{not} fully
encoded by the screening channel T-matrix in the single-particle sector, or
equivalently the two-point function. Instead, there exists a contribution
from connected four-point diagrams, corresponding to two-particle scattering
processes in the screening channel, which cannot be interpreted as resulting from a single-particle scattering amplitude.\cite{PhysRevB.88.245104} This is not in contradiction with the
famous Meir-Wingreen formula\cite{PhysRevLett.68.2512,*PhysRevB.50.5528} due
to the violation of the proportionate coupling condition.\cite%
{PhysRevB.76.113302} For the short ABK ring, the four-point contribution
becomes comparable to the two-point contribution well above the Kondo
temperature $T\gg T_{K}$, but can be approximately eliminated at
temperatures low compared to the bandwidth and the on-site repulsion of the
QD, $T\ll \min \left\{ t,U\right\} $, by applying the bias voltage and
probing the current in a particular fashion. (This does not mean the
four-point contribution is negligible for $T\ll \min \left\{ t,U\right\} $,
however.) One naturally wonders how this result generalizes to the closed
long ring at high and low temperatures, and how it possibly modifies early
predictions on conductance,\cite{PhysRevB.83.165310} which is again expected
to display qualitatively different behaviors for $L_{K}\gg L$ and $%
L_{K}\lesssim L$.

On the other hand, efforts to measure the $\pi /2$ phase shift are mainly
concentrated on two-path AB interferometer devices.\cite%
{PhysRevLett.100.226601,Nature.385.417,Science.290.779,PhysRevLett.88.076601,Nature.436.529,PhysRevLett.113.126601}
In these devices, electrons from the source lead propagate through two
possible paths (QD path and reference path) to the drain lead; the two paths
enclose a tunable AB phase $\varphi $, and a QD tuned into the Kondo regime
is embedded in the QD path. Most importantly, the complex transmission
amplitudes through the two paths $t_{d}e^{i\varphi }$ and $t_{ref}$ should
be independent of each other, and the total coherent transmission amplitude
at zero temperature $t_{sd}=t_{ref}+t_{d}e^{i\varphi }$ is the sum of the
individual amplitudes (the \textquotedblleft two-slit
condition\textquotedblright ), meaning multiple traversals of the ring are
negligible. Using a multi-particle scattering formalism, and assuming that
only single-particle scattering processes are coherent, Ref.~%
\onlinecite{PhysRevB.86.115129} calculates the conductance of such an
interferometer with an embedded Kondo QD in terms of the single-particle
T-matrix through the QD, and concludes that the AB oscillations are
suppressed by inelastic multi-particle scattering processes due to the Kondo
QD.

The two-path interferometer can in principle be realized through open AB
rings, where in contrast to closed rings, the propagating electrons may leak
into side leads attached to the ring. For a non-interacting QD, Ref.~%
\onlinecite{PhysRevB.66.115311} presents the criteria for an open long ring
to yield the intrinsic transmission phase through the QD: all lossy arms
with side leads should have a small transmission and a small reflection. A
small transmission suppresses multiple traversals of the ring and guarantees
the validity of the two-slit assumption, while a small reflection prevents
electrons from \textquotedblleft rattling\textquotedblright\ (tunneling back
and forth) across the QD. However, when the QD is in the Kondo regime, as
with the previously discussed closed AB rings, the transmission probability
through the QD\cite{PhysRevB.72.073311} and even the Kondo temperature\cite%
{PhysRevB.72.245313} may be sensitive to the AB phase and other details of
the geometry, hampering the detection of the intrinsic phase shift across
the QD. In addition, since the screening channels in the open ABK ring and
in the simple embedded QD geometry are usually not the same, it is not
obvious that the single-particle sector T-matrices coincide in the two
geometries. These issues are not addressed in Ref.~%
\onlinecite{PhysRevB.86.115129}, which simply assumes that the two-slit
condition is obeyed by the coherent processes, and that the T-matrix of the
open ABK ring is identical to that of the QD embedded between source and
drain leads. To our knowledge, it has been a mystery whether in certain
parameter regimes the open long ABK ring realizes the two-path
interferometer with a Kondo QD, where the Kondo temperature and the
transmission probability through the QD are independent of the details of
other parts of the ring, and the T-matrix of the ring truthfully reflects
the T-matrix of the QD.

The aforementioned problems in closed and open ABK rings prompt a unified
treatment of linear DC conductance in different mesoscopic geometries
containing an interacting QD. Much work has been done on generic mesoscopic
geometries,\cite{PhysRevB.71.035333,*PhysRevB.73.125338,PhysRevB.76.113302}
but in our formalism presented in this paper we aim to take the connected
contribution into account expressly, and refrain from making assumptions
about the geometry in question (such as parity symmetry).

We study a QD represented by an Anderson impurity, which is embedded in a
junction connecting an arbitrary number of Fermi liquid leads. The junction
is regarded as a black box characterized only by its scattering S-matrix and
its coupling with the QD, and all leads (including source, drain and
possibly side leads) are treated on equal footing. In parallel with Ref.~%
\onlinecite{PhysRevB.88.245104} we find that the linear DC conductance is
given by the sum of a \textquotedblleft disconnected\textquotedblright\ part
and a \textquotedblleft connected\textquotedblright\ part. The disconnected
part has the appearance of a linear response Landauer formula, where the
\textquotedblleft transmission amplitude\textquotedblright\ is linear in the
T-matrix of the screening channel in the single-particle sector, and indeed
reduces at zero temperature to a non-interacting transmission amplitude
appropriate for the local Fermi liquid theory. The connected part is again a
Fermi surface property, can be eliminated by proper application of bias
voltages, and is calculated perturbatively at weak coupling $T\gg T_{K}$, as
well as at strong coupling $T\ll T_{K}$ provided the local Fermi liquid
theory applies.

Our formalism is subsequently applied to long ABK rings. In the case of
closed rings, we show that for $T\gg T_{K}$, the high-temperature
conductance does exhibit qualitatively different behaviors as a function of
the AB phase for $T\gg v_{F}/L$ and $T\ll v_{F}/L$. In the case of open
rings, when the small transmission condition is met, we find the mesoscopic
fluctuations are suppressed, and the two-path interferometer behavior is
indeed recovered at low temperatures. If in addition the small reflection
condition is satisfied, the Kondo temperature of the QD and the complex
transmission amplitude through the QD are both unaffected by the details of
the ring. We then find the conductance at $T\gg T_{K}$ and $T\ll T_{K}$, and
in particular rigorously calculate the normalized visibility\cite%
{PhysRevB.86.115129} of the AB oscillations in the Fermi liquid regime. We
show that while the deviation of normalized visibility from unity is indeed
proportional to inelastic scattering as predicted by Ref.~%
\onlinecite{PhysRevB.88.245104}, the constant of proportionality depends on
non-universal particle-hole symmetry breaking potential scattering. Our
findings also suggest that the $\pi /2$ phase shift across the QD is
measurable in our two-path interferometer when the criteria of small
transmission and small reflection are fulfilled. We stress again that, while
we focus on long ABK rings in this paper, our general formalism is
applicable to a Kondo impurity embedded in an arbitrary non-interacting
multi-terminal mesoscopic structure.

The rest of this paper is outlined below. In Sec.~\ref{sec:model} we provide
a formulation of our generalized Anderson model with an interacting QD,
separate the screening channel from the non-screening ones, and discuss the
effective Kondo model in the local moment regime. In Sec.~\ref%
{sec:conductance} the linear DC conductance is calculated using Kubo
formula. Disconnected and connected contributions are examined separately,
along with the approximate elimination of the latter. Perturbation theories
in the weak-coupling and Fermi liquid regimes are employed in Sec.~\ref%
{sec:pert}; weak-coupling results applicable at high temperatures formally
resemble the short ring case. Sec.~\ref{sec:closed} applies the abstract
formalism to the closed long ring, and Sec.~\ref{sec:open} studies open long
rings and their potential utilization as two-path interferometers.
Conclusions and open questions are presented in Sec.~\ref{sec:outlook}. In
Appendix~\ref{sec:appcomp}, we make contact with early results by applying
our formalism in a few other mesoscopic systems. Appendix~\ref{sec:appdisc}
consists of details related to the calculation of disconnected
contributions. Appendix~\ref{sec:appNI} is a check of our formalism in the
case of a non-interacting QD. Appendix~\ref{sec:appFermi} focuses on the
Fermi liquid regime: we derive the T-matrix for the screening channel, and
perform another consistency check on our formalism by calculating the
connected contribution explicitly. Finally, Appendix~\ref{sec:appopen}
presents the non-interacting Schroedinger equations for the open long ring,
whose solutions are used in Sec.~\ref{sec:open}.

\section{Model\label{sec:model}}

Our generalized tight-binding Anderson model describes $N$ Fermi liquid
leads meeting at a junction containing a QD with an on-site Coulomb
repulsion. In addition to the QD, the junction comprises an arbitrary
configuration of non-interacting tight-binding sites. The full Hamiltonian
contains a non-interacting part, a QD part, and a coupling term between the
two:

\begin{subequations}
\begin{equation}
H=H_{0}+H_{T}+H_{d}\text{.}
\end{equation}

The non-interacting part is made up of two terms,

\begin{equation}
H_{0}=H_{0\text{,leads}}+H_{0\text{,junction}}\text{;}
\end{equation}%
the lead term

\begin{equation}
H_{0\text{,leads}}=-t\sum_{j=1}^{N}\sum_{n=0}^{\infty }\left( c_{j,n}^{\dag
}c_{j,n+1}+\text{h.c.}\right)
\end{equation}%
models the Fermi liquid leads as semi-infinite nearest-neighbor
tight-binding chains with hopping $t$, where $j$ is the lead index and $n$
is the site index. For simplicity all leads are assumed to be identical, and
we temporarily suppress the spin index. $H_{0\text{,junction}}$ is the
non-interacting part of the junction; it glues all leads together and often
includes additional sites (e.g. representing the arms of an ABK ring), but
does not include coupling to the interacting QD. In a typical open ABK ring
with electron leakage, two of the leads serve as source and drain
electrodes, while the remaining $N-2$ leads mimic the base contacts thorough
which electrons escape the junction. In experiments usually the current
flowing through the source or the drain is monitored, but the leakage
current can also be measured.

Assume that there are $M$ sites in the junction to which the QD is directly
coupled; hereafter we refer to these sites as the coupling sites. The
coupling to the QD can be written as

\begin{equation}
H_{T}=-\sum_{r=1}^{M}\left( t_{r}c_{C,r}^{\dag }d+\text{h.c.}\right)
\label{HT0}
\end{equation}%
where $d$ annihilates an electron on the QD, and $c_{C,r}^{\dag }$ creates
an electron on the $r$th coupling site. $c_{C,r}$ may coincide with $c_{j,0}$%
. In the simplest AB ring, there is only one physical AB phase, which may be
incorporated in either $H_{0\text{,junction}}$ or $H_{T}$. In more
complicated models both $H_{0\text{,junction}}$ and $H_{T}$ can depend on AB
phases.

Finally, the Hamiltonian of the interacting QD is given by

\begin{equation}
H_{d}=\epsilon _{d}d^{\dag }d+Un_{d\uparrow }n_{d\downarrow }
\end{equation}%
where $n_{d\sigma }=d_{\sigma }^{\dag }d_{\sigma }$. We assume $SU\left(
2\right) $ spin symmetry throughout the paper.

A generic system with $N=5$ and $M=3$ is sketched in Fig.~\ref{fig:sketch},
with details of the mesoscopic junction hidden. We will analyze more
concrete examples in Secs.~\ref{sec:closed} and \ref{sec:open}; additional
examples, previously studied in Refs.~%
\onlinecite{PhysRevB.82.165426,PhysRevB.88.245104,PhysRevLett.89.206602},
are provided in Appendix~\ref{sec:appcomp}.

\begin{figure}[ptb]
\includegraphics[width=0.6\textwidth]{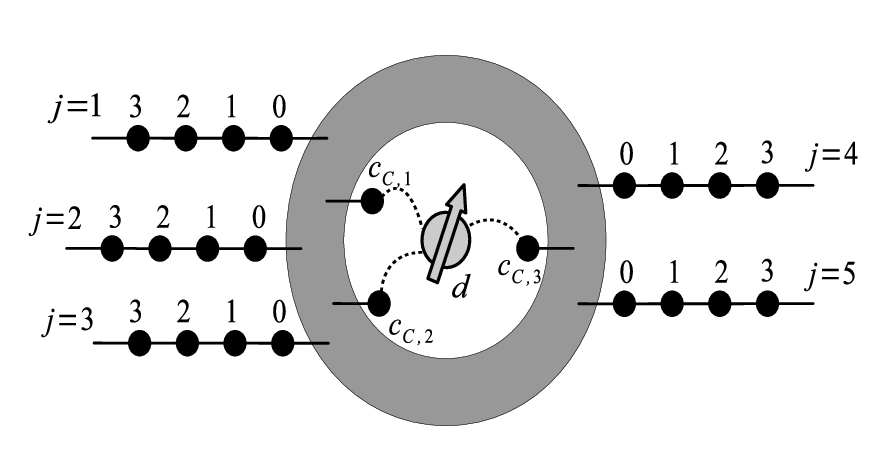}%
\caption{Sketch of a
generic system which allows the application of our formalism. Here $N=5$ and
$M=3$.\label{fig:sketch}}
\end{figure}

\subsection{Screening and non-screening channels}

While it is $H_{0\text{,junction}}$\ that ultimately determines the
properties of the junction, its details are actually not important in our
formalism. Instead, in the following we characterize the model by its
background scattering S-matrix and coupling site wave functions. Both
quantities are easily obtained from a given $H_{0\text{,junction}}$, and as
we show in Sec.~\ref{sec:conductance}, they play a central role in our quest
for the linear DC conductance.

To recast our model into the standard form of an interacting QD coupled to a
continuum of states, it is convenient to first diagonalize the
non-interacting part of the Hamiltonian $H_{0}$ by introducing the
scattering basis $q_{j,k}$:

\end{subequations}
\begin{equation}
H_{0}=\int_{0}^{\pi }\frac{dk}{2\pi }\sum_{j=1}^{N}\epsilon
_{k}q_{j,k}^{\dag }q_{j,k}\text{,}
\end{equation}%
where $\epsilon _{k}=-2t\cos k$ is the dispersion relation in the leads, and
for simplicity we let the lattice constant $a=1$. In addition to the
scattering states $q_{j,k}$, there may exist a number of bound states with
their energies outside of the continuum, but since their wave functions
decay exponentially away from the junction region, they do not affect linear
DC transport properties.

The scattering basis operator $q_{j,k}$ annihilates a scattering state
electron incident from lead $j$ with momentum $k$, and obeys the
anti-commutation relation $\left\{ q_{j,k},q_{j^{\prime },k^{\prime }}^{\dag
}\right\} =2\pi \delta _{jj^{\prime }}\delta \left( k-k^{\prime }\right) $.
The corresponding wave function has the following form on site $n$ in lead $%
j^{\prime }$,

\begin{subequations}
\begin{equation}
\chi _{j,k}\left( j^{\prime },n\right) =\delta _{jj^{\prime
}}e^{-ikn}+S_{j^{\prime }j}\left( k\right) e^{ikn}\text{;}  \label{bgwflead}
\end{equation}%
and on coupling site $r$,

\begin{equation}
\chi _{j,k}\left( r\right) =\Gamma _{rj}\left( k\right) \text{.}
\end{equation}%
In other words, for an electron incident from lead $j$, $S_{j^{\prime }j}$
is the background reflection or transmission amplitude in lead $j^{\prime }$%
, and $\Gamma _{rj}$ is the wave function on coupling site $r$. The
scattering S-matrix $S$ is unitary: $S^{\dag }S=1$.

From its wave function, $q_{j,k}$ can be related to $c_{j,n}$ and $c_{C,r}$:

\end{subequations}
\begin{subequations}
\begin{equation}
c_{j,n}=\int_{0}^{\pi }\frac{dk}{2\pi }\sum_{j^{\prime }=1}^{N}\left[ \delta
_{jj^{\prime }}e^{-ikn}+S_{jj^{\prime }}\left( k\right) e^{ikn}\right]
q_{j^{\prime },k}\text{,}  \label{leadsb}
\end{equation}%
and

\begin{equation}
c_{C,r}=\int_{0}^{\pi }\frac{dk}{2\pi }\sum_{j^{\prime }=1}^{N}\Gamma
_{rj^{\prime }}\left( k\right) q_{j^{\prime },k}\text{.}  \label{cplsitesb}
\end{equation}%
We now express $H_{T}$ in the scattering basis. Inserting Eq.~(\ref%
{cplsitesb}) into Eq.~(\ref{HT0}), we find the QD is only coupled to one
channel in the continuum, i.e. the screening channel:

\end{subequations}
\begin{equation}
H_{T}=-\int_{0}^{\pi }\frac{dk}{2\pi }V_{k}\left( \psi _{k}^{\dag }d+\text{%
h.c.}\right) \text{,}  \label{Andersonmodel}
\end{equation}%
where the screening channel operator $\psi _{k}^{\dag }$ is defined by

\begin{equation}
\psi _{k}^{\dag }=\frac{1}{V_{k}}\sum_{j=1}^{N}\sum_{r=1}^{M}t_{r}\Gamma
_{r,j}^{\ast }\left( k\right) q_{j,k}^{\dag }\text{,}
\end{equation}%
and the normalization factor $V_{k}>0$ is defined by

\begin{equation}
V_{k}^{2}=\sum_{j=1}^{N}\sum_{r,r^{\prime }=1}^{M}t_{r}t_{r^{\prime }}^{\ast
}\Gamma _{rj}^{\ast }\left( k\right) \Gamma _{r^{\prime }j}\left( k\right) =%
\tr\left[ \Gamma ^{\dag }\left( k\right) \lambda \Gamma \left( k\right) %
\right] \text{.}  \label{Vksquared}
\end{equation}%
This ensures $\left\{ \psi _{k},\psi _{k^{\prime }}^{\dag }\right\} =2\pi
\delta \left( k-k^{\prime }\right) $. Here we also introduce the $M\times M$
Hermitian QD coupling matrix $\lambda $,

\begin{equation}
\lambda_{rr^{\prime}}=t_{r}t_{r^{\prime}}^{\ast}\text{.}
\label{couplingmatlambda}
\end{equation}

It will be useful to define a series of non-screening channels $\phi _{l,k}$
orthogonal to $\psi $, where $l=1$, $\cdots $, $N-1$. The $\phi $ channels
are decoupled from the QD. In a compact notation we can write the
transformation from the scattering basis to the screening--non-screening
basis as

\begin{equation}
\Psi _{k}\equiv \left( 
\begin{array}{c}
\psi _{k} \\ 
\phi _{1,k} \\ 
\cdots \\ 
\phi _{N-1,k}%
\end{array}%
\right) =\mathbb{U}_{k}\left( 
\begin{array}{c}
q_{1,k} \\ 
q_{2,k} \\ 
\cdots \\ 
q_{N,k}%
\end{array}%
\right) \text{,}
\end{equation}%
where $\mathbb{U}$ is a unitary matrix. The first row of $\mathbb{U}$\ is
known:

\begin{equation}
\mathbb{U}_{1,j,k}=\frac{1}{V_{k}}\sum_{r=1}^{M}t_{r}^{\ast }\Gamma
_{rj}\left( k\right) \text{.}  \label{bbUmatrow1}
\end{equation}%
As long as $\mathbb{U}$ stays unitary, its remaining matrix elements can be
chosen freely without affecting physical observables. $\Psi _{k}$ now also
diagonalizes $H_{0}$,

\begin{equation}
H_{0}=\int_{0}^{\pi }\frac{dk}{2\pi }\epsilon _{k}\left( \psi _{k}^{\dag
}\psi _{k}+\sum_{l=1}^{N-1}\phi _{l,k}^{\dag }\phi _{l,k}\right) \text{;}
\end{equation}%
we shall also need the inverse transformation,

\begin{equation}
q_{j,k}=\mathbb{U}_{1,j,k}^{\ast }\psi _{k}+\sum_{l=1}^{N-1}\mathbb{U}%
_{l+1,j,k}^{\ast }\phi _{l,k}\text{.}  \label{scattoscrn}
\end{equation}

\subsection{Kondo model}

In the local moment regime of the Anderson model,\cite%
{ProgTheorPhys.32.37,hewson1997kondo} for $T\ll U$ we can perform the
Schrieffer-Wolff transformation\cite{PhysRev.149.491} on $\psi $ to obtain
an effective Kondo model with a reduced bandwidth and a momentum-dependent
coupling:

\begin{subequations}
\begin{equation}
H=H_{0}+\int \frac{dkdk^{\prime }}{\left( 2\pi \right) ^{2}}\left(
J_{kk^{\prime }}\psi _{k}^{\dag }\frac{\vec{\sigma}}{2}\psi _{k^{\prime
}}\cdot \vec{S}_{d}+K_{kk^{\prime }}\psi _{k}^{\dag }\psi _{k^{\prime
}}\right) \text{,}  \label{Kondomodel}
\end{equation}%
where all momenta are between $k_{F}-\Lambda _{0}$ and $k_{F}+\Lambda _{0}$
with Fermi wave vector $k_{F}$, $0<\Lambda _{0}\ll k_{F}$ is the initial
momentum cutoff, and the dispersion is linearized near the Fermi energy as $%
\epsilon _{k}=v_{F}\left( k-k_{F}\right) $. For the tight-binding model $%
v_{F}=2t\sin k_{F}$.

The interaction consists of a spin-flip term $J$,

\begin{equation}
J_{kk^{\prime}}=V_{k}V_{k^{\prime}}\left( j_{k}+j_{k^{\prime}}\right) \text{,%
}  \label{KondoJ}
\end{equation}

\begin{equation}
j_{k}=\frac{1}{\epsilon _{k}-\epsilon _{d}}+\frac{1}{U+\epsilon
_{d}-\epsilon _{k}}\approx j\text{,}
\end{equation}%
and a particle-hole symmetry breaking potential scattering term $K$,

\begin{equation}
K_{kk^{\prime}}=\frac{1}{4}V_{k}V_{k^{\prime}}\left(
\kappa_{k}+\kappa_{k^{\prime}}\right) \text{,}  \label{KondoK}
\end{equation}

\begin{equation}
\kappa _{k}=\frac{1}{\epsilon _{k}-\epsilon _{d}}-\frac{1}{U+\epsilon
_{d}-\epsilon _{k}}\approx \kappa \text{.}
\end{equation}%
The energy cutoff is initially $D_{0}\equiv v_{F}\Lambda _{0}$. When we
reduce the running energy cutoff from $D$ to $D+dD$\ ($0<-dD\ll D$)\ to
integrate out the high-energy degrees of freedom in the narrow strips of
energy $\left( -D,-D-dD\right) $ and $\left( D+dD,D\right) $, $K$ is exactly
marginal in the renormalization group (RG) sense, whereas $J$ is marginally
relevant and obeys the following RG equation:

\end{subequations}
\begin{equation}
-\frac{d\left( \nu J_{kk^{\prime }}\right) }{d\ln D}=\frac{1}{2}\left( \nu
J_{k,k_{F}+\Lambda }\nu J_{k_{F}+\Lambda ,k^{\prime }}+\nu
J_{k,k_{F}-\Lambda }\nu J_{k_{F}-\Lambda ,k^{\prime }}\right) \text{,}
\label{KondoJkkprRG}
\end{equation}%
or equivalently

\begin{equation}
-\frac{dj}{d\ln D}=\nu j^{2}\left( V_{k_{F}+\Lambda }^{2}+V_{k_{F}-\Lambda
}^{2}\right) \text{,}  \label{KondojRG}
\end{equation}%
where $\nu $ is the density of states per channel per spin, $\nu =1/\left(
2\pi v_{F}\right) $. Therefore, renormalization of the Kondo coupling is
controlled by the momentum-dependent normalization factor $V_{k}^{2}$,
defined in Eq.~(\ref{Vksquared}). $j$ is the only truly independently
renormalized coupling constant despite the appearance of Eq.~(\ref%
{KondoJkkprRG}); this follows from the fact that the screening channel is
the only channel coupled to the QD.\cite{PhysRevLett.89.206602}

The prototype Kondo model possesses a momentum-independent coupling
function, $J_{kk^{\prime }}\approx 2jV_{k_{F}}^{2}$. As a result,
spin-charge separation occurs and the Kondo interaction is found to be
exclusively in the spin sector.\cite{PhysRevB.48.7297} The charge sector is
nothing but a non-interacting theory with a particle-hole symmetry breaking
phase shift due to the potential scattering term $K$, while at very low
energy scales the spin sector renormalizes to a local Fermi liquid theory
with $\pi /2$ phase shift.

On the other hand, in a mesoscopic geometry $V_{k}^{2}$ often exhibit
fluctuations on a mesoscopic energy scale $E_{V}$. (More precisely, $E_{V}$
can be defined as the energy corresponding to the largest Fourier component
in the spectrum of $V_{k}^{2}$, but for both specific models discussed in
this paper we can simply read it off the analytic expression.) In the
presence of a characteristic length scale $L$, $E_{V}$ may be of the order
of the Thouless energy $v_{F}/L$, as is the case for\ the closed long ABK
ring in\ Sec.~\ref{sec:closed}; however this is not always true, with a
counterexample provided by the open long ABK ring in Sec.~\ref{sec:open}
where $E_{V}$ is of the order of the bandwidth $4t$. Well above $E_{V}$, $%
V_{k}^{2}$ appears featureless and can be approximated by its mean value $%
\overline{V_{k}^{2}}$ with respect to $k$.\ The Kondo temperature $T_{K}$
can be loosely defined as the energy cutoff at which the dimensionless
coupling $2\nu j\overline{V_{k}^{2}}$ becomes $O\left( 1\right) $. As
briefly sketched in Sec.~\ref{sec:intro}, there are two very different
parameter regimes of the Kondo temperature:\cite%
{PhysRevLett.89.206602,PhysRevB.83.165310}

a) The small Kondo cloud regime $T_{K}\gg E_{V}$. For $E_{V}\sim v_{F}/L$,
the size of the Kondo screening cloud $L_{K}\equiv v_{F}/T_{K}\ll L$; hence
the name. In this regime, the bare Kondo coupling is sufficiently large, so
that $2\nu j\overline{V_{k}^{2}}$ renormalizes to $O\left( 1\right) $ before
it \textquotedblleft senses\textquotedblright\ any mesoscopic fluctuation.
By approximating $V_{k}^{2}\approx \overline{V_{k}^{2}}$,\ Eq.~(\ref%
{KondojRG}) has a solution

\begin{equation}
j\left( D\right) \approx \frac{j_{0}}{1+2\nu j_{0}\overline{V_{k}^{2}}\ln 
\frac{D}{D_{0}}}\text{,}  \label{KondojRGsolsc}
\end{equation}%
where $j_{0}$ is the bare Kondo coupling constant at the initial energy
cutoff $D_{0}$. Eq.~(\ref{KondojRGsolsc}) gives the \textquotedblleft
background\textquotedblright\ Kondo temperature

\begin{equation}
T_{K}\approx T_{K}^{0}\equiv D_{0}\exp \left( -\frac{1}{2\nu j_{0}\overline{%
V_{k}^{2}}}\right) \text{,}  \label{TKKondobulk}
\end{equation}%
independent of the mesoscopic details of the geometry. For $T\ll T_{K}$, the
low-energy effective theory is also conjectured to be a Fermi liquid, but
the T-matrix (or the phase shift) of the screening channel is not yet known
with certainty.\cite{PhysRevLett.89.206602,EurophysLett.97.17006}

b) The very large Kondo cloud regime $T_{K}\ll E_{V}$. For $E_{V}\sim v_{F}/L
$, $L_{K}\gg L$. In this regime, the bare Kondo coupling is very small, and $%
j$ does not begin to renormalize significantly until the energy cutoff is
well below $E_{V}$.\ The variation of $V_{k}^{2}$ is hence negligible in the
resulting low energy theory, $V_{k}^{2}\approx V_{k_{F}}^{2}$, but $%
V_{k_{F}}^{2}$ may be significantly different from $\overline{V_{k}^{2}}$,
which means Kondo temperature is thus highly sensitive to the mesoscopic
details of $V_{k_{F}}^{2}$. Because $V_{k}^{2}$ is almost independent of $k$%
, we may map the low-energy theory in question onto the conventional Kondo
model, where conduction electrons are scattered by a point-like spin in real
space. (We stress that this mapping would not be possible for a strongly $k$%
-dependent $V_{k}^{2}$, which is the case for the small cloud
regime.) Following well-known results in the conventional Kondo model,\cite%
{hewson1997kondo} we see that the low-energy effective theory is a local Fermi liquid theory,
with parameters also sensitive to mesoscopic details.

\section{Linear DC conductance\label{sec:conductance}}

\begin{figure}[ptb]
\includegraphics[width=0.3\textwidth]{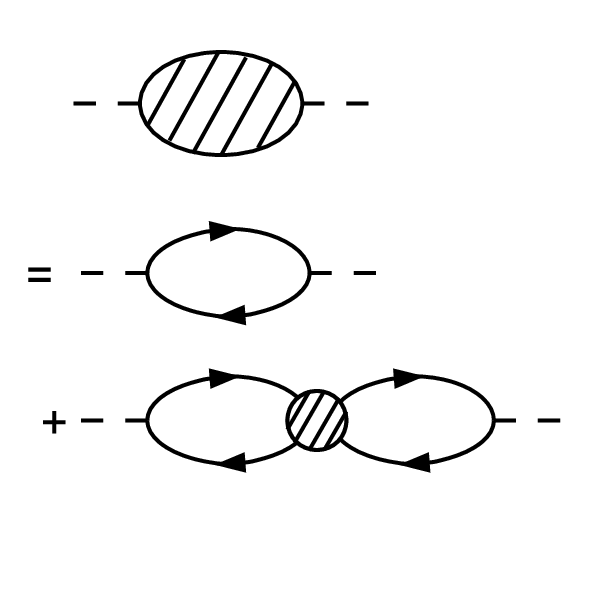}\caption{Disconnected
(self-energy) and connected (vertex correction) contributions to the
density-density correlation function, which is directly related to the
conductance through the Kubo formula Eq.~(\ref{Kubo}). The dashed lines
represent external legs at times $\bar{t}$ and $0$, the solid lines
represent fully dressed $\Psi$ fermion propagators, and the hatched circle
represents all connected 4-point vertices of the screening
channel.\label{fig:discconn}}
\end{figure}

In this section we calculate the DC conductance tensor of the system in
linear response theory, generalizing the results in Ref.~%
\onlinecite{PhysRevB.88.245104} to our multi-terminal setup. The result is
presented as the sum of a disconnected contribution and a connected one
(Fig.~\ref{fig:discconn}). By \textquotedblleft
disconnected\textquotedblright\ and \textquotedblleft
connected\textquotedblright , we are referring to the topology of the
corresponding Feynman diagrams: a disconnected contribution originates from
a Feynman diagram without any cross-links, and can always be written as the
product of two two-point functions. The disconnected contribution has a
simple Landauer form, and is quadratic in the T-matrix of the screening
channel $\psi $. The connected contribution is also shown to depend on
properties near the Fermi surface only, but it is usually difficult to
evaluate analytically except at high temperatures, or at low temperatures if
the Fermi-liquid perturbation theory is applicable. Nevertheless, just as
with the short ABK ring, the connected contribution can be approximately
eliminated at temperatures low compared to another mesoscopic energy scale $%
T\ll E_{\text{conn}}$.

\subsection{Kubo formula in terms of screening and non-screening channels}

The linear DC conductance tensor $G_{jj^{\prime }}$ is defined through $%
\left\langle I_{j}\right\rangle =\sum_{j^{\prime }}G_{jj^{\prime
}}V_{j^{\prime }}$, where $I_{j}$ is the current operator in lead $j$, and $%
V_{j^{\prime }}$ is the applied bias voltage on lead $j^{\prime }$. $I_{j}$\
is given by $I_{j}=-edN_{j}/d\bar{t}$, where $\bar{t}$ is the real time
variable, and

\begin{equation}
N_{j}\equiv \sum_{n=0}^{\infty }c_{j,n}^{\dag }c_{j,n}  \label{leaddensity0}
\end{equation}%
is the density operator in lead $j$. $G_{jj^{\prime }}$\ is then given by
the Kubo formula

\begin{equation}
G_{jj^{\prime }}=\frac{e^{2}}{h}\lim_{\Omega \rightarrow 0}\left( 2\pi
i\Omega \right) G_{jj^{\prime }}^{\prime }\left( \Omega \right) \text{,}
\label{Kubo}
\end{equation}%
where the retarded density-density correlation function is

\begin{equation}
G_{jj^{\prime }}^{\prime }\left( \Omega \right) \equiv -i\int_{0}^{\infty }d%
\bar{t}e^{i\Omega ^{+}\bar{t}}\left\langle \left[ N_{j}\left( \bar{t}\right)
,N_{j^{\prime }}\left( 0\right) \right] \right\rangle  \label{GjjprOmega}
\end{equation}%
and $\Omega ^{+}\equiv \Omega +i0^{+}$. The retarded correlation function
can be obtained by means of analytic continuation $i\omega _{p}\rightarrow
\Omega ^{+}$ from its imaginary time counterpart,

\begin{equation}
\mathcal{G}_{jj^{\prime }}^{\prime }\left( i\omega _{p}\right)
=-\int_{0}^{\beta }d\tau e^{i\omega _{p}\tau }\left\langle T_{\tau
}N_{j}\left( \tau \right) N_{j^{\prime }}\left( 0\right) \right\rangle \text{%
,}  \label{Gjjpriomegap}
\end{equation}%
where $\omega _{p}=2p\pi /\beta $ is a bosonic Matsubara frequency, $\beta
=1/T$ is the inverse temperature, and $T_{\tau }$ is the imaginary
time-ordering operator.

To calculate the correlation function we need the density operator $N_{j}$
written as bilinears of $\Psi $. This is achieved by the insertion of Eq.~(%
\ref{leadsb}) and then Eq.~(\ref{scattoscrn}) into Eq.~(\ref{leaddensity0}).
We find

\begin{equation}
N_{j}=\int_{0}^{\pi }\frac{dk_{1}dk_{2}}{\left( 2\pi \right) ^{2}}\Psi
_{k_{1}}^{\dag }\mathbb{M}_{k_{1}k_{2}}^{j}\Psi _{k_{2}}\text{,}
\label{leaddensity}
\end{equation}%
where for $l_{1}$, $l_{2}=1$, ..., $N$,

\begin{align}
\left( \mathbb{M}_{k_{1}k_{2}}^{j}\right) _{l_{1}l_{2}}& =\sum_{j_{1}j_{2}}%
\mathbb{U}_{l_{1},j_{1},k_{1}}\mathbb{U}_{l_{2},j_{2},k_{2}}^{\ast }\left[
\delta _{jj_{1}}\delta _{jj_{2}}\frac{1}{1-e^{i\left( k_{1}-k_{2}+i0\right) }%
}+S_{jj_{1}}^{\ast }\left( k_{1}\right) \delta _{jj_{2}}\frac{1}{%
1-e^{-i\left( k_{1}+k_{2}-i0\right) }}\right.  \notag \\
& \left. +\delta _{jj_{1}}S_{jj_{2}}\left( k_{2}\right) \frac{1}{%
1-e^{i\left( k_{1}+k_{2}+i0\right) }}+S_{jj_{1}}^{\ast }\left( k_{1}\right)
S_{jj_{2}}\left( k_{2}\right) \frac{1}{1-e^{i\left( k_{2}-k_{1}+i0\right) }}%
\right] \text{.}  \label{bbMmat}
\end{align}%
The matrix $\mathbb{M}$ obeys $\mathbb{M}_{k_{1}k_{2}}^{j}=\left( \mathbb{M}%
_{k_{2}k_{1}}^{j}\right) ^{\dag }$ which ensures $N_{j}$ is Hermitian.

\subsection{Disconnected part}

We substitute Eq.~(\ref{leaddensity}) into Eq.~(\ref{GjjprOmega}). The
disconnected part of the conductance is obtained by pairing up $\Psi $ and $%
\Psi ^{\dag }$ operators to form two two-point Green's functions:

\begin{equation}
G_{jj^{\prime }}^{\prime D}\left( \Omega \right) =-2i\int_{0}^{\infty }d\bar{%
t}e^{i\Omega ^{+}\bar{t}}\int_{0}^{\pi }\frac{dk_{1}dk_{2}}{\left( 2\pi
\right) ^{2}}\frac{dq_{1}dq_{2}}{\left( 2\pi \right) ^{2}}\tr \left[ \mathbb{%
M}_{k_{1}k_{2}}^{j}\mathbb{G}_{k_{2}q_{1}}^{>}\left( \bar{t}\right) \mathbb{M%
}_{q_{1}q_{2}}^{j^{\prime }}\mathbb{G}_{q_{2}k_{1}}^{<}\left( -\bar{t}%
\right) -\mathbb{M}_{q_{1}q_{2}}^{j^{\prime }}\mathbb{G}_{q_{2}k_{1}}^{>}%
\left( -\bar{t}\right) \mathbb{M}_{k_{1}k_{2}}^{j}\mathbb{G}%
_{k_{2}q_{1}}^{<}\left( \bar{t}\right) \right] \text{.}  \label{GjjprDgl}
\end{equation}%
Here the factor of $2$ in the second line is due to the spin degeneracy. The
greater and lesser Green's functions in the screening--non-screening basis
are defined as

\begin{subequations}
\begin{equation}
\mathbb{G}_{kq}^{>}\left( \bar{t}\right) \equiv-i\left( 
\begin{array}{cccc}
\left\langle \psi_{k}\left( \bar{t}\right) \psi_{q}^{\dag}\left( 0\right)
\right\rangle &  &  &  \\ 
& \left\langle \phi_{1,k}\left( \bar{t}\right) \phi_{1,q}^{\dag}\left(
0\right) \right\rangle &  &  \\ 
&  & \cdots &  \\ 
&  &  & \left\langle \phi_{N-1,k}\left( \bar{t}\right) \phi_{N-1,q}^{\dag
}\left( 0\right) \right\rangle%
\end{array}
\right) \text{,}
\end{equation}
and

\begin{equation}
\mathbb{G}_{kq}^{<}\left( \bar{t}\right) \equiv +i\left( 
\begin{array}{cccc}
\left\langle \psi _{q}^{\dag }\left( 0\right) \psi _{k}\left( \bar{t}\right)
\right\rangle &  &  &  \\ 
& \left\langle \phi _{1,q}^{\dag }\left( 0\right) \phi _{1,k}\left( \bar{t}%
\right) \right\rangle &  &  \\ 
&  & \cdots &  \\ 
&  &  & \left\langle \phi _{N-1,q}^{\dag }\left( 0\right) \phi
_{N-1,k}\left( \bar{t}\right) \right\rangle%
\end{array}%
\right) \text{.}
\end{equation}%
In equilibrium, fluctuation-dissipation theorem requires that $\mathbb{G}%
_{kq}^{>}\left( \omega \right) =2i\left[ 1-f\left( \omega \right) \right] 
\operatorname{Im} \mathbb{G}_{kq}^{R}\left( \omega \right) $, and $\mathbb{G}%
_{kq}^{<}\left( \omega \right) =-2if\left( \omega \right) \operatorname{Im} \mathbb{G%
}_{kq}^{R}\left( \omega \right) $, where $f\left( \omega \right) =1/\left(
e^{\beta \omega }+1\right) $ is the Fermi function. These equilibrium
relations result from the fact that $\mathbb{G}_{kq}^{R}\left( \omega
\right) =\mathbb{G}_{qk}^{R}\left( \omega \right) $ for the Anderson model
[see Eq.~(\ref{Dyson}) below].\cite{PhysRevB.88.245104} With these relations
Eq.~(\ref{GjjprDgl}) becomes

\end{subequations}
\begin{equation}
G_{jj^{\prime }}^{\prime D}\left( \Omega \right) =8\int \frac{d\omega
d\omega ^{\prime }}{\left( 2\pi \right) ^{2}}\frac{f\left( \omega \right)
-f\left( \omega ^{\prime }\right) }{\omega -\omega ^{\prime }+\Omega ^{+}}%
\int_{0}^{\pi }\frac{dk_{1}dk_{2}}{\left( 2\pi \right) ^{2}}\frac{%
dq_{1}dq_{2}}{\left( 2\pi \right) ^{2}}\tr \left[ \mathbb{M}_{k_{1}k_{2}}^{j}%
\operatorname{Im} \mathbb{G}_{k_{2}q_{1}}^{R}\left( \omega ^{\prime }\right) \mathbb{%
M}_{q_{1}q_{2}}^{j^{\prime }}\operatorname{Im} \mathbb{G}_{q_{2}k_{1}}^{R}\left(
\omega \right) \right] \text{.}  \label{GjjprDImIm}
\end{equation}%
We note that, in contrast to the case of Ref.~\onlinecite{PhysRevB.88.245104}%
, the momentum integral here is not necessarily real. Instead, its complex
conjugate takes the same form but with $\omega $ and $\omega ^{\prime }$
interchanged:

\begin{align}
& \int_{0}^{\pi }\frac{dk_{1}dk_{2}}{\left( 2\pi \right) ^{2}}\frac{%
dq_{1}dq_{2}}{\left( 2\pi \right) ^{2}}\tr \left[ \mathbb{M}_{k_{1}k_{2}}^{j}%
\operatorname{Im} \mathbb{G}_{k_{2}q_{1}}^{R}\left( \omega ^{\prime }\right) \mathbb{%
M}_{q_{1}q_{2}}^{j^{\prime }}\operatorname{Im} \mathbb{G}_{q_{2}k_{1}}^{R}\left(
\omega \right) \right] ^{\ast }  \notag \\
& =\int_{0}^{\pi }\frac{dk_{1}dk_{2}}{\left( 2\pi \right) ^{2}}\frac{%
dq_{1}dq_{2}}{\left( 2\pi \right) ^{2}}\tr \left[ \mathbb{M}_{k_{1}k_{2}}^{j}%
\operatorname{Im} \mathbb{G}_{k_{2}q_{1}}^{R}\left( \omega \right) \mathbb{M}%
_{q_{1}q_{2}}^{j^{\prime }}\operatorname{Im} \mathbb{G}_{q_{2}k_{1}}^{R}\left(
\omega ^{\prime }\right) \right] \text{.}
\end{align}%
Making use of this property, we can show that $\left[ G_{jj^{\prime
}}^{\prime D}\left( -\Omega \right) \right] ^{\ast }=G_{jj^{\prime
}}^{\prime D}\left( \Omega \right) $. Thus the disconnected contribution to
the DC conductance can be written as

\begin{equation}
G_{jj^{\prime }}^{D}=\frac{e^{2}}{h}\lim_{\Omega \rightarrow 0}\left( -2\pi
\Omega \right) \operatorname{Im}G_{jj^{\prime }}^{\prime D}\left( \Omega \right) 
\text{.}
\end{equation}%
We should realize, however, that taking the imaginary part of $G_{jj^{\prime
}}^{\prime D}$ is generally not equivalent to taking the $\delta $-function
part of $1/\left( \omega -\omega ^{\prime }+\Omega ^{+}\right) $\ in Eq.~(%
\ref{GjjprDImIm}).

For the Anderson model, it is not difficult to find the Dyson's equation for
the retarded Green's function by the equation-of-motion technique:

\begin{equation}
\mathbb{G}_{k_{2}q_{1}}^{R}\left( \omega \right) =2\pi \delta \left(
k_{2}-q_{1}\right) g_{k_{2}}^{R}\left( \omega \right) +\tau _{\psi
}g_{k_{2}}^{R}\left( \omega \right) \text{T}_{k_{2}q_{1}}\left( \omega
\right) g_{q_{1}}^{R}\left( \omega \right) \text{,}  \label{Dyson}
\end{equation}%
where the free retarded Green's function for $\psi $ and $\phi $ is

\begin{equation}
g_{k}^{R}\left( \omega \right) =\frac{1}{\omega ^{+}-\epsilon _{k}}\text{,}
\end{equation}%
and $\tau _{\psi }$ is the projection operator onto the screening channel
subspace. Again, only the Green's function of the screening channel is
modified by coupling to the QD. The retarded T-matrix of the screening
channel in the single-particle sector is related to the retarded two-point
function of the QD by

\begin{equation}
\text{T}_{k_{2}q_{1}}\left( \omega \right) =V_{k_{2}}G_{dd}^{R}\left( \omega
\right) V_{q_{1}}\text{,}
\end{equation}%
where $G_{dd}^{R}\left( \omega \right) \equiv -i\int_{0}^{\infty }d\bar{t}%
e^{i\omega ^{+}\bar{t}}\left\langle \left\{ d\left( \bar{t}\right) ,d^{\dag
}\left( 0\right) \right\} \right\rangle $.

From Eqs.~(\ref{GjjprDImIm}) and (\ref{Dyson}) we may express the
disconnected contribution to the linear DC conductance in the Landauer form:

\begin{equation}
G_{jj^{\prime }}^{D}=\frac{2e^{2}}{h}\int_{-2t}^{2t}d\epsilon _{p}\left[
-f^{\prime }\left( \epsilon _{p}\right) \right] \mathcal{T}_{jj^{\prime
}}^{D}\left( \epsilon _{p}\right) \text{,}  \label{GDTD}
\end{equation}%
where the disconnected \textquotedblleft transmission
probability\textquotedblright\ $\mathcal{T}_{jj^{\prime }}^{D}$\ is written
in terms of the absolute square of a \textquotedblleft transmission
amplitude\textquotedblright ,

\begin{eqnarray}
\mathcal{T}_{jj^{\prime }}^{D}\left( \epsilon _{p}\right) &=&\delta
_{jj^{\prime }}-\left\vert \left\{ S\left( p\right) \left[ 1-2i\pi \nu _{p}%
\text{T}_{pp}\left( \epsilon _{p}\right) \mathbb{U}_{p}^{\dag }\tau _{\psi }%
\mathbb{U}_{p}\right] \right\} _{jj^{\prime }}\right\vert ^{2}  \notag \\
&=&\delta _{jj^{\prime }}-\left\vert S_{jj^{\prime }}\left( p\right) +\frac{%
2i}{V_{p}^{2}}\left[ S\left( p\right) \Gamma ^{\dag }\left( p\right) \lambda
\Gamma \left( p\right) \right] _{jj^{\prime }}\left[ -\pi \nu _{p}\text{T}%
_{pp}\left( \epsilon _{p}\right) \right] \right\vert ^{2}\text{.}
\label{discprob}
\end{eqnarray}%
Again $\lambda $ is the QD coupling matrix defined in Eq.~(\ref%
{couplingmatlambda}) and $\nu _{p}$ is the density of states per channel per
spin for the tight-binding model

\begin{equation}
\nu _{p}=\frac{1}{4\pi t\sin p}\text{.}
\end{equation}%
The detailed derivation of Eq.~(\ref{discprob}) by contour methods is left
for Appendix \ref{sec:appdisc}. As a consistency check, we show in Appendix~%
\ref{sec:appNI} that for a non-interacting QD, $U=0$, Eq.~(\ref{discprob})
and solving the Schroedinger's equation yield the same transmission
probability.

At zero temperature, when the single-particle sector of the screening
channel T-matrix obeys the optical theorem and the inelastic part of the
T-matrix vanishes,\cite{PhysRevLett.93.107204} there is no connected
contribution and Eq.~(\ref{GDTD}) yields the full linear DC conductance.\cite%
{PhysRevB.88.245104} In this case a clear picture emerges from Eq.~(\ref%
{discprob}): the conductance is given by the Landauer formula with an
effective single-particle S-matrix, which is obtained from the original
S-matrix simply by imposing a phase shift on the screening channel,
corresponding to the particle-hole symmetry breaking potential scattering
and the elastic scattering by the Kondo singlet.\cite%
{PhysRevLett.61.1768,PhysRevB.82.165426}

Another useful representation of the disconnected probability, similar to
that in Ref.~\onlinecite{PhysRevB.88.245104}, is obtained by expanding Eq.~(%
\ref{discprob}):

\begin{subequations}
\begin{eqnarray}
\mathcal{T}_{jj^{\prime }}^{D}\left( \epsilon _{p}\right) &=&\mathcal{T}%
_{0,jj^{\prime }}\left( \epsilon _{p}\right) +\mathcal{Z}_{R,jj^{\prime
}}\left( \epsilon _{p}\right) \operatorname{Re} \left[ -\pi \nu _{p}\text{T}%
_{pp}\left( \epsilon _{p}\right) \right]  \notag \\
&&+\mathcal{Z}_{I,jj^{\prime }}\left( \epsilon _{p}\right) \operatorname{Im} \left[
-\pi \nu _{p}\text{T}_{pp}\left( \epsilon _{p}\right) \right] +\mathcal{Z}%
_{2,jj^{\prime }}\left( \epsilon _{p}\right) \left\vert -\pi \nu _{p}\text{T}%
_{pp}\left( \epsilon _{p}\right) \right\vert ^{2}\text{,}
\label{discprobKYA}
\end{eqnarray}%
with a background transmission term

\begin{equation}
\mathcal{T}_{0,jj^{\prime }}\left( \epsilon _{p}\right) =\delta _{jj^{\prime
}}-\left\vert S_{jj^{\prime }}\left( p\right) \right\vert ^{2}\text{,}
\label{discprobKYAbgt}
\end{equation}%
a term linear in the real part of the T-matrix, proportional to

\begin{equation}
\mathcal{Z}_{R,jj^{\prime }}\left( \epsilon _{p}\right) =\frac{4}{V_{p}^{2}}%
\operatorname{Im}\left\{ \left[ S\left( p\right) \Gamma ^{\dag }\left( p\right)
\lambda \Gamma \left( p\right) \right] _{jj^{\prime }}S_{jj^{\prime }}^{\ast
}\left( p\right) \right\} \text{,}  \label{discprobKYAZR}
\end{equation}%
a term linear in the imaginary part, proportional to

\begin{equation}
\mathcal{Z}_{I,jj^{\prime }}\left( \epsilon _{p}\right) =\frac{4}{V_{p}^{2}}%
\operatorname{Re}\left\{ \left[ S\left( p\right) \Gamma ^{\dag }\left( p\right)
\lambda \Gamma \left( p\right) \right] _{jj^{\prime }}S_{jj^{\prime }}^{\ast
}\left( p\right) \right\} \text{,}  \label{discprobKYAZI}
\end{equation}%
and a term quadratic in the T-matrix, proportional to

\begin{equation}
\mathcal{Z}_{2,jj^{\prime }}\left( \epsilon _{p}\right) =-\frac{4}{V_{p}^{4}}%
\left\vert \left[ S\left( p\right) \Gamma ^{\dag }\left( p\right) \lambda
\Gamma \left( p\right) \right] _{jj^{\prime }}\right\vert ^{2}\text{.}
\label{discprobKYAZ2}
\end{equation}

In the DC limit, the total current flowing out of the junction is zero, and
a uniform voltage applied to all leads does not result in any current; hence
the linear DC conductance satisfies current and voltage Kirchhoff's laws $%
\sum_{j}G_{jj^{\prime }}=\sum_{j^{\prime }}G_{jj^{\prime }}=0$. As a
comparison it is interesting to consider the sum of the disconnected
transmission probability, Eq.~(\ref{discprobKYA}), over $j$ or $j^{\prime }$%
. Using the unitarity of $S$ and Eq.~(\ref{couplingmatlambda}) it is not
difficult to find that

\end{subequations}
\begin{equation}
\sum_{j}\mathcal{T}_{jj^{\prime }}^{D}\left( \epsilon _{p}\right) =\left\{ 
\operatorname{Im} \left[ -\pi \nu _{p}\text{T}_{pp}\left( \epsilon _{p}\right) %
\right] -\left\vert -\pi \nu _{p}\text{T}_{pp}\left( \epsilon _{p}\right)
\right\vert ^{2}\right\} \frac{4}{V_{p}^{2}}\left[ \Gamma ^{\dag }\left(
p\right) \lambda \Gamma \left( p\right) \right] _{j^{\prime }j^{\prime }}%
\text{,}  \label{TDrowsum}
\end{equation}%
and

\begin{equation}
\sum_{j^{\prime }}\mathcal{T}_{jj^{\prime }}^{D}\left( \epsilon _{p}\right)
=\left\{ \operatorname{Im}\left[ -\pi \nu _{p}\text{T}_{pp}\left( \epsilon
_{p}\right) \right] -\left\vert -\pi \nu _{p}\text{T}_{pp}\left( \epsilon
_{p}\right) \right\vert ^{2}\right\} \frac{4}{V_{p}^{2}}\left[ S\left(
p\right) \Gamma ^{\dag }\left( p\right) \lambda \Gamma \left( p\right)
S^{\dag }\left( p\right) \right] _{jj}\text{.}  \label{TDcolsum}
\end{equation}%
As mentioned in Ref.~\onlinecite{PhysRevB.88.245104}, the quantity in curly
brackets in Eqs.~(\ref{TDrowsum}) and (\ref{TDcolsum}) measures the
deviation of the single-particle sector of the T-matrix from the optical
theorem.\cite{PhysRevLett.93.107204} In the case of a non-interacting QD or
the $T=0$ Fermi liquid theory of the Kondo limit, where the connected
contribution to the conductance vanishes, these row/column sum formulas
conform to our expectations: the T-matrix obeys the optical theorem, leading
to $\sum_{j}\mathcal{T}_{jj^{\prime }}^{D}=\sum_{j^{\prime }}\mathcal{T}%
_{jj^{\prime }}^{D}=0$, so that $\sum_{j}G_{jj^{\prime }}=\sum_{j^{\prime
}}G_{jj^{\prime }}=0$ is ensured.

\subsection{Connected part and its low-temperature elimination}

In this subsection we show that the connected contribution to the
conductance is again a Fermi surface contribution, and discuss how it can be
approximately eliminated at low temperatures. Following Ref.~%
\onlinecite{PhysRevB.88.245104} we construct a transmission probability for
this contribution. After a partial insertion of Eq.~(\ref{leaddensity}) into
Eq.~(\ref{Gjjpriomegap}), the connected part of the density-density
correlation function can be written as

\begin{equation}
\mathcal{G}_{jj^{\prime }}^{\prime C}\left( i\omega _{p}\right)
=\int_{0}^{\beta }d\tau e^{i\omega _{p}\tau }P_{jj^{\prime }}\left( \tau
,\tau \right) \text{,}
\end{equation}%
where the connected four-point function $P_{jj^{\prime }}$ with two temporal
arguments is

\begin{equation}
P_{jj^{\prime }}\left( \tau _{1},\tau _{2}\right) \equiv -\int_{0}^{\pi }%
\frac{dk_{1}dk_{2}}{\left( 2\pi \right) ^{2}}\left( \mathbb{M}%
_{k_{1}k_{2}}^{j}\right) _{11}\sum_{\sigma }\left\langle T_{\tau }\psi
_{k_{1}\sigma }^{\dag }\left( \tau _{1}\right) \psi _{k_{2}\sigma }\left(
\tau _{2}\right) N_{j^{\prime }}\left( 0\right) \right\rangle _{C}\text{;}
\end{equation}%
the subscript $C$ denotes connected diagrams. Note that only the screening
channel contributes to the connected part, as the non-screening channels are
free fermions. Using the equation-of-motion technique, it is easy to relate $%
P_{jj^{\prime }}$ to a partially amputated quantity:

\begin{equation}
P_{jj^{\prime}}\left( \tau,\tau\right) =\int_{0}^{\pi}\frac{dk_{1}dk_{2}}{%
\left( 2\pi\right) ^{2}}\left( \mathbb{M}_{k_{1}k_{2}}^{j}\right)
_{11}V_{k_{1}}V_{k_{2}}\int d\tau_{1}d\tau_{2}g_{k_{1}}\left(
\tau_{1}-\tau\right) g_{k_{2}}\left( \tau-\tau_{2}\right)
\sum_{\sigma}\left\langle T_{\tau}d_{\sigma}^{\dag}\left( \tau_{1}\right)
d_{\sigma}\left( \tau _{2}\right) N_{j^{\prime}}\left( 0\right)
\right\rangle _{C}\text{,}
\end{equation}
where

\begin{equation}
g_{k}\left( \tau \right) \equiv \left[ f\left( \tau \right) -\theta \left(
\tau \right) \right] e^{-\epsilon _{k}\tau }  \label{imagtimebaregf}
\end{equation}%
is the imaginary time free Green's function and $\theta \left( \tau \right) $
is the Heaviside unit-step function. With $\tau $ only appearing in free
propagators, we can perform the Fourier transform explicitly,

\begin{align}
\mathcal{G}_{jj^{\prime }}^{\prime C}\left( i\omega _{p}\right) & =\frac{1}{%
\beta }\sum_{\omega _{m}}P_{jj^{\prime }}\left( i\omega _{m},i\omega
_{m}+i\omega _{p}\right)  \notag \\
& =\frac{1}{\beta }\sum_{\omega _{m}}\int_{0}^{\pi }\frac{dk_{1}dk_{2}}{%
\left( 2\pi \right) ^{2}}\left( \mathbb{M}_{k_{1}k_{2}}^{j}\right)
_{11}g_{k_{1}}\left( i\omega _{m}\right) g_{k_{2}}\left( i\omega
_{m}+i\omega _{p}\right) V_{k_{1}}V_{k_{2}}  \notag \\
& \times \int d\tau _{1}d\tau _{2}e^{-i\omega _{m}\tau _{1}}e^{i\left(
\omega _{m}+\omega _{p}\right) \tau _{2}}\sum_{\sigma }\left\langle T_{\tau
}d_{\sigma }^{\dag }\left( \tau _{1}\right) d_{\sigma }\left( \tau
_{2}\right) N_{j^{\prime }}\left( 0\right) \right\rangle _{C}\text{.}
\end{align}%
One may now use the contour integration argument in Ref.~%
\onlinecite{PhysRevB.88.245104}.\cite{mahan2000many} The final result is
that the connected contribution to the DC conductance is expressed in terms
of a transmission probability $\mathcal{T}^{C}$ related to $P_{jj^{\prime }}$%
:

\begin{equation}
G_{jj^{\prime}}^{C}=\frac{2e^{2}}{h}\int_{-2t}^{2t}d\omega\left[ -f^{\prime
}\left( \omega\right) \right] \mathcal{T}_{jj^{\prime}}^{C}\left(
\omega\right) \text{,}
\end{equation}
where

\begin{equation}
\mathcal{T}_{jj^{\prime }}^{C}\left( \omega \right) =\lim_{\Omega
\rightarrow 0}\frac{\Omega ^{2}}{8}P_{jj^{\prime }}\left( \omega -i\eta
_{1},\omega +\Omega +i\eta _{2}\right) +\text{c.c.}
\end{equation}%
and

\begin{align}
& P_{jj^{\prime }}\left( \omega -i\eta _{1},\omega +\Omega +i\eta
_{2}\right)   \notag \\
& =\int_{0}^{\pi }\frac{dk_{1}dk_{2}}{\left( 2\pi \right) ^{2}}\left( 
\mathbb{M}_{k_{1}k_{2}}^{j}\right) _{11}g_{k_{1}}^{A}\left( \omega \right)
g_{k_{2}}^{R}\left( \omega +\Omega \right) V_{k_{1}}V_{k_{2}}\int d\tau
_{1}d\tau _{2}e^{-i\omega ^{-}\tau _{1}}e^{i\left( \omega ^{+}+\Omega
\right) \tau _{2}}\sum_{\sigma }\left\langle T_{\tau }d_{\sigma }^{\dag
}\left( \tau _{1}\right) d_{\sigma }\left( \tau _{2}\right) N_{j^{\prime
}}\left( 0\right) \right\rangle _{C}\text{.}
\end{align}%
Here $\eta _{1}$, $\eta _{2}\rightarrow 0^{+}$ are positive infinitesimal
numbers.

It is in fact possible to do the $k_{1}$ and $k_{2}$ integrals. Using Eqs.~(%
\ref{bbUmatrow1}), (\ref{bbMmat}) and finally (\ref{Gammaminusk}), we obtain

\begin{align}
& \int_{0}^{\pi }\frac{dk_{1}dk_{2}}{\left( 2\pi \right) ^{2}}\left( \mathbb{%
M}_{k_{1}k_{2}}^{j}\right) _{11}g_{k_{1}}^{A}\left( \epsilon _{p}\right)
g_{k_{2}}^{R}\left( \epsilon _{p}+\Omega \right) V_{k_{1}}V_{k_{2}}  \notag
\\
& =\sum_{r_{1}r_{2}}t_{r_{1}}^{\ast }t_{r_{2}}\int_{-\pi }^{\pi }\frac{%
dk_{1}dk_{2}}{\left( 2\pi \right) ^{2}}g_{k_{1}}^{A}\left( \epsilon
_{p}\right) g_{k_{2}}^{R}\left( \epsilon _{p}+\Omega \right) \Gamma
_{r_{1}j}\left( k_{1}\right) \Gamma _{r_{2}j}^{\ast }\left( k_{2}\right) 
\frac{1}{1-e^{i\left( k_{1}-k_{2}+i0\right) }}\text{.}
\end{align}%
Here domains of the momentum integrals are extended to $\left( -\pi ,\pi
\right) $ according to Eq.~(\ref{Gammaminusk}), which facilitates the
application of the residue method. As explained in Appendix~\ref{sec:appdisc}%
, the poles of $\Gamma \left( k_{1}\right) $ and $\Gamma ^{\ast }\left(
k_{2}\right) $ are not important in the DC limit $\Omega \rightarrow 0$.
Therefore, the $O\left( 1/\Omega \right) $ contribution is dominated by the
poles of the free Green's functions, and is given by

\begin{equation}
\int_{0}^{\pi }\frac{dk_{1}dk_{2}}{\left( 2\pi \right) ^{2}}\left( \mathbb{M}%
_{k_{1}k_{2}}^{j}\right) _{11}g_{k_{1}}^{A}\left( \epsilon _{p}\right)
g_{k_{2}}^{R}\left( \epsilon _{p}+\Omega \right) V_{k_{1}}V_{k_{2}}=\frac{%
2\pi i}{\Omega }\nu _{p}\left[ S\left( p^{\prime }\right) \Gamma ^{\dag
}\left( p^{\prime }\right) \lambda \Gamma \left( p\right) S^{\dag }\left(
p\right) \right] _{jj}+O\left( 1\right) \text{.}
\end{equation}%
where $\epsilon _{p}+\Omega \equiv \epsilon _{p^{\prime }}$, $0\leq p$, $%
p^{\prime }\leq \pi $. This leads to

\begin{equation}
\mathcal{T}_{jj^{\prime }}^{C}\left( \epsilon _{p}\right) =\nu _{p}\left[
S\left( p\right) \Gamma ^{\dag }\left( p\right) \lambda \Gamma \left(
p\right) S^{\dag }\left( p\right) \right] _{jj}\left[ \frac{i\pi }{4}%
\lim_{\Omega \rightarrow 0}\Omega \int d\tau _{1}d\tau _{2}e^{-i\omega
^{-}\tau _{1}}e^{i\left( \omega ^{+}+\Omega \right) \tau _{2}}\sum_{\sigma
}\left\langle T_{\tau }d_{\sigma }^{\dag }\left( \tau _{1}\right) d_{\sigma
}\left( \tau _{2}\right) N_{j^{\prime }}\left( 0\right) \right\rangle _{C}+%
\text{c.c.}\right] \text{.}
\end{equation}%
A similar manipulation can be done for the $N_{j^{\prime }}$ part of the
correlation function.

One can again consider the row and column sums of the tensor $\mathcal{T}%
^{C} $. Tracing over $j$ immediately yields

\begin{equation}
\sum_{j}\mathcal{T}_{jj^{\prime }}^{C}\left( \epsilon _{p}\right) =\nu
_{p}V_{p}^{2}\left[ \frac{i\pi }{4}\lim_{\Omega \rightarrow 0}\Omega \int
d\tau _{1}d\tau _{2}e^{-i\omega ^{-}\tau _{1}}e^{i\left( \omega ^{+}+\Omega
\right) \tau _{2}}\sum_{\sigma }\left\langle T_{\tau }d_{\sigma }^{\dag
}\left( \tau _{1}\right) d_{\sigma }\left( \tau _{2}\right) N_{j^{\prime
}}\left( 0\right) \right\rangle _{C}+\text{c.c.}\right] \text{;}
\end{equation}%
combining the last two equations, we have

\begin{equation}
\mathcal{T}_{jj^{\prime }}^{C}\left( \epsilon _{p}\right) -\frac{1}{V_{p}^{2}%
}\left[ S\left( p\right) \Gamma ^{\dag }\left( p\right) \lambda \Gamma
\left( p\right) S^{\dag }\left( p\right) \right] _{jj}\sum_{j^{\prime \prime
}}\mathcal{T}_{j^{\prime \prime }j^{\prime }}^{C}\left( \epsilon _{p}\right)
=0\text{.}  \label{connelimT}
\end{equation}

Let us now define $E_{\text{conn}}$ as the characteristic energy scale below
which both $S\left( p\right) $ and $\Gamma \left( p\right) $ vary slowly. By
definition $E_{\text{conn}}\lesssim E_{V}$; while $E_{\text{conn}}$ is not
necessarily the same as $E_{V}$, for the two ABK ring geometries considered
in this paper $E_{V}\sim E_{\text{conn}}$. For a mesoscopic structure with
characteristic length scale $L$, $E_{\text{conn}}$ is usually the Thouless
energy, $E_{\text{conn}}\sim v_{F}/L$; however, this is again not always the
case, and the open long ABK ring in Sec.~\ref{sec:open} provides a
counterexample where $E_{\text{conn}}$ is comparable to the bandwidth. Below 
$E_{\text{conn}}$, the function $\left[ S\left( p\right) \Gamma ^{\dag
}\left( p\right) \lambda \Gamma \left( p\right) S^{\dag }\left( p\right) %
\right] _{jj}/V_{p}^{2}$ is only weakly dependent on $p$.

Eq.~(\ref{connelimT}) suggests that we can approximately eliminate the
connected part of $G_{jj^{\prime }}$, provided the temperature is low
compared to $E_{\text{conn}}$.\cite{PhysRevB.88.245104} Consider the linear
combination

\begin{equation}
G_{jj^{\prime }}-\frac{1}{V_{k_{F}}^{2}}\left[ S\left( k_{F}\right) \Gamma
^{\dag }\left( k_{F}\right) \lambda \Gamma \left( k_{F}\right) S^{\dag
}\left( k_{F}\right) \right] _{jj}\sum_{j^{\prime \prime }}G_{j^{\prime
\prime }j^{\prime }}\equiv G_{jj^{\prime }}\text{;}
\end{equation}%
this corresponds to measuring the conductance by measuring the current in
lead $j$, plus a constant times the total current in all leads. (Note that
here we include both disconnected and connected contributions.) By
Kirchhoff's law, this linear combination must equal $G_{jj^{\prime }}$
itself. We write it as a sum of disconnected and connected contributions:

\begin{eqnarray}
G_{jj^{\prime }} &=&\left( G_{jj^{\prime }}^{D}-\frac{1}{V_{k_{F}}^{2}}\left[
S\left( k_{F}\right) \Gamma ^{\dag }\left( k_{F}\right) \lambda \Gamma
\left( k_{F}\right) S^{\dag }\left( k_{F}\right) \right] _{jj}\sum_{j^{%
\prime \prime }}G_{j^{\prime \prime }j^{\prime }}^{D}\right)  \notag \\
&&+\int d\epsilon _{p}\left[ -f^{\prime }\left( \epsilon _{p}\right) \right]
\left\{ \mathcal{T}_{jj^{\prime }}^{C}\left( \epsilon _{p}\right) -\frac{1}{%
V_{k_{F}}^{2}}\left[ S\left( k_{F}\right) \Gamma ^{\dag }\left( k_{F}\right)
\lambda \Gamma \left( k_{F}\right) S^{\dag }\left( k_{F}\right) \right]
_{jj}\sum_{j^{\prime \prime }}\mathcal{T}_{j^{\prime \prime }j^{\prime
}}^{C}\left( \epsilon _{p}\right) \right\} \text{.}
\end{eqnarray}%
For $T\ll E_{\text{conn}}$, by Eq.~(\ref{connelimT}), the quantity in curly
brackets approximately vanishes for $\left\vert \epsilon _{p}-\epsilon
_{k_{F}}\right\vert \lesssim T$, whereas the Fermi factor approximately
vanishes for $\left\vert \epsilon _{p}-\epsilon _{k_{F}}\right\vert \gg T$.
Therefore

\begin{equation}
G_{jj^{\prime }}\approx G_{jj^{\prime }}^{D}-\frac{1}{V_{k_{F}}^{2}}\left[
S\left( k_{F}\right) \Gamma ^{\dag }\left( k_{F}\right) \lambda \Gamma
\left( k_{F}\right) S^{\dag }\left( k_{F}\right) \right] _{jj}\sum_{j^{%
\prime \prime }}G_{j^{\prime \prime }j^{\prime }}^{D}\text{;}
\label{connelimG}
\end{equation}%
in other words, at $T\ll E_{\text{conn}}$\ it is possible to write the
conductance in terms of disconnected contributions alone.

Since Eq.~(\ref{connelimG}) contains only the disconnected contribution, we
may calculate it explicitly using Eqs.~(\ref{discprobKYA}) and (\ref%
{TDrowsum}). Since both $S\left( p\right) $ and $\Gamma \left( p\right) $
are slowly varying below the energy scale $E_{\text{conn}}$, we find the
conductance is approximately linear in the T-matrix,

\begin{eqnarray}
G_{jj^{\prime }} &\approx &\frac{2e^{2}}{h}\int d\epsilon _{p}\left[
-f^{\prime }\left( \epsilon _{p}\right) \right] \left\{ \mathcal{T}%
_{0,jj^{\prime }}\left( \epsilon _{k_{F}}\right) +\mathcal{Z}_{R,jj^{\prime
}}\left( \epsilon _{k_{F}}\right) \operatorname{Re}\left[ -\pi \nu _{p}\text{T}%
_{pp}\left( \epsilon _{p}\right) \right] \right.  \notag \\
&&\left. +\left[ \mathcal{Z}_{I,jj^{\prime }}\left( \epsilon _{k_{F}}\right)
+\mathcal{Z}_{2,jj^{\prime }}\left( \epsilon _{k_{F}}\right) \right] \operatorname{Im%
}\left[ -\pi \nu _{p}\text{T}_{pp}\left( \epsilon _{p}\right) \right]
\right\} \text{,}  \label{connelimG1}
\end{eqnarray}%
provided $T\ll E_{\text{conn}}$. Eq.~(\ref{connelimG1}) can also be obtained
by eliminating the connected part with the column sum Eq.~(\ref{TDcolsum})
instead of the row sum,

\begin{equation}
G_{jj^{\prime }}-\frac{1}{V_{k_{F}}^{2}}\left[ \Gamma ^{\dag }\left(
k_{F}\right) \lambda \Gamma \left( k_{F}\right) \right] _{j^{\prime
}j^{\prime }}\sum_{j^{\prime \prime }}G_{jj^{\prime \prime }}\text{,}
\end{equation}%
which corresponds to measuring the conductance by applying a small uniform
bias voltage in all leads, in addition to the small bias voltage in lead $%
j^{\prime }$.

Eq.~(\ref{connelimG1}) is the first central result of this paper. It
generalizes the result of the two-lead short ABK ring in Ref.~%
\onlinecite{PhysRevB.88.245104} to an arbitrary ABK ring, and expresses the
linear DC conductance as a linear function of the scattering channel
T-matrix, as long as the temperature is low compared to the mesoscopic
energy scale $E_{\text{conn}}$ at which $S\left( p\right) $ and $\Gamma
\left( p\right) $ varies significantly.

\section{Perturbation theories\label{sec:pert}}

\subsection{Weak-coupling perturbation theory}

Although we now understand that the connected part of the conductance can be
eliminated at low temperatures, this procedure may not be applicable in the
weak-coupling regime $T\gg T_{K}$. In this subsection we calculate the
linear DC conductance perturbatively in powers of $V_{k}^{2}/U$, again
generalizing the short ring results of Ref.~\onlinecite{PhysRevB.88.245104};
we expect the result to be valid in both small and large Kondo cloud regimes
as long as $T\gg T_{K}$ and the renormalized Kondo coupling constant remains
weak.

\subsubsection{Disconnected part}

We first find the disconnected part; the result is already given in Ref.~%
\onlinecite{PhysRevB.88.245104}, but for completeness we reproduce it here.
As implied by Eq.~(\ref{discprob}), our task\ amounts to calculating the
retarded T-matrix of the screening channel in the single-particle sector,
which is in turn achieved by calculating the two-point Green's function $%
-\left\langle T_{\tau }\psi _{k}\left( \tau \right) \psi _{k^{\prime
}}^{\dag }\left( 0\right) \right\rangle $. The pertinent Feynman diagrams to 
$O\left( J^{2}\right) $ and $O\left( K^{2}\right) $ are depicted in Fig.~\ref%
{fig:KondoFD}, and we find

\begin{equation}
\nu \text{T}_{kk^{\prime }}\left( \Omega \right) =\nu K_{kk^{\prime }}+\nu
^{2}\int d\epsilon _{q}\frac{1}{\Omega ^{+}-\epsilon _{q}}\left(
K_{kq}K_{qk^{\prime }}+\frac{3}{16}J_{kq}J_{qk^{\prime }}\right) \text{,}
\label{TmatKondo}
\end{equation}%
where again $\nu =1/\left( 2\pi v_{F}\right) $ for the model with a reduced
band. The factor of $3/16$ results from time-ordering and tracing over the
impurity spin, where we have used the following identity

\begin{equation}
\left\langle T_{\tau }S_{d}^{a}\left( \tau _{1}\right) S_{d}^{b}\left( \tau
_{2}\right) \right\rangle =\frac{1}{4}\delta ^{ab}\text{.}
\label{timeorderSd2}
\end{equation}

\begin{figure}[ptb]
\includegraphics[width=0.8\textwidth]{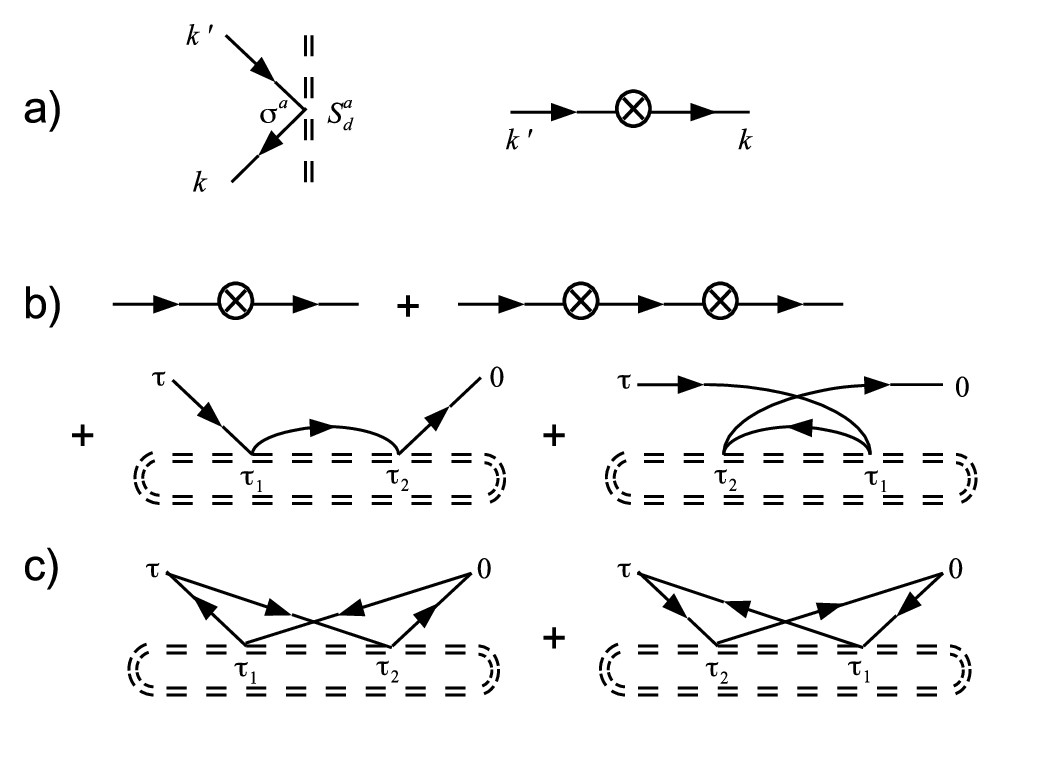}%
\caption{Diagrammatics of
weak-coupling perturbation theory. a) The vertices corresponding to the
Kondo coupling and the potential scattering in Eq.~(\ref{Kondomodel}). b)
Diagrams contributing to the T-matrix of the screening channel $\psi$
electrons up to $O\left(  J^{2} \right)  \sim O\left(  K^{2} \right)$. We
have traced over the impurity spin so that the double dashed lines (impurity spin propagators) form loops, and arranged the internal time variables
from left to right in increasing order. c) Connected diagrams contributing
to the linear DC conductance up to $O\left(  J^{2} \right) 
$.\label{fig:KondoFD}}
\end{figure}

The $O\left( K\right) $ and $O\left( K^{2}\right) $ terms, accounting for
the particle-hole symmetry breaking potential scattering due to the QD,
clearly obey the optical theorem $\operatorname{Im}\left[ -\pi \nu _{p}\text{T}%
_{pp}\left( \epsilon _{p}\right) \right] =\left\vert -\pi \nu _{p}\text{T}%
_{pp}\left( \epsilon _{p}\right) \right\vert ^{2}$. If $\kappa $ is
comparable to the renormalized value of $j$ then the $O\left( K\right) $
term dominates the T-matrix.

On the other hand, if we tune the QD to be particle-hole symmetric, $%
\epsilon _{d}=-U/2$ and $\kappa =0$, both\ terms containing $K$ will vanish,
and the $O\left( J^{2}\right) $ term becomes the lowest order contribution
to the T-matrix. For this term, one should also make a distinction between
the real principal value part and the imaginary $\delta $-function part. As
noticed in Ref.~\onlinecite{PhysRevB.82.165426} and reiterated in Ref.~%
\onlinecite{PhysRevB.88.245104}, the principal value part introduces
non-universalities due to its dependence on all energies in the reduced band 
$\left( -D,D\right) $; nevertheless it is merely an elastic potential
scattering term, and we neglect it in the following. Meanwhile, the $\delta $%
-function part is an inelastic effect stemming from the Kondo physics, as
can be seen from its violation of the optical theorem. (The T-matrix
apparently disobeys the optical theorem because it is restricted to the
single-particle sector, and the sum over intermediate states excludes
many-particle states.) Therefore, for a particle-hole symmetric QD, to $%
O\left( J^{2}\right) $ we have\cite{PhysRevB.88.245104}

\begin{equation}
-\pi \nu \text{T}_{pp}\left( \epsilon _{p}\right) =i\frac{3\pi ^{2}}{16}\nu
^{2}J_{pp}^{2}\text{.}  \label{TmatKondosym}
\end{equation}

The weak-coupling perturbation theory is famous for being infrared divergent
at $O\left( J^{3}\right) $,\cite{ProgTheorPhys.32.37,mahan2000many} but as
long as $T\gg T_{K}$, to logarithmic accuracy we can verify that the $%
O\left( J^{3}\right) $ corrections to the T-matrix can be absorbed into our $%
O\left( J^{2}\right) $ result by reinterpreting the bare Kondo coupling
constant $J_{pp}$\ as a renormalized one. The renormalization is governed by
Eq.~(\ref{KondoJkkprRG}), and cut off at either the \textquotedblleft
electron energy\textquotedblright\ $\left\vert \epsilon _{p}\right\vert $ or
the temperature $T$, whichever is larger. In other words, the Kondo coupling 
$J_{pp}$ in Eq.~(\ref{TmatKondosym}) should be replaced by $J_{pp}\left(
\max \left\{ \left\vert \epsilon _{p}\right\vert ,T\right\} \right) $, where
the argument in round brackets stands for the energy cutoff $D$ in Eq.~(\ref%
{KondoJkkprRG}) where the running coupling constant is evaluated.

\subsubsection{Connected part}

We now calculate the connected part to $O\left( J^{2}\right) $; the
calculation follows Ref.~\onlinecite{PhysRevB.88.245104} closely. Inserting
Eq.~(\ref{leaddensity}) into Eq.~(\ref{Gjjpriomegap}), we write the
connected part of the density-density correlation function in terms of a
four-point correlation function of $\psi $:

\begin{equation}
\mathcal{G}_{jj^{\prime }}^{\prime C}\left( i\omega _{p}\right)
=\int_{0}^{\pi }\frac{dk_{1}dk_{2}dq_{1}dq_{2}}{\left( 2\pi \right) ^{4}}%
\left( \mathbb{M}_{k_{1}k_{2}}^{j}\right) _{11}\left( \mathbb{M}%
_{q_{1}q_{2}}^{j^{\prime }}\right) _{11}\mathcal{G}%
_{k_{1}k_{2}q_{1}q_{2}}^{C}\left( i\omega _{p}\right) \text{,}
\label{GprCimag}
\end{equation}%
where

\begin{equation}
\mathcal{G}_{k_{1}k_{2}q_{1}q_{2}}^{C}\left( i\omega _{p}\right) \equiv
-\int_{0}^{\beta }d\tau e^{i\omega _{p}\tau }\sum_{\sigma \sigma ^{\prime
}}\left\langle T_{\tau }\psi _{k_{1}\sigma }^{\dag }\left( \tau \right) \psi
_{k_{2}\sigma }\left( \tau \right) \psi _{q_{1}\sigma ^{\prime }}^{\dag
}\left( 0\right) \psi _{q_{2}\sigma ^{\prime }}\left( 0\right) \right\rangle
_{C}\text{.}  \label{GCkkqqimag}
\end{equation}%
We insert Eqs.~(\ref{bbUmatrow1}) and (\ref{bbMmat}) into Eq.~(\ref{GprCimag}%
), and take the continuum limit, which is appropriate for the Kondo model.
Because in the wide band limit the most divergent contribution to $%
G_{jj^{\prime }}^{\prime C}\left( \Omega \right) $ is from $k_{1}\approx
k_{2}$ and $q_{1}\approx q_{2}$, we can expand the integrand around these
points,

\begin{align}
& \mathcal{G}_{jj^{\prime }}^{\prime C}\left( i\omega _{p}\right)  \notag \\
& =O\left( 1\right) +\int \frac{dk_{1}dk_{2}dq_{1}dq_{2}}{\left( 2\pi
\right) ^{4}}\frac{1}{V_{k_{1}}V_{k_{2}}V_{q_{1}}V_{q_{2}}}\mathcal{G}%
_{k_{1}k_{2}q_{1}q_{2}}^{C}\left( i\omega _{p}\right)
\sum_{r_{1}r_{2}r_{1}^{\prime }r_{2}^{\prime }}t_{r_{1}}^{\ast
}t_{r_{2}}t_{r_{1}^{\prime }}^{\ast }t_{r_{2}^{\prime }}  \notag \\
& \times \left\{ \Gamma _{r_{1}j}\left( k_{1}\right) \Gamma _{r_{2}j}^{\ast
}\left( k_{2}\right) \frac{1}{-i\left( k_{1}-k_{2}+i0\right) }+\left[ \Gamma
S^{\dag }\left( k_{1}\right) \right] _{r_{1}j}\left[ \Gamma S^{\dag }\left(
k_{2}\right) \right] _{r_{2}j}^{\ast }\frac{1}{-i\left(
k_{2}-k_{1}+i0\right) }\right\}  \notag \\
& \times \left\{ \Gamma _{r_{1}^{\prime }j^{\prime }}\left( q_{1}\right)
\Gamma _{r_{2}^{\prime }j^{\prime }}^{\ast }\left( q_{2}\right) \frac{1}{%
-i\left( q_{1}-q_{2}+i0\right) }+\left[ \Gamma S^{\dag }\left( q_{1}\right) %
\right] _{r_{1}^{\prime }j^{\prime }}\left[ \Gamma S^{\dag }\left(
q_{2}\right) \right] _{r_{2}^{\prime }j^{\prime }}^{\ast }\frac{1}{-i\left(
q_{2}-q_{1}+i0\right) }\right\} \text{.}  \label{GprCimag1}
\end{align}

The only non-vanishing diagrams at $O\left( J^{2}\right) $ are shown in
panel c) of Fig.~\ref{fig:KondoFD}:

\begin{eqnarray}
\mathcal{G}_{k_{1}k_{2}q_{1}q_{2}}^{C}\left( i\omega _{p}\right) &=&\frac{3}{%
8}J_{q_{2}k_{1}}J_{k_{2}q_{1}}\int_{0}^{\beta }d\tau e^{i\omega _{p}\tau
}\int_{0}^{\beta }d\tau _{1}d\tau _{2}g_{q_{2}}\left( -\tau _{1}\right)
g_{k_{1}}\left( \tau _{1}-\tau \right) g_{k_{2}}\left( \tau -\tau
_{2}\right) g_{q_{1}}\left( \tau _{2}\right)  \notag \\
&=&\frac{3}{8}J_{q_{2}k_{1}}J_{k_{2}q_{1}}\frac{1}{\beta }\sum_{\omega
_{n_{1}}}g_{q_{2}}\left( i\omega _{n_{1}}\right) g_{k_{1}}\left( i\omega
_{n_{1}}\right) g_{k_{2}}\left( i\omega _{n_{1}}+i\omega _{p}\right)
g_{q_{1}}\left( i\omega _{n_{1}}+i\omega _{p}\right) \text{;}
\end{eqnarray}%
we have again used Eq.~(\ref{timeorderSd2}). The frequency summation is
performed by deforming the complex plane contour and wrapping it around the
lines $\operatorname{Im} z=0$ and $\operatorname{Im} z=-\omega _{p}$. Analytic continuation
yields

\begin{align}
G_{k_{1}k_{2}q_{1}q_{2}}^{C}\left( \Omega \right) & =-\frac{%
3J_{q_{2}k_{1}}J_{k_{2}q_{1}}}{16\pi i}\int d\omega f\left( \omega \right)
\left\{ \left[ g_{q_{2}}^{R}\left( \omega \right) g_{k_{1}}^{R}\left( \omega
\right) -g_{q_{2}}^{A}\left( \omega \right) g_{k_{1}}^{A}\left( \omega
\right) \right] g_{k_{2}}^{R}\left( \omega +\Omega \right)
g_{q_{1}}^{R}\left( \omega +\Omega \right) \right.  \notag \\
& \left. +g_{q_{2}}^{A}\left( \omega -\Omega \right) g_{k_{1}}^{A}\left(
\omega -\Omega \right) \left[ g_{k_{2}}^{R}\left( \omega \right)
g_{q_{1}}^{R}\left( \omega \right) -g_{k_{2}}^{A}\left( \omega \right)
g_{q_{1}}^{A}\left( \omega \right) \right] \right\} \text{.}
\label{GCkkqqretKondo2}
\end{align}

Substituting Eqs.~(\ref{GCkkqqretKondo2}) and (\ref{KondoJ}) into Eq.~(\ref%
{GprCimag1}), we are able to evaluate the momentum integrals in the $\Lambda
\rightarrow \infty $ limit by contour methods. The $RRRR$ and $AAAA$ terms
vanish, and the $AARR$ terms combine to produce a Fermi surface factor $%
f^{\prime }\left( \omega \right) $:

\begin{align}
& G_{jj^{\prime }}^{\prime C}\left( \Omega \right)  \notag \\
& =O\left( 1\right) -\frac{1}{\Omega }\int d\omega \left[ -f^{\prime }\left(
\omega \right) \right] \frac{3\left( 2j\right) ^{2}}{16\pi i}%
\sum_{r_{1}r_{2}r_{1}^{\prime }r_{2}^{\prime }}t_{r_{1}}^{\ast
}t_{r_{2}}t_{r_{1}^{\prime }}^{\ast }t_{r_{2}^{\prime }}\left[ \Gamma
S^{\dag }\left( k_{F}+\frac{\omega }{v_{F}}\right) \right] _{r_{1}j}  \notag
\\
& \times \left[ \Gamma S^{\dag }\left( k_{F}+\frac{\omega +\Omega }{v_{F}}%
\right) \right] _{r_{2}j}^{\ast }\Gamma _{r_{1}^{\prime }j^{\prime }}\left(
k_{F}+\frac{\omega +\Omega }{v_{F}}\right) \Gamma _{r_{2}^{\prime }j^{\prime
}}^{\ast }\left( k_{F}+\frac{\omega }{v_{F}}\right) \frac{1}{v_{F}^{2}} 
\notag \\
& =O\left( 1\right) +\frac{1}{i\pi \Omega }\int d\epsilon _{p}\left[
-f^{\prime }\left( \epsilon _{p}\right) \right] \frac{3\pi ^{2}\nu
^{2}J_{pp}^{2}}{16}\mathcal{Z}_{2,jj^{\prime }}\left( \epsilon _{p}\right) 
\text{,}
\end{align}%
where we used Eq.~(\ref{discprobKYAZ2}). The connected contribution to the
conductance is now clearly a Fermi surface property:

\begin{equation}
G_{jj^{\prime }}^{C}=\frac{2e^{2}}{h}\int d\epsilon _{p}\left[ -f^{\prime
}\left( \epsilon _{p}\right) \right] \mathcal{T}_{jj^{\prime }}^{C\left(
2\right) }\left( \epsilon _{p}\right) \text{,}
\end{equation}%
where the connected transmission probability is

\begin{equation}
\mathcal{T}_{jj^{\prime }}^{C}\left( \epsilon _{p}\right) =\mathcal{Z}%
_{2,jj^{\prime }}\left( \epsilon _{p}\right) \frac{3\pi ^{2}}{16}\nu
^{2}J_{pp}^{2}\text{.}  \label{TC2Kondo}
\end{equation}%
This is formally identical to the short ring result, and is of the same
order of magnitude [$O\left( J^{2}\right) $] as the disconnected
contribution for a particle-hole symmetric QD.\cite{PhysRevB.88.245104} In
fact, if leads $j$ and $j^{\prime }$ are not directly coupled to each other
(i.e. they become decoupled when their couplings with the QD are turned off;
the simplest example is a QD embedded between source and drain leads)\cite%
{JPhysCondensMatter.16.R513}, we have $S_{jj^{\prime }}=0$, and the
disconnected contribution for a particle-hole symmetric QD is $O\left(
J^{4}\right) $. In this case the $O\left( J^{2}\right) $\ connected
contribution dominates.

Just as with the T-matrix, when we calculate the connected part further to $%
O\left( J^{3}\right) $ to logarithmic accuracy, the result can be absorbed
into Eq.~(\ref{TC2Kondo}) if the coupling constant $J$ is understood as
fully renormalized according to Eq.~(\ref{KondoJkkprRG}), with its
renormalization cut off by $\left\vert \epsilon _{p}\right\vert $ or $T$.

\subsubsection{Total conductance}

We write the total conductance at $T\gg T_{K}$ as a background term and a
correction due to the QD:

\begin{equation}
G_{jj^{\prime }}=\frac{2e^{2}}{h}\int d\epsilon _{p}\left[ -f^{\prime
}\left( \epsilon _{p}\right) \right] \left[ \mathcal{T}_{0,jj^{\prime
}}\left( \epsilon _{p}\right) +\delta \mathcal{T}_{jj^{\prime }}\left(
\epsilon _{p}\right) \right] \text{.}  \label{GjjprKondo}
\end{equation}

If the QD is well away from particle-hole symmetry, $\kappa $ can be of the
same order of magnitude as $j\left( T\right) $ even when the latter is fully
renormalized to the given temperature. In this case, the $T\gg T_{K}$\
correction to the background conductance will be dominated by the potential
scattering term; the connected contribution is negligible. The expression
for $\delta \mathcal{T}$ is

\begin{equation}
\delta \mathcal{T}_{jj^{\prime }}\left( \epsilon _{p}\right) \approx -%
\mathcal{Z}_{R,jj^{\prime }}\left( \epsilon _{p}\right) \pi \nu K_{pp}\text{.%
}  \label{deltaTjjprpot}
\end{equation}

If, however, the QD is particle-hole symmetric, the connected contribution
becomes important. Inserting Eq.~(\ref{TmatKondosym}) into Eq.~(\ref%
{discprobKYA}) and combining with (\ref{TC2Kondo}), we find the Kondo-type
correction to $O\left( J^{2}\right) $ at $T\gg T_{K}$

\begin{equation}
\delta \mathcal{T}_{jj^{\prime }}\left( \epsilon _{p}\right) =\left[ 
\mathcal{Z}_{I,jj^{\prime }}\left( \epsilon _{p}\right) +\mathcal{Z}%
_{2,jj^{\prime }}\left( \epsilon _{p}\right) \right] \frac{3\pi ^{2}}{16}\nu
^{2}J_{pp}^{2}\text{.}  \label{deltaTjjprKondo}
\end{equation}%
Again, in the RG improved perturbation theory, $J_{pp}$ in Eq.~(\ref%
{deltaTjjprKondo}) should be replaced by $J_{pp}\left( \max \left(
\left\vert \epsilon _{p}\right\vert ,T\right) \right) $, indicating that $%
J_{pp}$ is the renormalized value at the running energy cutoff $\max \left(
\left\vert \epsilon _{p}\right\vert ,T\right) $. This expression is valid as
long as $T\gg T_{K}$, irrespective of whether the system is in small or
large Kondo cloud regime.

Eq.~(\ref{deltaTjjprKondo}) is formally similar to the previously obtained
short ring result.\cite{PhysRevB.88.245104} It should be noted, however,
that the energy dependence of $\mathcal{Z}_{I}$, $\mathcal{Z}_{2}$ and $%
J^{2} $ is possibly much stronger than the short ring case, and the thermal
averaging in Eq.~(\ref{GjjprKondo}) can lead to very different results in
small and large Kondo cloud regimes. For instance, if $E_{\text{conn}}\ll
T_{K}\ll T$ (which may happen in the small cloud regime), the Fermi factor
in Eq.~(\ref{GjjprKondo}) averages over many peaks in $\mathcal{T}$, $%
\mathcal{Z}_{I}$, $\mathcal{Z}_{2}$\ and $V^{2}$ that are associated with
the underlying mesoscopic structure. In this case connected part elimination
is not applicable. On the other hand, if $T_{K}\ll T\ll E_{\text{conn}}$,
the variation of $\mathcal{T}$, $\mathcal{Z}_{I}$, $\mathcal{Z}_{2}$\ or $%
V^{2}$\ is negligible on the scale of $T$, and the Fermi factor in Eq.~(\ref%
{GjjprKondo}) may be approximated by a $\delta $ function. This leads to

\begin{equation}
G_{jj^{\prime }}=\frac{2e^{2}}{h}\left\{ \mathcal{T}_{0,jj^{\prime }}\left(
\epsilon _{k_{F}}\right) +\left[ \mathcal{Z}_{I,jj^{\prime }}\left( \epsilon
_{k_{F}}\right) +\mathcal{Z}_{2,jj^{\prime }}\left( \epsilon _{k_{F}}\right) %
\right] \frac{3\pi ^{2}}{16}\nu ^{2}J_{k_{F}k_{F}}^{2}\right\} \text{,}
\end{equation}%
which agrees with our prescription of eliminating the connected part, Eq.~(%
\ref{connelimG1}).

\subsection{Fermi liquid perturbation theory}

It is also interesting to consider temperatures low compared to the Kondo
temperature $T\ll T_{K}$. Since our formalism does not by itself provide a
low-energy effective theory of the small cloud regime for $T_{K}\gg E_{V}$,
we focus on the very large Kondo cloud regime $T_{K}\ll E_{V}$, where as
explained in Sec.~\ref{sec:model} the low-energy effective theory is simply
a Fermi liquid theory. If we further assume $T\ll E_{\text{conn}}$, then we
can simply eliminate the connected contribution to the conductance with Eq.~(%
\ref{connelimG1}).

To use Eq.~(\ref{connelimG1}) we need the low-energy T-matrix for the
screening channel in the single-particle sector in the Fermi liquid regime,
which is again well known.\cite{PhysRevB.48.7297,PhysRevB.77.180404} As
mentioned in Sec.~\ref{sec:intro}, the strong-coupling single particle wave
function at zero temperature is obtained by imposing a phase shift on the
weak-coupling wave function. This phase shift $\delta _{\psi \psi }$ results
from both elastic scattering off the Kondo singlet and particle-hole
symmetry breaking potential scattering:

\begin{equation}
\delta _{\psi \psi ,\sigma }=\sigma \frac{\pi }{2}+\delta _{P}\text{,}
\label{deltapsipsi}
\end{equation}%
where $\sigma =\pm 1$ for spin-up/spin-down electrons. To the lowest order
in potential scattering $O\left( K\right) $, we have\cite{PhysRevB.82.165426}

\begin{equation}
\tan \delta _{P}=-\pi \nu K_{k_{F}k_{F}}\text{.}  \label{tandeltaP}
\end{equation}

Let us introduce the phase-shifted screening channel $\tilde{\psi}$, which
is then related to the original screening channel $\psi $ via a scattering
basis transformation:

\begin{equation}
\psi _{k,\sigma }=\int_{k_{F}-\Lambda }^{k_{F}+\Lambda }\frac{dp}{2\pi }%
\left( \frac{i}{k-p+i0}-e^{2i\delta _{\psi \psi ,\sigma }}\frac{i}{k-p-i0}%
\right) \tilde{\psi}_{p,\sigma }\text{.}  \label{psitildescatt}
\end{equation}%
Using the definition of the T-matrix in the single-particle sector Eq.~(\ref%
{Dyson}) and the transformation Eq.~(\ref{psitildescatt}) one can show that
retarded T-matrices for $\psi $ and $\tilde{\psi}$ are related by

\begin{equation}
\text{T}_{kk^{\prime },\sigma }\left( \omega \right) =\frac{i}{2\pi \nu }%
\left( e^{2i\delta _{\psi \psi ,\sigma }}-1\right) +e^{2i\delta _{\psi \psi
,\sigma }}\text{\~{T}}_{kk^{\prime },\sigma }\left( \omega \right) \text{.}
\label{TvsTtilde}
\end{equation}%
Since $\tilde{\psi}$ diagonalizes the strong-coupling fixed point
Hamiltonian, by definition \~{T}$_{k_{F}k_{F},\sigma }\left( \omega
=0\right) =0$ at zero temperature.

\begin{figure}[ptb]
\includegraphics[width=0.4\textwidth]{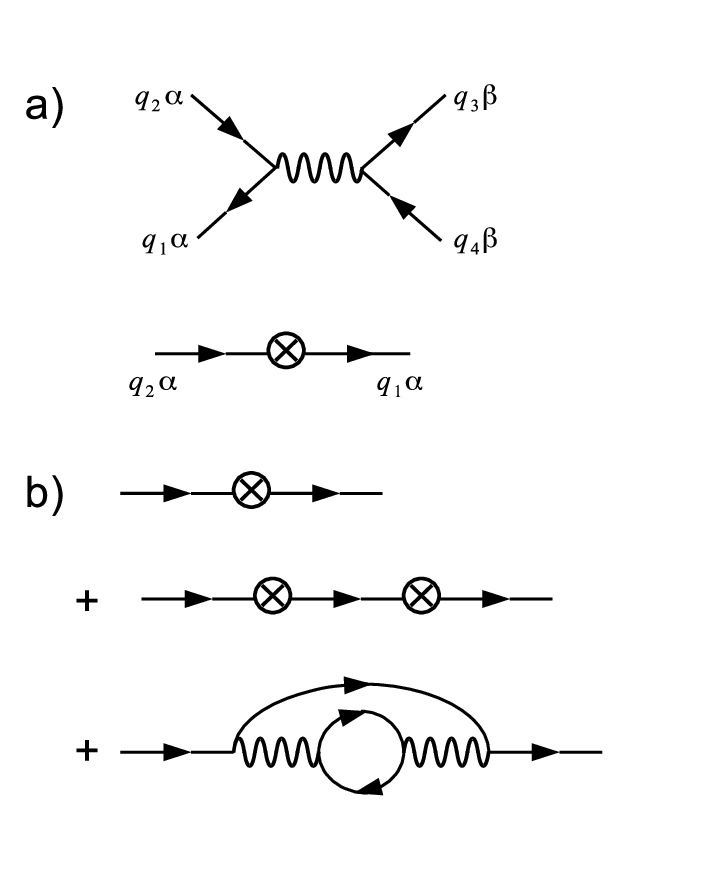}%
\caption{Diagrammatics of Fermi liquid perturbation theory. a) The two vertices given by the leading irrelevant operator, Eq.~(\ref{FLLIO}). b) Diagrams contributing to the
T-matrix of $\tilde\psi$ electrons up to $O\left( 1/T_{K}^{2} \right)  $. The propagators are those of the phase-shifted screening channel operators $\tilde\psi$.\label{fig:FermiFD1}}
\end{figure}

The leading irrelevant operator perturbing the strong-coupling fixed point
is localized at the QD (with a spatial extent of $v_{F}/T_{K}$), and
quadratic in spin current.\cite{PhysRevB.48.7297,condmat0501007} It is most
conveniently written in terms of $\tilde{\psi}$:

\begin{eqnarray}
H_{int} &=&\frac{2\pi v_{F}^{2}}{T_{K}}\colon \tilde{\psi}_{\alpha }^{\dag }%
\tilde{\psi}_{\alpha }\tilde{\psi}_{\beta }^{\dag }\tilde{\psi}_{\beta
}\colon \left( x=0\right)   \notag \\
&&-\frac{v_{F}^{2}}{T_{K}}\left[ \colon \tilde{\psi}_{\alpha }^{\dag }\left(
i\frac{d}{dx}-k_{F}\right) \tilde{\psi}_{\alpha }+\left( -i\frac{d}{dx}%
-k_{F}\right) \tilde{\psi}_{\alpha }^{\dag }\tilde{\psi}_{\alpha }\colon %
\right] \left( x=0\right) \text{.}  \label{FLLIO}
\end{eqnarray}%
Here $::$ denotes normal ordering, and sums over repeated spin indices $%
\alpha $ and $\beta $ are implied. The $\tilde{\psi}$ operators has been
unfolded, so that their wave functions are now defined on the entire real
axis instead of the positive real axis. The two terms in $H_{int}$ are
illustrated in panel a) of Fig.~\ref{fig:FermiFD1} as a four-point vertex
and a two-point one.\ Both terms share a single coupling constant of $%
O\left( 1/T_{K}\right) $, because the leading irrelevant operator written in
the $\tilde{\psi}$ basis must be particle-hole symmetric by definition. The
on-shell retarded T-matrix for $\tilde{\psi}$ in the single-particle sector
is calculated to $O\left( 1/T_{K}^{2}\right) $ in Ref.~%
\onlinecite{PhysRevB.48.7297}:

\begin{equation}
-\pi \nu \text{\~{T}}_{pp}\left( \epsilon _{p}\right) =\frac{\epsilon _{p}}{%
T_{K}}+i\frac{3\epsilon _{p}^{2}+\pi ^{2}T^{2}}{2T_{K}^{2}}\text{.}
\label{TtildeAL}
\end{equation}%
For completeness we give a derivation of this result in Appendix~\ref{sec:appFermi}. It is diagramatically represented by Fig.~\ref{fig:FermiFD1} panel b).

Substituting Eqs.~(\ref{deltapsipsi}), (\ref{TvsTtilde}) and (\ref{TtildeAL}%
) into Eq.~(\ref{connelimG1}), we eliminate the connected contribution, and
obtain the $T\ll T_{K}$ conductance in the very large Kondo cloud regime:

\begin{eqnarray}
G_{jj^{\prime }} &\approx &\frac{2e^{2}}{h}\left\{ \mathcal{T}_{0,jj^{\prime
}}\left( \epsilon _{k_{F}}\right) -\mathcal{Z}_{R,jj^{\prime }}\left(
\epsilon _{k_{F}}\right) \left( \frac{1}{2}-\frac{\pi ^{2}T^{2}}{T_{K}^{2}}%
\right) \sin 2\delta _{P}\right.  \notag \\
&&\left. +\left[ \mathcal{Z}_{I,jj^{\prime }}\left( \epsilon _{k_{F}}\right)
+\mathcal{Z}_{2,jj^{\prime }}\left( \epsilon _{k_{F}}\right) \right] \left(
\cos ^{2}\delta _{P}-\frac{\pi ^{2}T^{2}}{T_{K}^{2}}\cos 2\delta _{P}\right)
\right\} \text{.}  \label{connelimG2}
\end{eqnarray}
We note that the connected contribution can in fact be evaluated directly in the Fermi liquid theory. This is also done in Appendix~\ref{sec:appFermi}, and provides further verification of our scheme of eliminating the connected contribution.

Predictions of the conductance at high temperatures [Eq.~(\ref{GjjprKondo})]
and at low temperatures [Eq.~(\ref{connelimG2})] together constitute the
second main result of this paper. We emphasize once more that, while Eq.~(%
\ref{GjjprKondo}) is valid as long as $T\gg T_{K}$, Eq.~(\ref{connelimG2})
is expected to be justified provided $T\ll T_{K}\ll E_{V}$, so that the
Fermi liquid theory applies, and also $T\ll E_{\text{conn}}$, so that the
connected contribution can be eliminated.

For clarity we tabulate various regimes of energy scales discussed so far
(Table~\ref{tab:table1}). Note again that the connected contribution to
conductance can be eliminated when $T\ll E_{\text{conn}}$. In general we
have\ $E_{\text{conn}}\lesssim E_{V}$, but we assume in this table that $E_{%
\text{conn}}\sim E_{V}$, which is the case with the systems to be discussed
in this paper.

\begin{table}[b]
\caption{\label{tab:table1}Different regimes of energy scales discussed in this paper. $T$, $T_{K}$ and $E_{V}$ are respectively the temperature, the Kondo temperature, and the energy scale over which $V_{k}^{2}$ varies significantly. We also assume $E_{\text{conn}}\sim E_{V}$, where $E_{\text{conn}}$ is the energy scale over which $S$ and $\Gamma$ vary significantly. For the low-temperature conductance in the small Kondo cloud regime, see discussion in Sec.~\ref{sec:outlook}.
} 
\begin{ruledtabular}
\begin{tabular}{cp{4cm}p{4cm}p{4cm}}
&
\textrm{Weak-coupling perturbation theory applies}&
$T_{K}$\textrm{ depends on mesoscopic details}&
\textrm{Connected part elimination possible}\\
\colrule
$T\gg T_{K}\gg E_{V}$ & \textrm{Yes} & \textrm{No} & \textrm{No}\\
$T\gg E_{V}\gg T_{K}$ & \textrm{Yes} & \textrm{Yes} & \textrm{No}\\
$E_{V}\gg T\gg T_{K}$ & \textrm{Yes} & \textrm{Yes} & \textrm{Yes}\\
\end{tabular}
\begin{tabular}{cp{4cm}p{4cm}p{4cm}}
&
\textrm{Fermi liquid perturbation theory applies}&
$T_{K}$\textrm{ depends on mesoscopic details}&
\textrm{Connected part elimination possible}\\
\colrule
$T\ll T_{K}\ll E_{V}$ & \textrm{Yes} & \textrm{Yes} & \textrm{Yes}\\
$T\ll E_{V}\ll T_{K}$ & \textrm{?} & \textrm{No} & \textrm{Yes}\\
$E_{V}\ll T\ll T_{K}$ & \textrm{?} & \textrm{No} & \textrm{No}\\
\end{tabular}
\end{ruledtabular}
\end{table}

\section{Closed long ABK rings\label{sec:closed}}

\begin{figure}[ptb]
\includegraphics[width=0.8\textwidth]{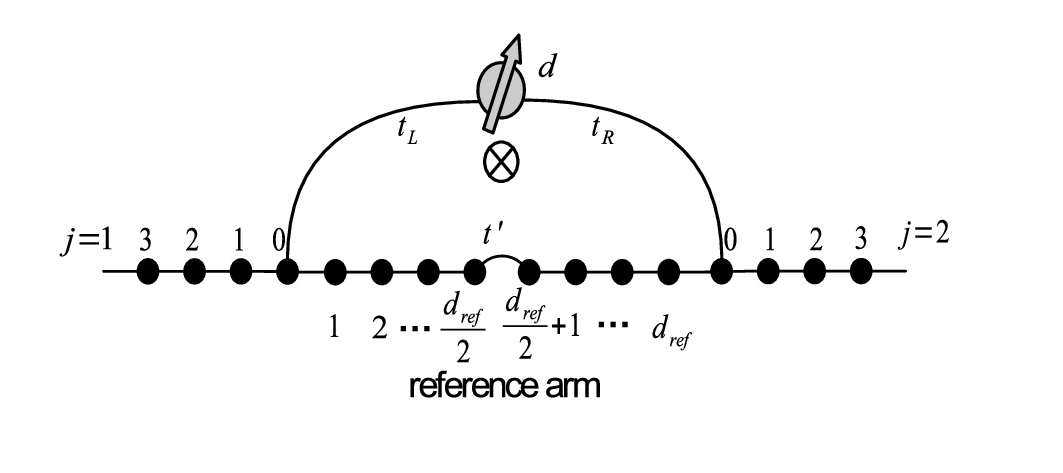}%
\caption{Geometry of the
long ABK ring with short upper arms and a pinched reference
arm.\label{fig:longring1}}
\end{figure}

In this section we apply our general formalism to the simplest model of a
closed long ABK ring, studied in Ref.~\onlinecite{PhysRevB.83.165310} (Fig.~%
\ref{fig:longring1}): the QD is coupled directly to the source and drain
leads, and a long reference arm connects the two leads smoothly. A weak link
with hopping $t^{\prime }$ splits the reference arm into two halves of equal
length $d_{ref}/2$ where $d_{ref}$ is an even integer. As opposed to Ref.~%
\onlinecite{PhysRevB.83.165310}, however, we use gauge invariance to assign
the AB phase to the QD tunnel couplings rather than the weak link: $%
t_{1}\equiv t_{L}e^{i\varphi /2}$ and $t_{2}\equiv t_{R}e^{-i\varphi /2}$.
We assume no additional non-interacting long arms connecting the QD with the
source and drain leads, because multiple traversal processes in such long QD
arms will lead to interference effects\cite{RevModPhys.82.2257} independent
of the AB phase, complicating the problem.\cite{PhysRevB.83.165310} The
Hamiltonian representing this model takes the form

\begin{eqnarray}
H_{0\text{,junction}} &=&-t\left[ \left(
\sum_{n=1}^{d_{ref}/2-1}+\sum_{n=d_{ref}/2+1}^{d_{ref}-1}\right)
c_{ref,n}^{\dag }c_{ref,n+1}+\text{h.c.}\right]  \notag \\
&&-t\left( c_{ref,1}^{\dag }c_{1,0}+c_{ref,d_{ref}}^{\dag }c_{2,0}+\text{h.c.%
}\right)  \notag \\
&&-t^{\prime }\left( c_{ref,d_{ref}/2}^{\dag }c_{ref,d_{ref}/2+1}+\text{h.c.}%
\right) \text{;}
\end{eqnarray}%
the coupling sites are $c_{C,r=1}\equiv c_{1,0}$ and $c_{C,r=2}\equiv
c_{2,0} $.

We first repeat the Kondo temperature analysis in Ref.~%
\onlinecite{PhysRevB.83.165310} in order to distinguish between small and
large Kondo cloud regimes, then carefully study the conductance at high and
low temperatures, taking into account the previously neglected connected
contribution.

\subsection{Kondo temperature}

The background S-matrix for this model is identical to the short ABK ring%
\cite{PhysRevB.82.165426} up to overall phases, due to the smooth connection
between reference arm and leads:

\begin{equation}
S\left( k\right) =e^{ikd_{ref}}\left( 
\begin{array}{cc}
\tilde{r}\left( k\right) & \tilde{t}\left( k\right) \\ 
\tilde{t}\left( k\right) & \tilde{r}\left( k\right)%
\end{array}%
\right) \text{;}  \label{SMatkYE}
\end{equation}%
where the S-matrix elements $\tilde{r}$ and $\tilde{t}$ for the weak link are

\begin{equation}
\tilde{r}\left( k\right) =-\frac{1-\tilde{\tau}^{2}}{e^{-2ik}-\tilde{\tau}%
^{2}}\text{, }\tilde{t}\left( k\right) =-\frac{2i\tilde{\tau}\sin k}{%
e^{-2ik}-\tilde{\tau}^{2}}\text{,}
\end{equation}%
and we introduce the shorthand $\tilde{\tau}=t^{\prime }/t$. The wave
function is also straightforward to find:

\begin{equation}
\Gamma _{jj^{\prime }}\left( k\right) =\delta _{jj^{\prime }}+S_{jj^{\prime
}}\left( k\right) \text{.}  \label{GammakYE}
\end{equation}

In the wide band limit, $\tilde{r}$ and $\tilde{t}$ are approximately
independent of $k$ in the reduced band $k_{F}-\Lambda _{0}<k<k_{F}+\Lambda
_{0}$ where the momentum cutoff $\Lambda _{0}\ll 1$. This allows us to
approximate them by their Fermi surface values, $\tilde{r}\left( k\right)
\approx \tilde{r}=\left\vert \tilde{r}\right\vert e^{i\theta }$ and $\tilde{t%
}\left( k\right) \approx \tilde{t}=\pm i\left\vert \tilde{t}\right\vert
e^{i\theta }$ (the $\pm \pi /2$ phase difference is required by unitarity of 
$S$); without loss of generality we focus on the $\tilde{t}=i\left\vert 
\tilde{t}\right\vert e^{i\theta }$ case.

From Eq.~(\ref{GammakYE}) one conveniently obtains the normalization factor

\begin{eqnarray}
V_{k}^{2} &=&2\left( t_{L}^{2}+t_{R}^{2}\right) \left[ 1+\left\vert \tilde{r}%
\right\vert \cos \left( kd_{ref}+\theta \right) -\gamma \left\vert \tilde{t}%
\right\vert \cos \varphi \sin \left( kd_{ref}+\theta \right) \right]  \notag
\\
&=&2\left( t_{L}^{2}+t_{R}^{2}\right) \left[ 1+\sqrt{1-\left\vert \tilde{t}%
\right\vert ^{2}\left( 1-\gamma ^{2}\cos ^{2}\varphi \right) }\cos \left(
kd_{ref}+\theta ^{\prime }\right) \right] \text{,}  \label{VksqYE}
\end{eqnarray}%
where $\gamma =2t_{L}t_{R}/\left( t_{L}^{2}+t_{R}^{2}\right) $ measures the
degree of symmetry of coupling to the QD. In the second line we have used $%
\left\vert \tilde{t}\right\vert ^{2}+\left\vert \tilde{r}\right\vert ^{2}=1$
and introduced another phase $\theta ^{\prime }$, where $\theta ^{\prime
}-\theta $ is a function of $\gamma $, $\left\vert \tilde{t}\right\vert $
and $\varphi $ but independent of~$k$. We note that this expression is also
applicable in the continuum limit, where the lattice constant $a\rightarrow
0 $ (we have previously set $a=1$) but the arm length $d_{ref}a$ is fixed.
In that case $d_{ref}$ should be understood as the arm length $d_{ref}a$.

For long rings and filling factors not too small $k_{F}d_{ref}\gg 1$, $%
V_{k}^{2}$ oscillates around $2\left( t_{L}^{2}+t_{R}^{2}\right) $ as a
function of $k$, and has its extrema at $k_{n}=\left( n\pi -\theta ^{\prime
}\right) /d_{ref}$ where $n$ takes integer values. The only characteristic
energy scale for $V_{k}^{2}$ is therefore the peak/valley spacing $\Delta
=v_{F}\pi /d_{ref}$, and $E_{V}\sim E_{\text{conn}}\sim \Delta $. As in Ref.~%
\onlinecite{PhysRevB.83.165310} we define the reduced band such that $\Delta
\ll D_{0}\equiv v_{F}\Lambda _{0}$, and the reduced band initially contains
many oscillations.

In the small Kondo cloud regime $T_{K}\gg \Delta $, one may assume the
oscillations of $V_{k}^{2}$ are smeared out when the energy cutoff is being
reduced from $D_{0}$, which is still well above $T_{K}$: $V_{k}^{2}\simeq 
\overline{V_{k}^{2}}=2\left( t_{L}^{2}+t_{R}^{2}\right) $. This means $T_{K}$
in this regime is approximately the background Kondo temperature\ $T_{K}^{0}$
defined in Eq.~(\ref{TKKondobulk}), independent of the position of the Fermi
level at the energy scale $\Delta $, and also independent of the magnetic
flux.

\begin{figure}
\includegraphics[width=0.6\textwidth]{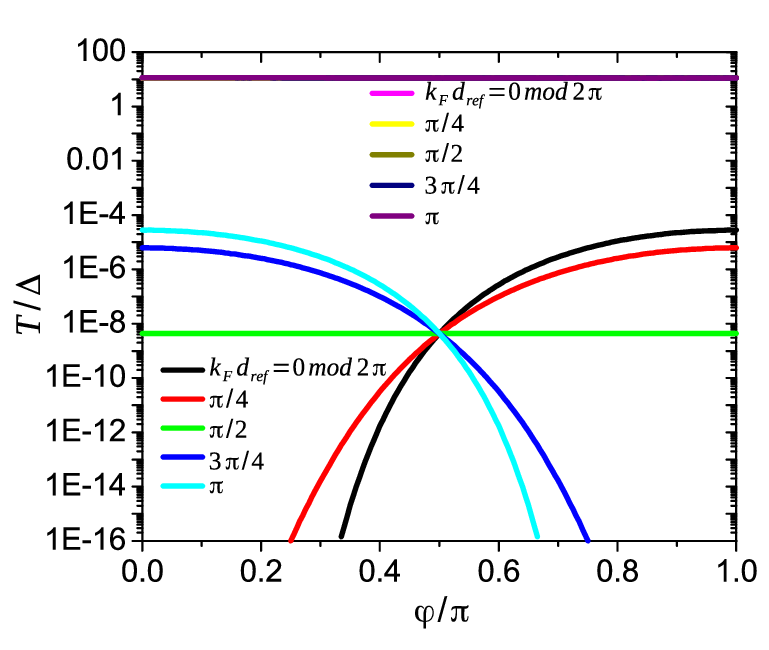}%
\caption{Kondo temperature $T_{K}$ for the closed long ABK ring, calculated by numerical integration of the weak coupling RG equation Eq.~(\ref{KondojRG}), plotted against the
AB phase $\varphi$. $T_{K}\left( \varphi \right)$ is an even function of $\varphi$
and has a period of $2\pi$, so only $0\leq \varphi \leq \pi$ is shown. System
parameters are: $d_{ref}=60$, $\theta =\pi/2$, $\left| \tilde{r} \right|
=0$, $t_{L}=t_{R}$, $D_{0}=10$. The curves with $T_{K}\gg \Delta$ (small Kondo cloud regime) have a large bare Kondo coupling $\left( t_{L}^{2}+ t_{R}^{2}\right) j_{0}/\pi =0.15$, whereas the curves with $T_{K}\ll \Delta$ (large Kondo cloud regime) have a much smaller bare Kondo coupling $\left( t_{L}^{2}+ t_{R}^{2}\right) j_{0}/\pi=0.02$. In the small cloud regime $T_{K}$ is almost independent of $\varphi$ and $k_{F}$, as the curves are flat and overlapping with each other. In the large cloud regime, however, $T_{K}$ highly sensitive to both $\varphi$ and $k_{F}$.\label{fig:longringTK}}
\end{figure}

On the other hand, in the large cloud regime $T_{K}\lesssim \Delta $, now
that the Kondo temperature is largely determined by the value of $V_{k}^{2}$
in a very narrow range of energies around the Fermi level, the mesoscopic $k$
oscillations become much more important. When the running energy cutoff $D$
is above the peak/valley spacing $\Delta $, the renormalization of $j$ is
controlled by $\overline{V_{k}^{2}}$ as in Eq.~(\ref{KondojRGsolsc}). Once $%
D $ is reduced below $\Delta $, we may approximate the renormalization of $j$
as being dominated by $V_{k_{F}}^{2}$. This leads to the following
estimation of the Kondo temperature:

\begin{equation}
T_{K}\simeq \Delta \exp \left[ -\frac{1}{2V_{k_{F}}^{2}\nu j\left( \Delta
\right) }\right] =\Delta \left( \frac{T_{K}^{0}}{\Delta }\right) ^{\left[ 1+%
\sqrt{1-\left\vert \tilde{t}\right\vert ^{2}\left( 1-\gamma ^{2}\cos
^{2}\varphi \right) }\cos \left( k_{F}d_{ref}+\theta ^{\prime }\right) %
\right] ^{-1}}\text{.}
\end{equation}%
It is clear that $T_{K}$ can be significantly dependent on the AB phase $%
\varphi $ in this regime. In particular, $T_{K}$ varies from $\sim \sqrt{%
T_{K}^{0}\Delta }$ (\textquotedblleft on resonance\textquotedblright )\ to
practically $0$ (\textquotedblleft off resonance\textquotedblright )\ as $%
\varphi $ is tuned between $0$ and $\pi $, when the Fermi energy is located
on a peak or in a valley $k_{F}=k_{n}$, the background transmission is
perfect $\left\vert \tilde{t}\right\vert =1$, and coupling to the QD is
symmetric $\gamma =1$; see Fig.~\ref{fig:longringTK}.\cite%
{PhysRevB.83.165310} (The special case $V_{k_{F}}^{2}=0$ corresponds to a
pseudogap problem $\nu V_{k}^{2}\propto \left( k-k_{F}\right) ^{2}$, and the
stable RG fixed point can be the local moment fixed point or the asymmetric
strong coupling fixed point, depending on the degree of particle-hole
symmetry.\cite{PhysRevB.70.094502,*PhysRevB.70.214427}) As a general rule,
stronger transmission through the pinch $\left\vert \tilde{t}\right\vert $
and greater symmetry of coupling $\gamma $\ result in stronger interference
between the two tunneling paths through the device, and hence increases the
tunability of the Kondo temperature by the magnetic flux.

\subsection{High-temperature conductance}

We now calculate the conductance at $T\gg T_{K}$ by perturbation theory.
Following the discussion in Ref.~\onlinecite{PhysRevB.83.165310}, we
consider the case of a particle-hole symmetric QD $\kappa =0$ and $%
K_{kk^{\prime }}=0$, and also ignore the elastic real part of the potential
scattering generated at $O\left( J^{2}\right) $.\cite{PhysRevB.88.245104}
These assumptions allow us to adopt Eq.~(\ref{deltaTjjprKondo}) for the $%
O\left( J^{2}\right) $ correction to the transmission probability:

\begin{equation}
\delta \mathcal{T}_{jj^{\prime }}\left( \epsilon _{k}\right) =3\pi ^{2}\nu
^{2}j^{2}\left( V_{k}^{2}\operatorname{Re}\left\{ \left[ S\left( k\right) \Gamma
^{\dag }\left( k\right) \lambda \Gamma \left( k\right) \right] _{jj^{\prime
}}S_{jj^{\prime }}^{\ast }\left( k\right) \right\} -\left\vert \left[
S\left( k\right) \Gamma ^{\dag }\left( k\right) \lambda \Gamma \left(
k\right) \right] _{jj^{\prime }}\right\vert ^{2}\right) \text{,}
\label{deltaTht}
\end{equation}%
where we have used Eqs.~(\ref{discprobKYAZI}) and (\ref{discprobKYAZ2}).

Note that Eq.~(\ref{deltaTht}) does not depend on details of the
non-interacting part of the ring Hamiltonian $H_{0,\text{junction}}$. For a
parity-symmetric geometry with two leads and two coupling sites ($N=M=2$),
when coupling to the QD is also symmetric ($t_{L}=t_{R}$) and time-reversal
symmetry is present ($\varphi =0$ or $\pi $), we can further show that the
sign of the $O\left( J^{2}\right) $ transmission probability correction is
determined by the sign of $1-2\left\vert \mathcal{T}_{0,12}\right\vert $, a
property discussed in Ref.~\onlinecite{PhysRevB.88.245104} at the end of
Sec.~IV C. Indeed, parity symmetry implies that $S_{11}=S_{22}$, $%
S_{12}=S_{21}$, $\Gamma _{11}=\Gamma _{22}$, $\Gamma _{12}=\Gamma _{21}$;
hence it is not difficult to find from Eq.~(\ref{deltaTht}) that

\begin{equation}
\frac{1}{4}\left[ \mathcal{\delta T}_{11}\left( \epsilon _{k}\right) +%
\mathcal{\delta T}_{22}\left( \epsilon _{k}\right) -\mathcal{\delta T}%
_{12}\left( \epsilon _{k}\right) -\mathcal{\delta T}_{21}\left( \epsilon
_{k}\right) \right] =\frac{3}{8}\pi ^{2}\nu ^{2}J_{kk}^{2}\left[
1-2\left\vert S_{12}\left( k\right) \right\vert ^{2}\right] \text{.}
\end{equation}%
The left-hand side correspond to a particular way to measure the
conductance, namely parity-symmetric bias voltage and parity-symmetric
current probes, or $y=1/2$ in Sec.~V of Ref.~\onlinecite{PhysRevB.88.245104}.

We now return to the long ring geometry without assumptions about $t_{L}$, $%
t_{R}$ and $\varphi $. Plugging Eqs.~(\ref{SMatkYE}) and (\ref{GammakYE})
into Eq.~(\ref{deltaTht})\ we find

\begin{equation}
\mathcal{\delta T}_{11}\left( \epsilon _{k}\right) =-3\pi ^{2}\nu ^{2}j^{2}%
\left[ C_{0}\left( k\right) +C_{1}\left( k\right) \cos \varphi +C_{2}\left(
k\right) \cos 2\varphi \right] \text{,}  \label{deltaTYE}
\end{equation}%
where the coefficients $C_{0}\left( k\right) $, $C_{1}\left( k\right) $ and $%
C_{2}\left( k\right) $ are independent of $\varphi $ but are usually
complicated functions of $k$:

\begin{subequations}
\begin{align}
C_{0}\left( k\right) & =\left( t_{L}^{4}+t_{R}^{4}\right) \left\vert \tilde{t%
}\right\vert ^{2}\left[ 1+2\left\vert \tilde{r}\right\vert \cos \left(
kd_{ref}+\theta \right) +\left\vert \tilde{r}\right\vert ^{2}\cos 2\left(
kd_{ref}+\theta \right) \right]  \notag \\
& -2t_{L}^{2}t_{R}^{2}\left[ 3-4\left\vert \tilde{t}\right\vert
^{2}+4\left\vert \tilde{r}\right\vert ^{3}\cos \left( kd_{ref}+\theta
\right) +\left( \left\vert \tilde{r}\right\vert ^{4}+\left\vert \tilde{t}%
\right\vert ^{4}\right) \cos 2\left( kd_{ref}+\theta \right) \right] \text{,}
\end{align}

\begin{align}
C_{1}\left( k\right) & =4\left\vert \tilde{t}\right\vert t_{L}t_{R}\left(
t_{L}^{2}+t_{R}^{2}\right) \sin \left( kd_{ref}+\theta \right)  \notag \\
& \times \left[ \left\vert \tilde{r}\right\vert ^{2}\cos 2\left(
kd_{ref}+\theta \right) +\left\vert \tilde{r}\right\vert \left( \left\vert 
\tilde{r}\right\vert ^{2}-\left\vert \tilde{t}\right\vert ^{2}\right) \cos
\left( kd_{ref}+\theta \right) -\left\vert \tilde{t}\right\vert ^{2}\right] 
\text{,}
\end{align}

\begin{equation}
C_{2}\left( k\right) =2\left\vert \tilde{t}\right\vert
^{2}t_{L}^{2}t_{R}^{2}\left\{ 1+2\left\vert \tilde{r}\right\vert \cos \left(
kd_{ref}+\theta \right) +\left\vert \tilde{r}\right\vert ^{2}\cos 2\left(
kd_{ref}+\theta \right) \right\} \text{.}
\end{equation}%
In the special case of a smooth reference arm $\left\vert \tilde{r}%
\right\vert =0$ and $\left\vert \tilde{t}\right\vert =1$, the Kondo-type
correction becomes especially simple:

\end{subequations}
\begin{align}
\mathcal{\delta T}_{11}\left( \epsilon _{k}\right) & =-3\pi ^{2}\nu ^{2}j^{2}%
\left[ t_{L}^{2}+t_{R}^{2}-2t_{L}t_{R}\sin \left( kd_{ref}+\theta +\varphi
\right) \right]  \notag \\
& \times \left[ t_{L}^{2}+t_{R}^{2}-2t_{L}t_{R}\sin \left( kd_{ref}+\theta
-\varphi \right) \right] \text{.}
\end{align}%
As in Refs.~\onlinecite{PhysRevB.83.165310,PhysRevB.88.245104}, only the
first and the second harmonics of the AB phase $\varphi $ appear in the
correction to the transmission probability $\mathcal{\delta T}_{11}$.

We may perform the thermal averaging in Eq.~(\ref{GjjprKondo}) at this
stage. The Fermi factor $-f^{\prime }\left( \epsilon _{k}\right) $ ensures
only the energy range $\left\vert \epsilon _{k}\right\vert \lesssim T$
contributes significantly to the conductance; in this energy range the
renormalization of $j$ is cut off by $T$.

In the small Kondo cloud regime, $T\gg T_{K}$ means $T\gg \Delta $ so that\
we can average over many peaks of $\mathcal{\delta T}_{jj^{\prime }}\left(
\epsilon _{k}\right) $, so we may drop all rapidly oscillating Fourier
components in Eq.~(\ref{deltaTYE}). This leads to

\begin{equation}
\delta G\approx -\frac{2e^{2}}{h}3\pi ^{2}\nu ^{2}\left[ j\left( T\right) %
\right] ^{2}\left[ \left( t_{L}^{4}+t_{R}^{4}\right) \left\vert \tilde{t}%
\right\vert ^{2}-2t_{L}^{2}t_{R}^{2}\left( 3-4\left\vert \tilde{t}%
\right\vert ^{2}\right) +2\left\vert \tilde{t}\right\vert
^{2}t_{L}^{2}t_{R}^{2}\cos 2\varphi \right] \text{.}  \label{deltaGYEsc}
\end{equation}%
We see that the first harmonic in $\varphi $ approximately drops out upon
thermal averaging.

\begin{figure}
\includegraphics[width=0.4\textwidth]{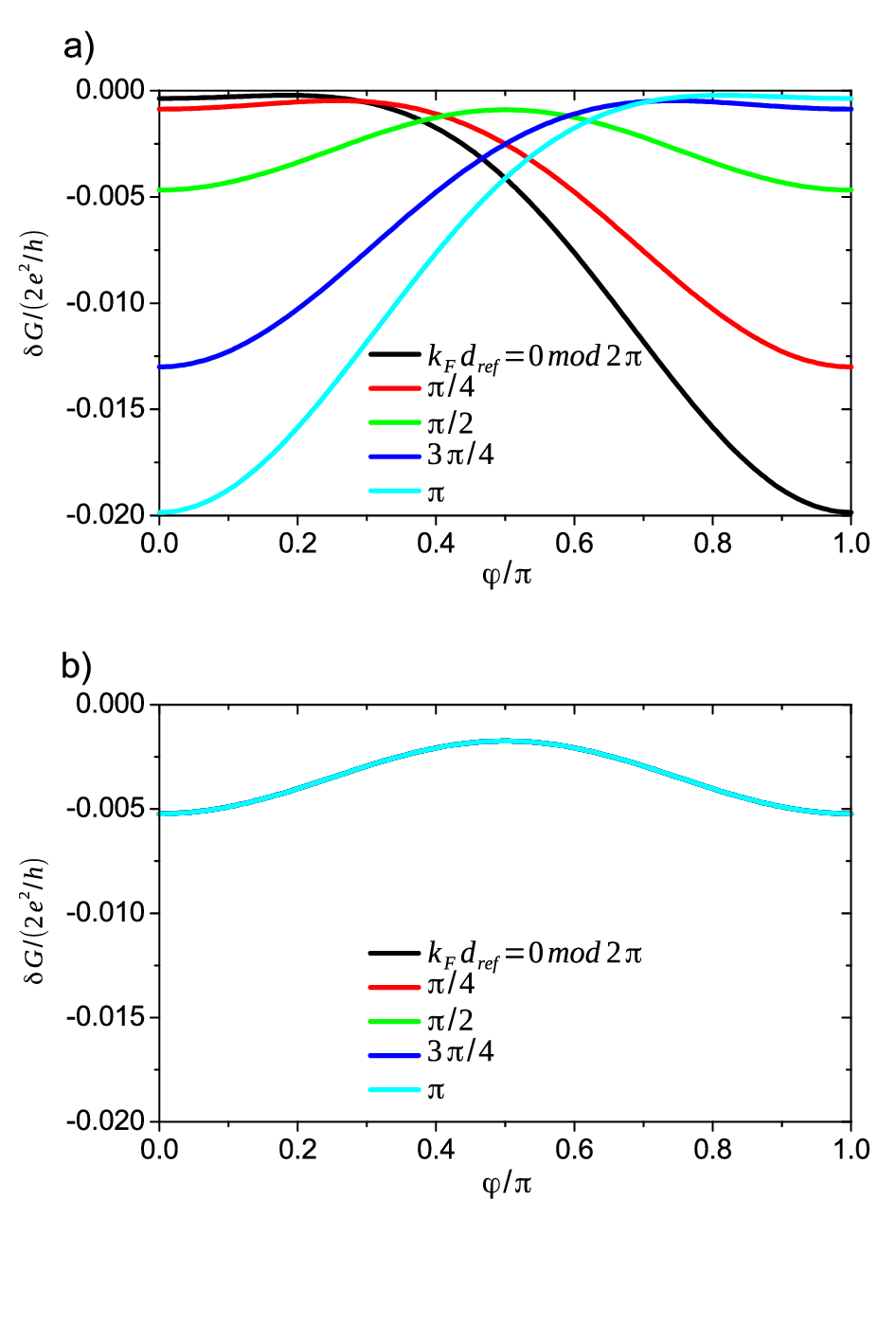}%
\caption{Kondo-type correction to the conductance $\delta G$
at $T\gg T_{K}$ for the closed long ABK ring with a particle-hole symmetric QD, calculated
by RG improved perturbation theory Eq.~(\ref{deltaTYE}), plotted against the
AB phase $\varphi$. Again only $0\leq \varphi \leq \pi$ is shown. System
parameters are: $d_{ref}=60$, $\theta =\pi/2$, $\left| \tilde{r} \right|
=0$, $t_{L}=t_{R}$, $\left( t_{L}^{2}+ t_{R}^{2}\right) j_{0}/\pi
=0.02$ at $D_{0}=10$ (i.e. the system is in the large cloud regime). $T/\Delta=0.0955$ in panel a) and $T/\Delta=19.1$ in panel b). For $T\ll \Delta$ the conductance shows considerable $k_F$ dependence, while for $T\gg \Delta$ such dependence essentially vanishes and curves at different $k_F$ overlap. Also, for $T\gg \Delta$ the first harmonic $\cos \varphi$ drops out as predicted by Eq.~(\ref{deltaGYEsc}), and $\delta G \left( \varphi \right)$ has a period of $\pi$. \label{fig:longringht}}
\end{figure}

On the other hand, in the large Kondo cloud regime, for $T\gg T_{K}$ it is
possible to have either $T\gg \Delta $ or $T\ll \Delta $. In the former case
Eq.~(\ref{deltaGYEsc}) continues to hold. In the latter case $\mathcal{%
\delta T}_{jj^{\prime }}\left( \epsilon _{k}\right) $ has little variation
in the energy range $\left\vert \epsilon _{k}\right\vert \lesssim T$, so it
is appropriate to replace $-f^{\prime }\left( \epsilon _{k}\right) $ with a $%
\delta $ function at the Fermi level; thus

\begin{equation}
\delta G\approx -\frac{2e^{2}}{h}3\pi ^{2}\nu ^{2}\left[ j\left( T\right) %
\right] ^{2}\left[ C_{0}\left( k_{F}\right) +C_{1}\left( k_{F}\right) \cos
\varphi +C_{2}\left( k_{F}\right) \cos 2\varphi \right] \text{.}
\label{deltaGYElc}
\end{equation}%
Fig.~\ref{fig:longringht} illustrates these two different cases for the
large Kondo cloud regime. We note that our $T\gg T_{K}$ results, Eq.~(\ref%
{deltaGYEsc}) for $T\gg \Delta $ and Eq.~(\ref{deltaGYElc}) for $T\ll \Delta 
$, are different from those of Ref.~\onlinecite{PhysRevB.83.165310}. We
believe the discrepancy is due to the fact that only single-particle
scattering processes are taken into consideration by Ref. %
\onlinecite{PhysRevB.83.165310}; the connected contribution to the
conductance is omitted, despite being of comparable magnitude with the
disconnected contribution.

\subsection{Fermi liquid conductance}

It is also interesting to calculate the conductance in the $T\ll T_{K}$
limit in the very large Kondo cloud regime, starting from Eq.~(\ref%
{connelimG2}). We make the assumption that the particle-hole symmetry
breaking potential scattering is negligible, $\delta _{P}=0$, as in Ref.~%
\onlinecite{PhysRevB.83.165310}. Inserting Eqs.~(\ref{SMatkYE}) and (\ref%
{GammakYE}) into Eqs.~(\ref{discprobKYAZI}) and (\ref{discprobKYAZ2}), we
find the total conductance has the form

\begin{equation}
G=\frac{2e^{2}}{h}\left[ T_{s}+\left( \left\vert \tilde{t}\right\vert
^{2}-T_{s}\right) \left( \frac{\pi T}{T_{K}}\right) ^{2}\right] \text{,}
\label{GYEFL}
\end{equation}%
where the $T=0$ transmission probability is

\begin{align}
T_{s}& =\left\vert e^{ik_{F}d_{ref}}\tilde{t}-\frac{2}{V_{k_{F}}^{2}}\left[
t_{L}e^{i\frac{\varphi }{2}}\left( e^{ik_{F}d_{ref}}\tilde{r}+1\right)
+t_{R}e^{-i\frac{\varphi }{2}}e^{ik_{F}d_{ref}}\tilde{t}\right] \right. 
\notag \\
& \left. \times \left[ t_{L}e^{-i\frac{\varphi }{2}}e^{ik_{F}d_{ref}}\tilde{t%
}+t_{R}e^{i\frac{\varphi }{2}}\left( 1+e^{ik_{F}d_{ref}}\tilde{r}\right) %
\right] \right\vert ^{2}\text{.}
\end{align}%
While Eq.~(\ref{GYEFL}) is ostensibly in agreement with Eq.~(69) of Ref.~%
\onlinecite{PhysRevB.83.165310}, we suspect that there are two oversights in
the derivation of the latter: at finite temperature, Ref.~%
\onlinecite{PhysRevB.83.165310} neglects the connected contribution to the
conductance, and also replaces the thermal factor $-f^{\prime }\left(
\epsilon _{p}\right) $ with a $\delta $ function in Eq.~(\ref{connelimG1}).
These two discrepancies cancel each other, leading to the same $T\ll T_{K}$
result as ours.

\section{Open long ABK rings\label{sec:open}}

We turn to the open long ABK ring, with strong electron leakage due to side
leads coupled to the arms of the ring, where our multi-terminal formalism
shows its full capacity.

\begin{figure}[ptb]
\includegraphics[width=0.8\textwidth]{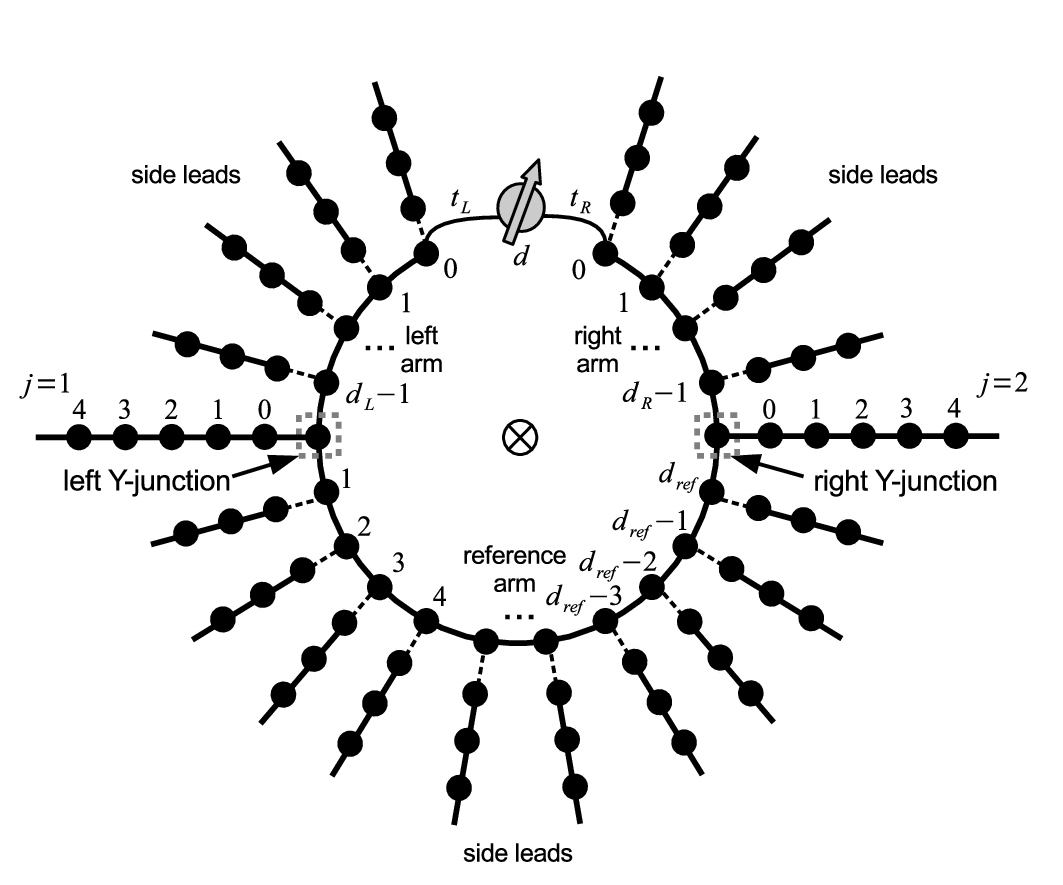}%
\caption{Geometry of the
open long ABK ring. Side leads are appended to the QD arms and the reference
arms, which are all of comparable lengths.\label{fig:longring2}}
\end{figure}

In our geometry shown in Fig.~\ref{fig:longring2}, the source lead branches
into two paths at the left Y-junction, a QD\ path of length $d_{L}+d_{R}$
and a reference path of length $d_{ref}$. These two paths converge at the
right Y-junction at the end of the drain lead. An embedded QD in the Kondo
regime separates the QD path into two arms of lengths $d_{L}$ and $d_{R}$.
To open up the ring we attach additional non-interacting side leads to all
sites inside the ring other than the two central sites in the Y-junctions
and QD.\cite%
{PhysRevB.66.115311,PhysRevB.72.073311,PhysRevB.72.245313,PhysRevB.86.115129}
The side leads, numbering $d_{L}+d_{R}+d_{ref}$ in total, are assumed to be
identical to the main leads (source and drain), except that the first link
on every side lead (connecting site $0$ of the side lead to its base site in
the ring) is assumed to have a hopping strength $t_{x}$ which is generally
different than the bulk hopping $t$. The Hamiltonian representing this model
is therefore

\begin{eqnarray}
H_{0\text{,junction}} &=&-t\left( \sum_{n=0}^{d_{L}-2}c_{L,n}^{\dag
}c_{L,n+1}+\sum_{n=0}^{d_{R}-2}c_{R,n}^{\dag
}c_{R,n+1}+\sum_{n=1}^{d_{ref}-1}c_{ref,n}^{\dag }c_{ref,n+1}+\text{h.c.}%
\right)  \notag \\
&&-\left[ \left( t_{JL}^{L}c_{1,0}^{\dag }+t_{JQ}^{L}c_{L,d_{L}-1}^{\dag
}+t_{JR}^{L}c_{ref,1}^{\dag }\right) c_{JL}+\text{h.c.}\right]  \notag \\
&&-\left[ \left( t_{JL}^{R}c_{2,0}^{\dag }+t_{JQ}^{R}c_{R,d_{R}-1}^{\dag
}+t_{JR}^{R}c_{ref,d_{ref}}^{\dag }\right) c_{JR}+\text{h.c.}\right]  \notag
\\
&&-t_{x}\left( \sum_{n=0}^{d_{L}-1}c_{L,n}^{\dag }c_{n,0}^{\left( L\right)
}+\sum_{n=0}^{d_{R}-1}c_{R,n}^{\dag }c_{n,0}^{\left( R\right)
}+\sum_{n=1}^{d_{ref}}c_{ref,n}^{\dag }c_{n,0}^{\left( ref\right) }+\text{%
h.c.}\right) \text{,}
\end{eqnarray}%
where $c_{JL\left( R\right) }$ is the annihilation operator on the central
site of the left (right) Y-junction, and $c_{n,0}^{\left( \alpha \right) }$\
is the annihilation operator on site $0$ of the side lead attached to the $n$%
th site on arm $\alpha $, $\alpha =L$, $R$ and $ref$. The coupling sites are 
$c_{C,r=1}\equiv c_{L,0}$ and $c_{C,r=2}\equiv c_{R,0}$, and again we let
the couplings to the QD be $t_{1}=t_{L}e^{i\frac{\varphi }{2}}$ and $%
t_{2}=t_{R}e^{-i\frac{\varphi }{2}}$.

Our hope is that in certain parameter regimes the open long ring provides a
realization of the two-path interferometer, where the two-slit interference
formula applys:

\begin{equation}
G_{sd}=G_{ref}+G_{d}+2\sqrt{\eta _{v}}\sqrt{G_{ref}G_{d}}\cos \left( \varphi
+\varphi _{t}\right) \text{,}  \label{twoslit}
\end{equation}%
where $G_{ref}$ is the conductance through the reference arm with the QD arm
sealed off, and $G_{d}$ is the conductance through the QD with the reference
arm sealed off. $\varphi $ is as before the AB phase,\ and $\varphi _{t}$ is
the accumulated non-magnetic phase difference of the two paths (\textit{%
including} the $\pi /2$ transmission phase through the QD). $\eta _{v}$ is
the unit-normalized visibility of the AB oscillations; $\eta _{v}=1$ at zero
temperature if all transport processes are coherent.\cite{PhysRevB.86.115129}
In the two-path interferometer regime, $\varphi _{t}$ reflects the intrinsic
transmission phase through the QD, provided that the geometric phases of the
two paths are the same (e.g. identical path lengths in a continuum model),
no external magnetic field is applied to the QD, and the particle-hole
symmetry breaking phase shift is zero.

For non-interacting embedded QDs well outside of the Kondo regime, small
transmission through the lossy arms is known to suppress multiple traversals
of the ring and ensure that the transmission amplitudes in two paths are
mutually independent.\cite{PhysRevB.66.115311} We show below that in our
interferometer with a Kondo QD, the same criterion renders the mesoscopic
fluctuations of the normalization factor $V_{k}^{2}$ negligible, and paves
the way to the two-slit condition $t_{sd}=t_{ref}+t_{d}e^{i\varphi }$. If we
additionally have small reflection by the lossy arms, then both the Kondo
temperature of the system and the intrinsic transmission amplitude through
the QD are the same as their counterparts for a QD directly embedded between
the source and the drain. At finite temperature $T\ll T_{K}$, we recover and
generalize the single-channel Kondo results of Ref.~%
\onlinecite{PhysRevB.86.115129} for the normalized visibility $\eta _{v}$
and the transmission phase $\varphi _{t}$.

\subsection{Wave function on a single lossy arm}

\begin{figure}[ptb]
\includegraphics[width=0.5\textwidth]{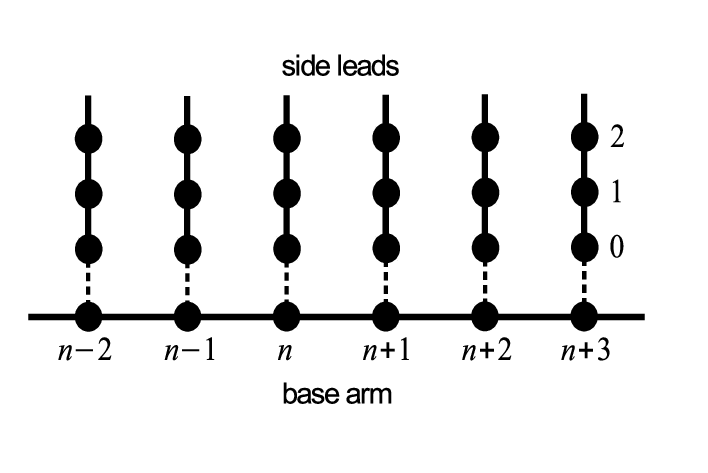}%
\caption{A single lossy
arm attached to side leads.\label{fig:lossyarm}}
\end{figure}

To solve for the background S-matrix $S$ and the wave function matrix $%
\Gamma $ of the open ring, we first analyze a single lossy arm attached to
side leads, depicted in Fig.~\ref{fig:lossyarm}.\cite{PhysRevB.66.115311}

Consider an arbitrary site labeled $n$ on this arm; let the wave function on
this site be $\phi _{n}$, and let incident and scattered amplitudes on the
side lead attached to this site be $A_{n}^{s}$ and $B_{n}^{s}$. The wave
function on site $l$ ($l\geq 0$)\ on the side lead is then written as $%
A_{n}^{s}e^{-ikl}+B_{n}^{s}e^{ikl}$. The Schroedinger's equations are

\begin{subequations}
\begin{equation}
t\left( A_{n}^{s}e^{ik}+B_{n}^{s}e^{-ik}\right) =t_{x}\phi _{n}\text{,}
\end{equation}

\begin{equation}
\left( -2t\cos k\right) \phi _{n}=-t\phi _{n-1}-t\phi _{n+1}-t_{x}\left(
A_{n}^{s}+B_{n}^{s}\right) \text{.}
\end{equation}%
Eliminating $B_{n}^{s}$, we find

\end{subequations}
\begin{equation}
\left( -2\cos k+\frac{t_{x}^{2}}{t^{2}}e^{ik}\right) \phi _{n}=-\phi
_{n-1}-\phi _{n+1}+e^{ik}\left( 2i\sin k\right) \frac{t_{x}}{t}A_{n}^{s}%
\text{.}  \label{teethScheq}
\end{equation}%
This means if $A_{n}^{s}=0$, i.e.\ no electron is incident from the side
lead $n$, we can write the wave function on the $n$th site on the arm as

\begin{equation}
\phi _{n}=C_{L}\eta _{1}^{n}+C_{R}\eta _{2}^{n}\text{,}  \label{lossteeth}
\end{equation}%
where $C_{L,R}$ are constants independent of $n$ and $k$. $\eta _{1,2}$ are
roots of the characteristic equation

\begin{equation}
\eta ^{2}+\left( -2\cos k+\frac{t_{x}^{2}}{t^{2}}e^{ik}\right) \eta +1=0%
\text{,}
\end{equation}%
so that $\eta _{1}\eta _{2}=1$. Hereafter we choose the convention $%
\left\vert \eta _{1}\right\vert <1$. When $t_{x}/t\ll \sin k$, to the lowest
nontrivial order in $t_{x}/t$,

\begin{equation}
\eta _{1}\approx e^{ik}\left( 1-\frac{t_{x}^{2}}{t^{2}}\frac{e^{ik}}{2i\sin k%
}\right) \text{,}
\end{equation}%
and thus $\left\vert \eta _{1}\right\vert ^{2}\approx 1-t_{x}^{2}/t^{2}$.

Eq.~(\ref{lossteeth}) bypasses the difficulty of solving for each $\phi _{n}$
individually: on the same arm the constants $C_{L}$ and $C_{R}$ only change
where the side lead incident amplitude $A_{n}^{s}\neq 0$.

Let us now quantify the conditions of small transmission and small
reflection. Connecting external leads smoothly to both ends of a lossy arm
of length $d_{A}\gg 1$, we may write the scattering state wave function
incident from one end as

\begin{equation}
\left\{ 
\begin{array}{c}
e^{-ikn}+\tilde{R}e^{ikn}\left( \text{left lead, }n=0,1,2,\cdots \right) \\ 
C_{L}\eta _{1}^{n}+C_{R}\eta _{1}^{-n}\left( \text{lossy arm, }n=1,\cdots
,d_{A}\right) \\ 
\tilde{T}e^{ikn}\left( \text{right lead, }n=0,1,2,\cdots \right)%
\end{array}%
\right. \text{;}
\end{equation}%
the Schroedinger's equation then yields

\begin{subequations}
\begin{equation}
1+\tilde{R}=C_{L}+C_{R}\text{,}
\end{equation}

\begin{equation}
e^{ik}+\tilde{R}e^{-ik}=C_{L}\eta _{1}+C_{R}\eta _{1}^{-1}\text{,}
\end{equation}

\begin{equation}
\tilde{T}=C_{L}\eta _{1}^{d_{A}+1}+C_{R}\eta _{1}^{-d_{A}-1}\text{,}
\end{equation}

\begin{equation}
\tilde{T}e^{-ik}=C_{L}\eta _{1}^{d_{A}}+C_{R}\eta _{1}^{-d_{A}}\text{.}
\end{equation}%
It is now straightforward to find the transmission and reflection
coefficients:

\end{subequations}
\begin{subequations}
\begin{equation}
\tilde{T}=\frac{e^{ik}\left( e^{2ik}-1\right) \eta _{1}^{d_{A}}\left( \eta
_{1}^{2}-1\right) }{1-\eta _{1}^{2d_{A}+2}+2e^{ik}\eta _{1}\left( \eta
_{1}^{2d_{A}}-1\right) +e^{2ik}\left( \eta _{1}^{2}-\eta
_{1}^{2d_{A}}\right) }\text{,}
\end{equation}

\begin{equation}
\tilde{R}=\frac{e^{ik}\left( \eta _{1}^{2d_{A}}-1\right) \left[ e^{ik}\left(
1+\eta _{1}^{2}\right) -\left( e^{2ik}+1\right) \eta _{1}\right] }{1-\eta
_{1}^{2d_{A}+2}+2e^{ik}\eta _{1}\left( \eta _{1}^{2d_{A}}-1\right)
+e^{2ik}\left( \eta _{1}^{2}-\eta _{1}^{2d_{A}}\right) }\text{.}
\end{equation}

At $k=0$ or $\pi $ we always have trivially$\left\vert \tilde{R}\right\vert
=1$ and $\left\vert \tilde{T}\right\vert =0$; we therefore focus on energies
that are not too close to the band edges, so that $\sin k$ is not too small.
In this case, under the long arm assumption $d_{A}\gg 1$, the small
transmission condition $\left\vert \tilde{T}\right\vert \ll 1$ is satisfied
if and only if $\left\vert \eta _{1}\right\vert ^{d_{A}}\ll 1$, and the
small reflection condition $\left\vert \tilde{R}\right\vert \ll 1$ is
satisfied if and only if $t_{x}\ll t$.\cite{PhysRevB.66.115311}

\subsection{Background S-matrix and coupling site wave functions}

We now return to the open long ring model to solve for $S$ and $\Gamma $
with the aid of Eq.~(\ref{lossteeth}). Let us denote the incident amplitude
vector by

\end{subequations}
\begin{equation}
\left( A_{1},A_{2},A_{0}^{\left( L\right) },\cdots ,A_{d_{L}-1}^{\left(
L\right) },A_{1}^{\left( ref\right) },\cdots ,A_{d_{L}}^{\left( ref\right)
},A_{0}^{\left( R\right) },\cdots ,A_{d_{R}-1}^{\left( R\right) }\right) ^{T}%
\text{;}
\end{equation}%
here $A_{n}^{\left( \alpha \right) }$\ is the incident amplitude in the side
lead attached to the $n$th site on arm $\alpha $. We are interested in the
normalization factor $V_{k}^{2}$ and the source-lead component of the
conductance tensor $G_{12}$; for this purpose, according to Eqs.~(\ref%
{Vksquared}) and (\ref{connelimG2}), the first two rows of the S-matrix $S$
and the full coupling site wave function matrix $\Gamma $ must be found. In
other words, we need to express the scattered amplitudes in the main leads $%
B_{1}$, $B_{2}$, as well as the wave functions at the coupling sites $\Gamma
_{1}$ and $\Gamma _{2}$, in terms of incident amplitudes. Some details of
this straightforward calculation are given in Appendix~\ref{sec:appopen},
and we skip to the solution now.

If we assume $d_{L}\sim d_{R}\sim d_{ref}/2\gg 1$ (comparable arm lengths
and path lengths in the long ring) and $\left\vert \eta _{1}\right\vert
^{d_{L}}\ll 1$ (small transmission criterion), to $O\left( \left\vert \eta
_{1}\right\vert ^{d_{L}}\right) $ we have

\begin{subequations}
\label{longringBGamma}
\begin{align}
B_{1}& =S_{L11}^{\prime }A_{1}+S_{L13}^{\prime }S_{R31}^{\prime }\eta
_{1}^{d_{ref}-1}A_{2}-\sum_{n=0}^{d_{L}-1}e^{ik}\left( 2i\sin k\right) \frac{%
t_{x}}{t}\frac{\eta _{1}^{n+1}-\eta _{1}^{-n-1}}{\eta _{1}-\eta _{1}^{-1}}%
S_{L12}^{\prime }\eta _{1}^{d_{L}}A_{n}^{\left( L\right) }  \notag \\
& +\sum_{n=1}^{d_{ref}}e^{ik}\left( 2i\sin k\right) \frac{t_{x}}{t}\frac{%
S_{R33}^{\prime }\eta _{1}^{d_{ref}-n}+\eta _{1}^{-d_{ref}+n}}{\eta
_{1}-\eta _{1}^{-1}}S_{L13}^{\prime }\eta _{1}^{d_{ref}-1}A_{n}^{\left(
ref\right) }\text{,}
\end{align}

\begin{align}
B_{2} &
=S_{L31}^{\prime}S_{R13}^{\prime}\eta_{1}^{d_{ref}-1}A_{1}+S_{R11}^{%
\prime}A_{2}+\sum_{n=1}^{d_{ref}}e^{ik}\left( 2i\sin k\right) \frac{t_{x}}{t}%
\frac{\eta_{1}^{-n+1}+S_{L33}^{\prime}\eta_{1}^{n-1}}{\eta _{1}-\eta_{1}^{-1}%
}S_{R13}^{\prime}\eta_{1}^{d_{ref}-1}A_{n}^{\left( ref\right) }  \notag \\
& -\sum_{n=0}^{d_{R}-1}e^{ik}\left( 2i\sin k\right) \frac{t_{x}}{t}\frac{%
\eta_{1}^{n+1}-\eta_{1}^{-n-1}}{\eta_{1}-\eta_{1}^{-1}}S_{R12}^{\prime
}\eta_{1}^{d_{R}}A_{n}^{\left( R\right) }\text{,}
\end{align}

\begin{align}
\Gamma _{1}& =S_{L21}^{\prime }\eta _{1}^{\left( d_{L}-1\right) }\left(
1-\eta _{1}^{2}\right) A_{1}-\sum_{n=0}^{d_{L}-1}e^{ik}\left( 2i\sin
k\right) \frac{t_{x}}{t}\left( \eta _{1}^{-d_{L}+n+1}+S_{L22}^{\prime }\eta
_{1}^{d_{L}-n-1}\right) \eta _{1}^{d_{L}}A_{n}^{\left( L\right) }  \notag \\
& -\sum_{n=1}^{d_{ref}}e^{ik}\left( 2i\sin k\right) \frac{t_{x}}{t}\left(
S_{R33}^{\prime }\eta _{1}^{d_{ref}-n}+\eta _{1}^{-d_{ref}+n}\right)
S_{L23}^{\prime }\eta _{1}^{d_{L}}\eta _{1}^{d_{ref}-1}A_{n}^{\left(
ref\right) }\text{,}
\end{align}

\begin{align}
\Gamma _{2}& =S_{R21}^{\prime }\eta _{1}^{\left( d_{R}-1\right) }\left(
1-\eta _{1}^{2}\right) A_{2}-\sum_{n=1}^{d_{ref}}e^{ik}\left( 2i\sin
k\right) \frac{t_{x}}{t}\left( \eta _{1}^{-n+1}+S_{L33}^{\prime }\eta
_{1}^{n-1}\right) S_{R23}^{\prime }\eta _{1}^{d_{R}}\eta
_{1}^{d_{ref}-1}A_{n}^{\left( ref\right) }  \notag \\
& -\sum_{n=0}^{d_{R}-1}e^{ik}\left( 2i\sin k\right) \frac{t_{x}}{t}\left(
\eta _{1}^{-d_{R}+n+1}+S_{R22}^{\prime }\eta _{1}^{d_{R}-n-1}\right) \eta
_{1}^{d_{R}}A_{n}^{\left( R\right) }\text{.}
\end{align}%
Here the $3\times 3$ matrices $S_{L}^{\prime }$ and $S_{R}^{\prime }$ are
defined in Eqs.~(\ref{longringSprL}) and (\ref{longringSprR}); they are
generally not unitary. Being properties of the Y-junctions, they are
independent of the amplitudes ($A$, $B$ etc.) and arm lengths ($d_{L}$, $%
d_{R}$ and $d_{ref}$), as can be seen from e.g. Eq.~(\ref{longringSprL1}).
In the limit $t_{x}/t=0$, $S_{L}^{\prime }$ and $S_{R}^{\prime }$ turn into
the usual unitary S-matrices $S_{L}$ and $S_{R}$.

\subsection{Kondo temperature and conductance}

To the lowest nontrivial order in $\left\vert \eta \right\vert ^{d_{L}}$,
Eq.~(\ref{longringBGamma}) leads to the following simple results after some
algebra:

\end{subequations}
\begin{equation}
V_{k}^{2}=-\left( \eta_{1}-\eta_{1}^{\ast}\right) \left( 2i\sin k\right)
\left( t_{L}^{2}+t_{R}^{2}\right) \text{,}  \label{longringVksq}
\end{equation}

\begin{equation}
S_{12}\left( k\right) =S_{L13}^{\prime}S_{R31}^{\prime}\eta_{1}^{d_{ref}-1}%
\text{,}  \label{longringS12}
\end{equation}

\begin{equation}
\left[ S\left( k\right) \Gamma ^{\dag }\left( k\right) \lambda \Gamma \left(
k\right) \right] _{12}=t_{L}t_{R}e^{i\varphi }\left( 2i\sin k\right) \left(
\eta _{1}-\eta _{1}^{-1}\right) S_{L12}^{\prime }S_{R21}^{\prime }\eta
_{1}^{d_{L}+d_{R}}\text{.}  \label{longringSdot12}
\end{equation}%
In obtaining Eq.~(\ref{longringSdot12}) we have used the algebraic identity

\begin{equation}
S_{L21}^{\prime \ast }S_{L11}^{\prime }\left\vert \left( 1-\eta
_{1}^{2}\right) \right\vert ^{2}=\left( 4\sin ^{2}k\right) \left( \frac{t_{x}%
}{t}\right) ^{2}\left[ \frac{\left\vert \eta _{1}\right\vert ^{2}\eta
_{1}^{\ast }S_{L12}^{\prime }}{\eta _{1}-\eta _{1}^{\ast }}-\frac{\left\vert
\eta _{1}\right\vert ^{2}}{1-\left\vert \eta _{1}\right\vert ^{2}}\left(
S_{L22}^{\prime \ast }S_{L12}^{\prime }+S_{L23}^{\prime \ast
}S_{L13}^{\prime }\right) \right] \text{;}
\end{equation}%
in the limit $t_{x}\rightarrow 0$ this is just a statement of the S-matrix
unitarity.

\begin{figure}
\includegraphics[width=0.6\textwidth]{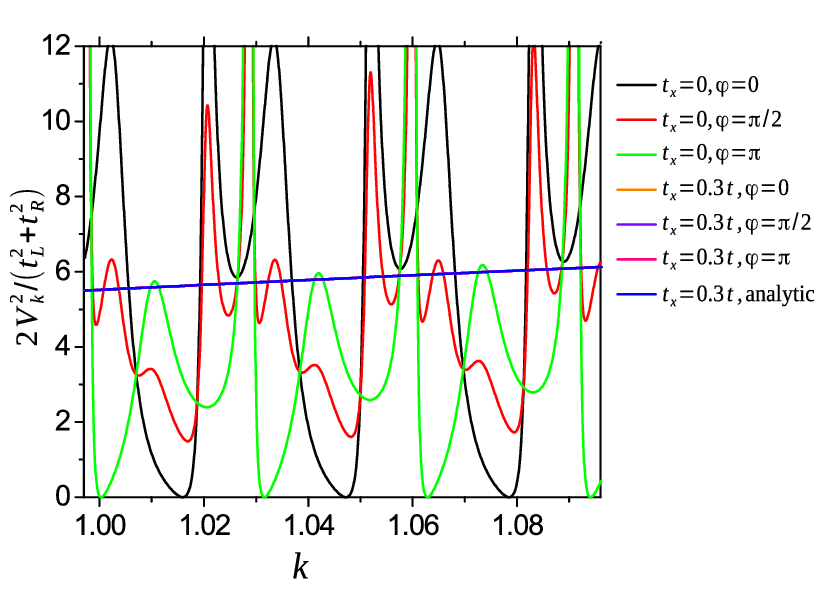}%
\caption{Normalization factor
$V_{k}^{2}$ from Eq.~(\ref{Vksquared}) as a function of $k$ for different AB
phases $\varphi$ in the open long ABK ring, obtained by solving the full
tight-binding model. We focus on a small slice of momentum $\left| k-\pi/3
\right|<0.05$. Two values of $t_{x}$ are considered: $t_{x}=0$ corresponding
to the closed ring without electron leakage, and $t_{x}=0.3t$ corresponding
to strong leakage along and small transmission across the arms. System
parameters are: $d_{L}=d_{R}=d_{ref}/2=100$,
$t^{L,R}_{JL}=t^{L,R}_{JQ}=t^{L,R}_{JR}=t$, and symmetric QD coupling
$t_{L}=t_{R}$. For comparison we have also plotted the analytic prediction
Eq.~(\ref{longringVksq}) for $t_{x}=0.3t$, which agrees quantitatively with
the full tight-binding solution. While $V_{k}^{2}$ for the closed ring is
extremely sensitive to $k_{F}$ and $\varphi$, the sensitivity is strongly
suppressed by electron leakage, and curves for different $\varphi$ overlap when $t_{x}=0.3t$. Since $V_{k}^{2}$ controls the
renormalization of the Kondo coupling, the Kondo temperature of the open
long ABK ring is not sensitive to mesoscopic details in the small
transmission limit.\label{fig:openVksq}}
\end{figure}

Eq.~(\ref{longringVksq}) tells us that, in the small transmission limit, the
normalization factor $V_{k}^{2}$ exhibits little mesoscopic fluctuation, so
that $E_{V}\sim t$; furthermore, it does not depend on the AB phase $\varphi 
$ at all (see Fig.~\ref{fig:openVksq}). When we also impose the small
reflection condition $t_{x}\ll t$, $\eta _{1}\approx e^{ik}$ and we find $%
V_{k}^{2}\approx \left( 4\sin ^{2}k\right) \left( t_{L}^{2}+t_{R}^{2}\right) 
$; this is precisely the normalization factor for a QD embedded between
source and drain leads. We recall from Eq.~(\ref{KondojRG}) that the
normalization of Kondo coupling is governed by $V_{k}^{2}$. Therefore, at
least for our simple model of an interacting QD, the conditions of small
transmission and small reflection combine to reduce the Kondo temperature of
the open long ABK ring to that of the simple embedded geometry, independent
of the details of the ring or the AB flux.

Proceeding with the small transmission assumption, we observe that since $%
E_{V}\sim t$, the distinction between small and large Kondo cloud regimes is
no longer applicable. This is presumably because the Kondo cloud leaks into
the side leads in the open ring, and is no longer confined in a mesoscopic
region as in the closed ring. The low-energy theory of our model is
therefore the usual local Fermi liquid. At zero temperature, the connected
contribution to the conductance vanishes, and the conductance $G_{12}$ is
proportional to the disconnected transmission probability Eq.~(\ref{discprob}%
) at the Fermi energy:

\begin{equation}
-\mathcal{T}_{12}^{D}\left( \epsilon _{k_{F}}\right) =\left\vert
S_{L13}^{\prime }S_{R31}^{\prime }\eta _{1}^{d_{ref}-1}+\frac{2t_{L}t_{R}}{%
t_{L}^{2}+t_{R}^{2}}e^{i\varphi }\frac{\eta _{1}-\eta _{1}^{-1}}{\eta
_{1}-\eta _{1}^{\ast }}S_{L12}^{\prime }S_{R21}^{\prime }\eta
_{1}^{d_{L}+d_{R}}\frac{1}{2}\left( e^{2i\delta _{P}}+1\right) \right\vert
^{2}\text{,}  \label{longringTD12}
\end{equation}%
where $\eta _{1}$, $S_{L}^{\prime }$ and $S_{R}^{\prime }$ are all evaluated
at the Fermi surface, and we have used Eqs.~(\ref{deltapsipsi}) and (\ref%
{TvsTtilde}). In the small reflection limit, Eq.~(\ref{longringTD12}) becomes

\begin{equation}
-\mathcal{T}_{12}^{D}\left( \epsilon _{k_{F}}\right) =\left\vert
S_{L13}S_{R31}\eta _{1}^{d_{ref}-1}+e^{i\varphi }S_{L12}S_{R21}\eta
_{1}^{d_{L}+d_{R}-2}t_{QD}\right\vert ^{2}\text{,}
\end{equation}%
where $S_{L,R}\equiv S_{L,R}^{\prime }\left( t_{x}\rightarrow 0\right) $ are
the aforementioned S-matrices of the Y-junctions, and

\begin{equation}
t_{QD}\equiv e^{2ik}\frac{2t_{L}t_{R}}{t_{L}^{2}+t_{R}^{2}}\frac{1}{2}\left(
e^{2i\delta _{P}}+1\right)
\end{equation}%
is the transmission amplitude through an embedded QD in the Kondo limit [see
Eq. (\ref{tQD})].

It is clear from Eq.~(\ref{longringTD12}) that the two-slit condition $%
t_{sd}=t_{ref}+t_{d}e^{i\varphi }$ is satisfied at zero temperature.
Furthermore, under the small reflection condition, both $%
t_{ref}=S_{L13}S_{R31}\eta _{1}^{d_{ref}-1}$ and $t_{d}=S_{L12}S_{R21}\eta
_{1}^{d_{L}+d_{R}-2}t_{QD}$ have straightforward physical interpretations;
in particular $t_{d}$ can be factorized into a part $t_{QD}$\ which is the
intrinsic transmission amplitude through QD, and a part due completely to
the rest of the QD arm and the two Y-junctions.

We now consider the finite temperature conductance, assuming realistically
that $T\ll t$ and $T_{K}\ll t$. If we further assume that the two
Y-junctions are non-resonant, so that $S_{L}^{\prime }$ and $S_{R}^{\prime }$
change significantly as functions of energy only on the scale of the
bandwidth $4t$, then\ mesoscopic fluctuations are entirely absent from Eq.~(%
\ref{longringSdot12}), i.e. $E_{\text{conn}}\sim t$. (It is worth mentioning
that $E_{\text{conn}}$ can be much less than $t$\ if the Y-junctions allow
resonances, e.g. when the central site of each Y-junction is weakly coupled
to all three legs; however, $E_{V}\sim t$ even in this case.) Since $T\ll E_{%
\text{conn}}$, we can comfortably eliminate the connected contribution and
use Eq.~(\ref{connelimG2}). At $T\ll T_{K}$ the total Fermi liquid regime
conductance $G\left( T,\varphi \right) \equiv -G_{12}$ is found to $O\left(
T/T_{K}\right) ^{2}$:

\begin{subequations}
\begin{align}
& G\left( T,\varphi \right) \equiv G_{ref}+G_{d}+2\sqrt{G_{ref}G_{d}}\left\{
\cos \left( \varphi +\theta +\delta _{P}\right) -\left( \frac{\pi T}{T_{K}}%
\right) ^{2}\right.  \notag \\
& \left. \times \left[ \frac{\cos \left( \varphi +\theta +\delta _{P}\right) 
}{2\cos ^{2}\delta _{P}}-\tan \delta _{P}\sin \left( \varphi +\theta +\delta
_{P}\right) \right] \right\} \text{.}  \label{longringG12Fermi}
\end{align}%
Here the conductance through the reference path is defined as

\begin{equation}
G_{ref}\equiv \frac{2e^{2}}{h}\left\vert S_{L13}^{\prime }S_{R31}^{\prime
}\eta _{1}^{d_{ref}-1}\right\vert ^{2}\text{,}
\end{equation}%
the conductance through the QD path is defined as

\begin{equation}
G_{d}\left( T\right) =G_{d}^{\left( 0\right) }\left[ \cos ^{2}\delta
_{P}-\left( \frac{\pi T}{T_{K}}\right) ^{2}\cos 2\delta _{P}\right]
\end{equation}%
with its $T=0$ and $\delta _{P}=0$ value

\begin{equation}
G_{d}^{\left( 0\right) }\equiv \frac{2e^{2}}{h}\frac{4t_{L}^{2}t_{R}^{2}}{%
\left( t_{L}^{2}+t_{R}^{2}\right) ^{2}}\left\vert \frac{\eta _{1}-\eta
_{1}^{-1}}{\eta _{1}-\eta _{1}^{\ast }}S_{L12}^{\prime }S_{R21}^{\prime
}\eta _{1}^{d_{L}+d_{R}}\right\vert ^{2}\text{,}
\end{equation}%
and finally the non-magnetic phase difference between the QD path and the
reference path (\textit{including} the QD) in the absence of $\delta _{P}$ is

\begin{equation}
\theta =\arg \left( \frac{\eta _{1}-\eta _{1}^{-1}}{\eta _{1}-\eta
_{1}^{\ast }}\eta _{1}^{d_{L}+d_{R}-d_{ref}+1}\frac{S_{L12}^{\prime
}S_{R21}^{\prime }}{S_{L13}^{\prime }S_{R31}^{\prime }}\right) \text{.}
\end{equation}%
Once again, $\eta _{1}$, $S_{L}^{\prime }$ and $S_{R}^{\prime }$ are all
evaluated at the Fermi surface.

For $T\gg T_{K}$, we discuss two different scenarios: the particle-hole
symmetric case and the strongly particle-hole asymmetric case. In the
particle-hole symmetric case, as explained in Sec.~\ref{sec:pert}, the
potential scattering term $K$ vanishes, and the $O\left( J^{2}\right) $\
connected contribution plays an important role. Inserting Eqs.~(\ref%
{longringVksq})--(\ref{longringSdot12}) into Eqs.~(\ref{deltaTjjprKondo})
and (\ref{GjjprKondo}), we find the total high-temperature conductance in
the particle-hole symmetric case to be

\end{subequations}
\begin{equation}
G\left( T,\varphi \right) =G_{ref}+G_{d}+2\frac{\sqrt{3}}{4}\frac{\pi }{\ln 
\frac{T}{T_{K}}}\sqrt{G_{ref}G_{d}}\cos \left( \varphi +\theta \right) \text{%
.}  \label{longringG12Kondosym}
\end{equation}%
Here the conductance through the reference path $G_{ref}$ is given as
before, while the conductance through the QD path has the usual logarithmic
temperature dependence,

\begin{equation}
G_{d}\left( T\right) \equiv \frac{3}{16}\frac{\pi ^{2}}{\ln ^{2}\frac{T}{%
T_{K}}}G_{d}^{\left( 0\right) }\text{.}
\end{equation}%
We have taken into account the renormalization of the Kondo coupling, Eq.~(%
\ref{KondojRGsolsc}); thermal averaging cuts off the logarithm at $T$. For
our slowly varying $V_{k}^{2}$ given by Eq.~(\ref{longringVksq}), $\overline{%
V_{k}^{2}}$ is simply the Fermi surface value $V_{k_{F}}^{2}$, and the Kondo
temperature is defined by Eq.~(\ref{TKKondobulk}).

Comparing Eqs.~(\ref{longringG12Fermi}) and (\ref{longringG12Kondosym}) and
noting that $\delta _{P}=0$, we find that there is no phase shift between $%
T\ll T_{K}$ and $T\gg T_{K}$ in the presence of particle-hole symmetry,
which is consistent with e.g. Fig. 4(d) of Ref.~%
\onlinecite{PhysRevLett.84.3710}. We also observe that the particle-hole
symmetric normalized visibility $\eta _{v}$, defined in Eq.~(\ref{twoslit}),
has a characteristic logarithmic behavior at $T\gg T_{K}$:

\begin{equation}
\eta _{v}=\frac{3}{16}\frac{\pi ^{2}}{\ln ^{2}\frac{T}{T_{K}}}\text{.}
\end{equation}

\begin{figure}
\includegraphics[width=0.6\textwidth]{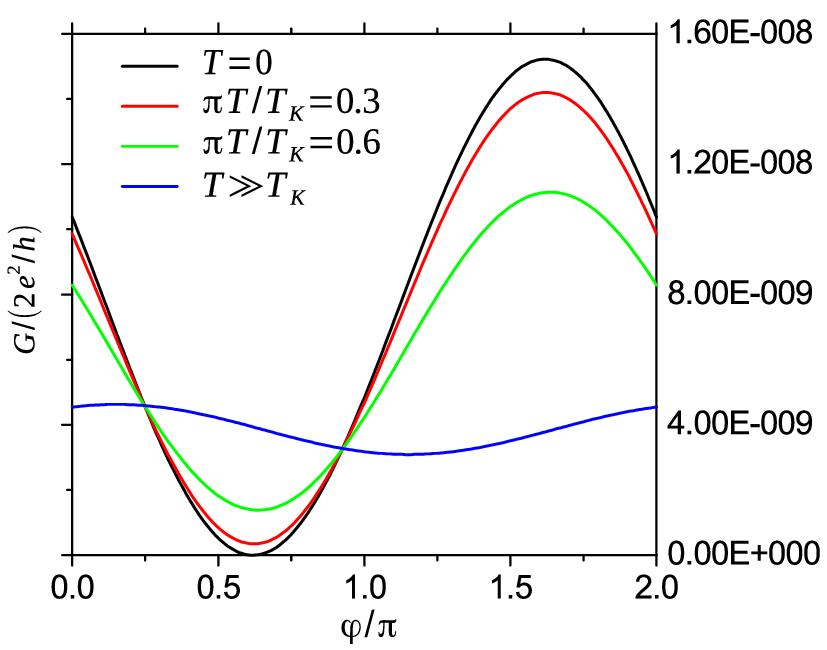}%
\caption{Low-temperature and high-temperature conductances $G$ as functions
of AB phase $\varphi$ in the open long ABK ring with a particle-hole asymmetric QD, calculated with Eqs.~(\ref{longringG12Fermi})
and (\ref{longringG12Kondoasym}). We assume
$T_{K}\ll t$ so that the thermal averaging in the high temperature case is trivial. System
parameters are: $t_{x}=0.3t$, $k_{F}=\pi/3$, $d_{L}=d_{R}=d_{ref}/2=100$,
$t^{L,R}_{JL}=t^{L,R}_{JQ}=t^{L,R}_{JR}=t$, and particle-hole symmetry
breaking phase shift $\delta_{P}=0.1$. A phase shift of approximately
$\pi/2$ is clearly visible as the temperature is lowered and Kondo
correlations become important.\label{fig:phaseshift}}
\end{figure}

On the other hand, to demonstrate the $\pi /2$ phase shift due to Kondo
physics, it is more useful to consider the case of relatively strong
particle-hole asymmetry $\kappa \sim j\left( T\right) $ at $T\gg T_{K}$. The leading
contribution to the conductance from potential scattering is $O\left(
K\right) $, and the leading contribution from the Kondo coupling is $O\left(
J^{2}\right) $; therefore, $\kappa \sim j\left( T\right) \ $indicates that
we may neglect the Kondo coupling altogether at temperature $T$. To the
lowest order in potential scattering $O\left( K\right) $, Eq.~(\ref%
{deltaTjjprpot}) applies; also, since $T\ll t$, the thermal averging in Eq.~(%
\ref{GjjprKondo}) becomes trivial. Using the relation between $K$ and $%
\delta _{P}$, Eq.~(\ref{tandeltaP}), we finally obtain

\begin{equation}
G\left( T,\varphi \right) =G_{ref}+2\sqrt{G_{ref}G_{d}^{\left( 0\right) }}%
\tan \delta _{P}\sin \left( \varphi +\theta \right) \text{.}
\label{longringG12Kondoasym}
\end{equation}%
Comparing Eqs.~(\ref{longringG12Fermi}) and (\ref{longringG12Kondoasym}), it
becomes evident that transmission through the QD undergoes a $\pi /2+\delta
_{P}$ phase shift from $T\gg T_{K}$ to $T\ll T_{K}$ as the Kondo
correlations are switched on; see Fig.~\ref{fig:phaseshift}. We remark that the strongly particle-hole asymmetric case represents the situation without Kondo correlation whereas the particle-hole symmetric case does not. This is because in the latter case the leading QD contribution to the conductance is $O\left( J^2 \right)$, which is of Kondo origin as we have discussed above Eq.~(\ref{TmatKondosym}).

We can again make a direct comparison with Eq.~(\ref{twoslit}). While $%
\theta $ itself is not necessarily $\pi /2$, $\varphi _{t}$ is
experimentally observed with respect to its value with Kondo correlations
turned off, so we should define the reference value $\varphi _{t}^{\left(
0\right) }$ by e.g. comparing with Eq.~(\ref{longringG12Kondoasym}):

\begin{equation}
\varphi _{t}^{\left( 0\right) }=\theta -\frac{\pi }{2}\text{.}
\end{equation}%
Therefore, to $O\left( T/T_{K}\right) ^{2}$ we readily obtain the following
results for the $T\ll T_{K}$ transmission phase and the normalized
visibility:

\begin{equation}
\varphi _{t}-\varphi _{t}^{\left( 0\right) }=\frac{\pi }{2}+\delta
_{P}-\left( \frac{\pi T}{T_{K}}\right) ^{2}\tan \delta _{P}\text{,}
\label{transphase}
\end{equation}

\begin{equation}
\eta _{v}=1-\left( \frac{\pi T}{T_{K}}\right) ^{2}\frac{1}{\cos ^{2}\delta
_{P}}\text{.}  \label{etav}
\end{equation}%
These are in agreement with the $\left\vert \delta _{P}\right\vert \ll 1$, $%
T=0$ and $\delta _{P}=0$, $T\ll T_{K}$\ results of Ref.~%
\onlinecite{PhysRevB.86.115129}, which assumes $\varphi _{t}^{\left(
0\right) }=0$, i.e. the non-magnetic phase difference between the two paths
is zero without Kondo correlations. Note that in obtaining the $T$
dependence in Eq.~(\ref{etav}) it is crucial to include the connected
contribution to conductance.

We stress that our $O\left( T/T_{K}\right) ^{2}$ results for the
transmission phase across the QD and the normalized visibility, Eqs.~(\ref%
{transphase}) and (\ref{etav}), are both exact in $\delta _{P}$, which is
non-universal and encompasses the effects of all particle-hole symmetry
breaking perturbation. In particular, the $\left( T/T_{K}\right) ^{2}$
coefficients were not reported previously.

\section{Conclusion and open questions\label{sec:outlook}}

In this paper, we generalized the method developed in Ref.~%
\onlinecite{PhysRevB.88.245104} to calculate the linear DC conductance
tensor of a generic multi-terminal Anderson model with an interacting QD.
The linear DC conductance of the system has a disconnected contribution of
the Landauer form, and a connected contribution which is also a Fermi
surface property. At temperatures low compared to the mesoscopic energy
scale below which the background S-matrix and the coupling site wave
functions vary slowly, $T\ll E_{\text{conn}}$, the connected contribution
can be approximately eliminated using properties of the conductance tensor;
the elimination procedure physically corresponds to probing the current
response or applying the bias voltages in a particular manner. At
temperatures high compared to the Kondo temperature $T\gg T_{K}$ this
connected part is computed explicitly to $O\left( J^{2}\right) $, and found
to be of the same order of magnitude as the disconnected part in the case of
a particle-hole symmetric QD.

With this method we scrutinize both closed and open long ABK ring models. We
find modifications to early results on the closed ring with a long reference
arm of length $L$: the high-temperature conductance at $T\gg T_{K}$ should
have qualitatively distinct behaviors for $T\gg v_{F}/L$ and $T\ll v_{F}/L$.
In the open ring we conclude that the two-path interferometer is realized
when the arms on the ring have weak transmission and weak reflection, and
demonstrate the possibility to observe in this device the $\pi /2$ phase
shift due to Kondo physics, and the suppression of AB oscillation visibility
due to inelastic scattering.

One question we have not so far addressed is the low-energy physics in the
small Kondo cloud regime, $T\ll T_{K}$ and $E_{V}\ll T_{K}$; here $E_{V}$ is
the energy scale below which the normalization factor $V_{k}^{2}$
controlling the Kondo temperature varies slowly, and $E_{V}\gtrsim E_{\text{%
conn}}$. We assume $E_{V}$ and $E_{\text{conn}}$ are of the same order of
magnitude, a condition satisfied by both the closed ring ($E_{V}\sim E_{%
\text{conn}}\sim v_{F}/L$) and the open ring with non-resonant Y-junctions ($%
E_{V}\sim E_{\text{conn}}\sim t$). For temperatures above the mesoscopic
energy scale $E_{V}\ll T\ll T_{K}$, we are no longer able to eliminate the
connected contribution. However, because $T\gg E_{V}$ one can argue that
physics associated with the energy scale $E_{V}$ is smeared out by thermal
fluctuations, and the mesoscopic system behaves as a bulk system with
parameters showing no mesoscopic fluctuations.\cite{EurophysLett.97.17006}
On the other hand, below the mesoscopic energy scale $T\ll E_{V}\ll T_{K}$,
since for $T\ll E_{\text{conn}}$ our formalism predicts that the connected
part can be eliminated, the knowledge of the screening channel T-matrix in
the single-particle sector alone is adequate for us to predict the
conductance. Ref.~\onlinecite{EurophysLett.97.17006} again offers an
appealing hypothesis: the low-energy effective theory is again a Fermi
liquid theory, with the T-matrix governed by Kondo physics at short range $%
\sim O\left( L_{K}\right) =O\left( v_{F}/T_{K}\right) $ and modulated by
mesoscopic fluctuations at long range $\sim O\left( L\right) $. This
scenario leads to a quasiparticle spectrum which is in qualitative agreement
with slave boson mean field theory.\cite{EurophysLett.97.17006} The Fermi
liquid picture is often analyzed by a renormalized perturbation theory of
the quasiparticles, where the bare parameters of the QD are replaced by
renormalized values; in particular the large $U$ between bare electrons is
replaced by a small renormalized $\tilde{U}$ between quasiparticles.\cite%
{hewson1997kondo} In the small Kondo cloud regime, we expect that the real
space geometry in the renormalized perturbation theory resembles that of the
bare theory;\cite{PhysRevLett.89.206602} thus a perturbation theory
calculation in $U$ in our formalism is potentially useful in understanding
the low-energy physics, as long as $U$ is interpreted as the effective $%
\tilde{U}$. It will be interesting to test this $E_{V}\ll T_{K}$ picture,
along with our perturbative predictions on conductance in other parameter
regimes in this paper, against results obtained from the numerical RG
algorithm.\cite{PhysRevLett.84.3710,PhysRevLett.87.156803,PhysRevB.77.180404}

There is also an issue regarding the assumption of a single-level QD in the
Kondo regime. To experimentally detect the $\pi /2$ phase shift in an AB
interferometer, one typically sweeps the plunger gate voltage on the QD, and
monitors the phase shift between consecutive Coulomb blockade peaks. A $\pi
/2$ plateau should be observed at $T\lesssim T_{K}$ near the center of each
Coulomb valley deep in the local moment regime, with an odd number of
electrons on the QD.\cite{PhysRevLett.84.3710,PhysRevLett.113.126601}
However, one needs to adopt a multi-level QD model to quantitatively
reproduce the experimental results, in particular the phase shift lineshape
asymmetry relative to the center of a valley, and also possibly a phase
lapse inside the valley.\cite{PhysRevLett.90.106602,PhysRevLett.113.126601}
A generalization of the current formalism to the multi-level case is
necessary in order to quantify the influence of the interferometer on the
measured transmission phase shift through a\ realistic QD.

Another natural open problem is the extension to the multi-channel Kondo
physics. In our generalized Anderson model, separation of screening and
non-screening channels is achieved in a single-level QD, and there is only
one effective screening channel. Exotic physics emerges in the presence of
two or more screening channels, realizable in e.g. a many-QD system.\cite%
{PhysRevB.48.7297,PhysRevLett.90.136602,Nature.446.167} In the 2-channel
Kondo effect with identical couplings to two channels, for example, the
low-energy physics is governed by a non-Fermi liquid RG fixed point: at zero
temperature a single particle scattered by the impurity can only enter a
many-body state, and there are no elastic single-particle scattering events.%
\cite{PhysRevLett.93.107204} Ref.~\onlinecite{PhysRevB.86.115129} discusses
the manifestations of the 2-channel Kondo physics in the two-path
interferometer, but again makes the two-slit assumption without examining
its validity. Therefore an extension of our approach to the multi-channel
case will be useful to justify the two-slit assumption in the open long ring
and thus the 2-channel predictions of Ref.~\onlinecite{PhysRevB.86.115129}.

\begin{acknowledgments}
This work was supported in part by NSERC of Canada, Discovery Grant
36318-2009 (ZS). ZS is grateful to Prof. Ian Affleck for initiating the
project, extensive discussions and proofreading the manuscript. YK
gratefully acknowledges the kind hospitality of UBC Department of Physics
and Astronomy during his visit. The authors would also like to acknowledge
fruitful discussions with Prof. L. I. Glazman, as well as valuable comments
and suggestions of two anonymous reviewers.
\end{acknowledgments}

\appendix

\section{Comparison with early results\label{sec:appcomp}}

\subsection{Short ABK ring}

\begin{figure}[ptb]
\includegraphics[width=0.6\textwidth]{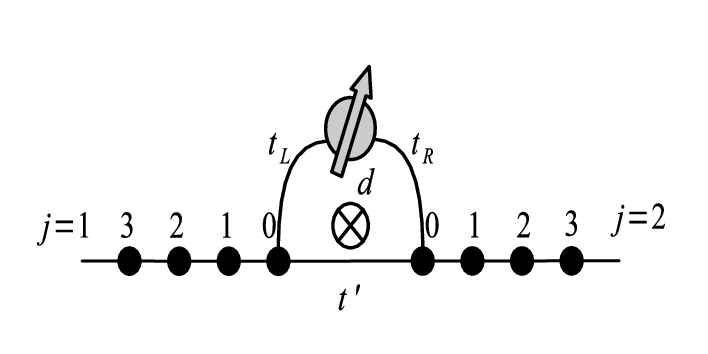}%
\caption{The short ABK
ring studied in Refs.~\onlinecite{PhysRevB.82.165426,PhysRevB.88.245104}.\label{fig:shortring}}
\end{figure}

Our formalism can be applied to the short ABK ring\cite%
{PhysRevB.82.165426,PhysRevB.88.245104} shown in Fig.~\ref{fig:shortring}.
There are two leads ($N=2$) and two coupling sites ($M=2$); $H_{0\text{%
,junction}}=-t^{\prime }\left( c_{1,0}^{\dag }c_{2,0}+\text{h.c.}\right) $.
The coupling sites coincide with the $0$th sites of the leads, $%
c_{C,r=1}\equiv c_{1,0}$, $c_{C,r=2}\equiv c_{2,0}$; also the AB phase is on
the couplings to the QD, $t_{1}=t_{L}e^{i\frac{\varphi }{2}}$ and $%
t_{2}=t_{R}e^{-i\frac{\varphi }{2}}$. We again let $\tilde{\tau}=t^{\prime
}/t$.

It is straightforward to obtain the background S-matrix and coupling site
wave function matrix:

\begin{equation}
S\left( k\right) =-\frac{1}{1-\tilde{\tau}^{2}e^{2ik}}\left( 
\begin{array}{cc}
e^{2ik}\left( 1-\tilde{\tau}^{2}\right) & e^{ik}\tilde{\tau}\left(
e^{2ik}-1\right) \\ 
e^{ik}\tilde{\tau}\left( e^{2ik}-1\right) & e^{2ik}\left( 1-\tilde{\tau}%
^{2}\right)%
\end{array}%
\right) \text{,}
\end{equation}

\begin{equation}
\Gamma \left( k\right) =-\frac{1}{1-\tilde{\tau}^{2}e^{2ik}}\left( 
\begin{array}{cc}
e^{2ik}-1 & 2ie^{2ik}\tilde{\tau}\sin k \\ 
2ie^{2ik}\tilde{\tau}\sin k & e^{2ik}-1%
\end{array}%
\right) \text{.}
\end{equation}%
With Eqs.~(\ref{KondojRG}), (\ref{deltaTjjprKondo}) and (\ref{connelimG2}),
one reproduces all analytic results in Refs.~%
\onlinecite{PhysRevB.82.165426,PhysRevB.88.245104}, including the Kondo
temperature, the high- and low-temperature conductance, and the elimination
of connected contribution at low temperatures.

The limit $t^{\prime }=0$ is useful as a benchmark against long ring
geometries so we study it in some more detail. In this limit we recover the
simplest geometry where a QD is embedded between source and drain leads.\cite%
{JPhysCondensMatter.16.R513} The normalization factor is

\begin{equation}
V_{k}^{2}=4\left( t_{L}^{2}+t_{R}^{2}\right) \sin ^{2}k\text{,}
\end{equation}%
and the zero-temperature transmission amplitude through the QD is given by
Eqs.~(\ref{discprob}), (\ref{deltapsipsi}), (\ref{TvsTtilde}) and (\ref%
{TtildeAL}):

\begin{eqnarray}
&&t_{QD}\equiv \frac{2i}{V_{k}^{2}}\left[ S\left( k\right) \Gamma ^{\dag
}\left( k\right) \lambda \Gamma \left( k\right) \right] _{12}\left[ -\pi \nu 
\text{T}_{kk}\left( \epsilon _{k}\right) \right]  \notag \\
&=&e^{2ik}\frac{2t_{L}t_{R}}{t_{L}^{2}+t_{R}^{2}}\frac{1}{2}\left(
e^{2i\delta _{P}}+1\right) \text{.}  \label{tQD}
\end{eqnarray}

\subsection{Finite quantum wire}

Another special case is the finite wire (or semi-transparent Kondo box)
geometry in Fig.~\ref{fig:simonaffleck}\ where the reference arm is absent;%
\cite{PhysRevLett.89.206602} again $N=M=2$. The left and right QD arms and
coupling sites are subject to gate voltages:

\begin{figure}[ptb]
\includegraphics[width=0.8\textwidth]{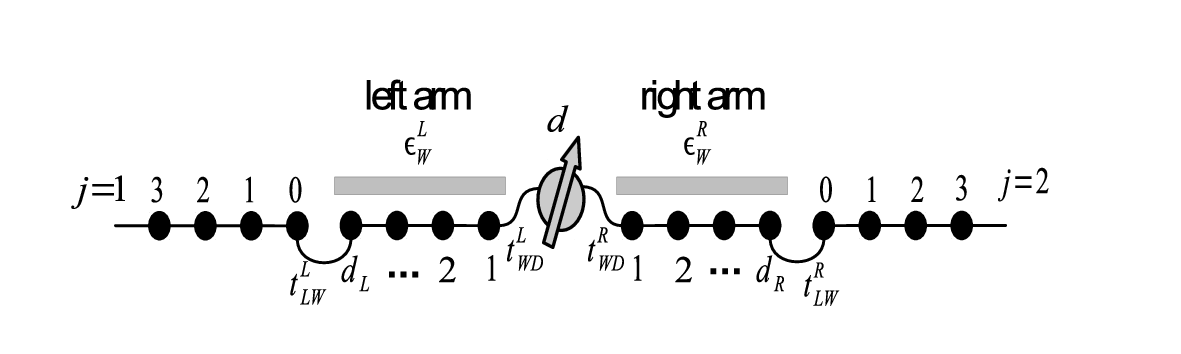}%
\caption{The finite
quantum wire geometry studied in Ref.~\onlinecite{PhysRevLett.89.206602}.\label{fig:simonaffleck}}
\end{figure}

\begin{eqnarray}
H_{0\text{,junction}} &=&-t\left( \sum_{n=1}^{d_{L}-1}c_{L,n}^{\dag
}c_{L,n+1}+\sum_{n=1}^{d_{R}-1}c_{R,n}^{\dag }c_{R,n+1}+\text{h.c.}\right) 
\notag \\
&&+\left( \epsilon _{W}^{L}\sum_{n=1}^{d_{L}}c_{L,n}^{\dag }c_{L,n}+\epsilon
_{W}^{R}\sum_{n=1}^{d_{R}}c_{R,n}^{\dag }c_{R,n}\right)  \notag \\
&&-\left( t_{LW}^{L}c_{L,d_{L}}^{\dag }c_{1,0}+t_{LW}^{R}c_{R,d_{R}}^{\dag
}c_{2,0}+\text{h.c.}\right) \text{.}
\end{eqnarray}%
The coupling sites are the first sites of the QD arms, $c_{C,r=1}\equiv
c_{L,1}$, $c_{C,r=2}\equiv c_{R,1}$; $t_{1}=t_{WD}^{L}$ and $%
t_{2}=t_{WD}^{R} $.

The two leads are decoupled without the QD, so $S$ and $\Gamma $ are both
diagonal. In this system we have

\begin{equation}
S_{11}\left( k\right) =-\frac{e^{ik}\sin k_{L}\left( d_{L}+1\right) -\gamma
_{L}^{2}\sin k_{L}d_{L}}{e^{-ik}\sin k_{L}\left( d_{L}+1\right) -\gamma
_{L}^{2}\sin k_{L}d_{L}}\text{,}
\end{equation}%
and

\begin{equation}
\Gamma _{11}\left( k\right) =-\frac{2i\gamma _{L}\sin k\sin k_{L}}{%
e^{-ik}\sin k_{L}\left( d_{L}+1\right) -\gamma _{L}^{2}\sin k_{L}d_{L}}\text{%
,}
\end{equation}%
where $k_{L}$ is determined by the gate voltage $\epsilon _{W}^{L}$,

\begin{equation}
-2t\cos k_{L}+\epsilon _{W}^{L}=-2t\cos k\text{,}
\end{equation}%
and $\gamma _{L}=t_{LW}^{L}/t$. $S_{22}$ and $\Gamma _{22}$ can be obtained
simply by substituting $L$ with $R$. Again, these results allow us to
reproduce the (weak coupling)\ Kondo temperature, the high-temperature
conductance, as well as the low-temperature conductance in the large Kondo
cloud regime. (We did not quantitatively discuss the low-temperature
conductance in the small cloud regime in this paper; see Sec.~\ref%
{sec:outlook}.)

\section{Details of the disconnected contribution\label{sec:appdisc}}

In this appendix we present the detailed derivation of Eq.~(\ref{discprob})
[or equivalently Eq.~(\ref{discprobKYA})] from Eq.~(\ref{GjjprDImIm}). The
calculations are similar to those in Appendix B of Ref.~%
\onlinecite{PhysRevB.88.245104}, but an important difference is that here we
cannot simply take the $\delta $-function part and neglect the principal
value part in Eq.~(\ref{GjjprDImIm}). Instead, most of the momentum
integrals are evaluated by means of contour integration.

From Eq.~(\ref{Dyson})

\begin{align}
-2\operatorname{Im} \mathbb{G}_{k_{2}q_{1}}^{R}\left( \omega \right) & =\left( 2\pi
\right) ^{2}\delta \left( k_{2}-q_{1}\right) \delta \left( \omega -\epsilon
_{k_{2}}\right) +i\tau _{\psi }  \notag \\
& \times \left[ g_{k_{2}}^{R}\left( \omega \right) V_{k_{2}}G_{dd}^{R}\left(
\omega \right) V_{q_{1}}g_{q_{1}}^{R}\left( \omega \right)
-g_{k_{2}}^{A}\left( \omega \right) V_{k_{2}}G_{dd}^{A}\left( \omega \right)
V_{q_{1}}g_{q_{1}}^{A}\left( \omega \right) \right] \text{.}
\end{align}%
We denote the three terms above as $0$, $R$ and $A$ respectively. Inserting
into Eq.~(\ref{GjjprDImIm}), we find $3$ types of contributions to the
disconnected part:

\begin{align}
G_{jj^{\prime }}^{\prime D}\left( \Omega \right) & =G_{jj^{\prime
},00}^{\prime D}\left( \Omega \right) +\left[ G_{jj^{\prime },0R}^{\prime
D}\left( \Omega \right) +G_{jj^{\prime },0A}^{\prime D}\left( \Omega \right)
+G_{jj^{\prime },R0}^{\prime D}\left( \Omega \right) +G_{jj^{\prime
},A0}^{\prime D}\left( \Omega \right) \right]  \notag \\
& +\left[ G_{jj^{\prime },RA}^{\prime D}\left( \Omega \right) +G_{jj^{\prime
},RR}^{\prime D}\left( \Omega \right) +G_{jj^{\prime },AR}^{\prime D}\left(
\Omega \right) +G_{jj^{\prime },AA}^{\prime D}\left( \Omega \right) \right] 
\text{;}
\end{align}%
The $00$ term is the background transmission, the first pair of square
brackets is linear in the T-matrix of the screening channel, and the second
pair of square brackets is quadratic in the T-matrix.

Due to the multiplying factor of $\Omega $ in Eq.~(\ref{Kubo}), $O\left(
1/\Omega \right) $ terms in $G_{jj^{\prime }}^{\prime D}$ contribute to the
linear DC conductance, while $O\left( 1\right) $ and other terms which are
regular in the DC limit $\Omega \rightarrow 0$ do not contribute. (We can
check explicitly that there are no $O\left( 1/\Omega ^{2}\right) $ or
higher-order divergences.) Therefore, in the DC limit we are only interested
in the $O\left( 1/\Omega \right) $ part of $G_{jj^{\prime }}^{\prime D}$.

\subsection{Properties of the S-matrix and the wave functions}

Before actually doing the calculations it is useful to examine the
properties of the background S-matrix and the wave functions in our
tight-binding model, since we rely on these properties to transform the
momentum integrals into contour integrals and evaluate them.

First consider the analytic continuation $k\rightarrow -k$. The wave
function \textquotedblleft incident\textquotedblright\ from lead $j$ at
momentum $-k$ takes the following form on lead $j^{\prime }$ [cf. Eq. (\ref%
{bgwflead})],

\begin{equation}
\chi _{j,-k}\left( j^{\prime },n\right) =\delta _{jj^{\prime
}}e^{ikn}+S_{j^{\prime }j}\left( -k\right) e^{-ikn}\text{;}
\end{equation}%
and on coupling site $r$,

\begin{equation*}
\chi _{j,-k}\left( r\right) =\Gamma _{rj}\left( -k\right) \text{.}
\end{equation*}%
This wave function should be a linear combination of the scattering state
wave functions at momentum $k$ which form a complete basis. The linear
coefficients are obtained from S-matrix unitarity:

\begin{equation}
\chi _{j,-k}\left( j^{\prime },n\right) =\sum_{j^{\prime \prime
}}S_{jj^{\prime \prime }}^{\ast }\left( k\right) \chi _{j^{\prime \prime
},k}\left( j^{\prime },n\right) =\delta _{jj^{\prime }}e^{ikn}+S_{jj^{\prime
}}^{\ast }\left( k\right) e^{-ikn}\text{,}
\end{equation}%
and the same coefficients apply to the coupling sites:

\begin{equation*}
\chi _{j,-k}\left( r\right) =\sum_{j^{\prime \prime }}S_{jj^{\prime \prime
}}^{\ast }\left( k\right) \Gamma _{rj^{\prime \prime }}\left( k\right) \text{%
.}
\end{equation*}%
Hence

\begin{equation}
S\left( -k\right) =S^{\dag }\left( k\right) \text{,}  \label{SMatminusk}
\end{equation}

\begin{equation}
\Gamma \left( -k\right) =\Gamma \left( k\right) S^{\dag }\left( k\right) 
\text{.}  \label{Gammaminusk}
\end{equation}
Eq.~(\ref{SMatminusk}) is known as the Hermitian analyticity of the S-matrix.%
\cite{JPhysA.40.2485}

Another useful property is the location of poles of $S\left( k\right) \equiv
S\left( z=e^{ik}\right) $ on the $z$ complex plane. Our analysis closely
follows Ref.~\onlinecite{perelomov1998quantum} which deals with the case of
quadratic dispersion.

Consider one pole of the S-matrix $k\equiv k_{1}+ik_{2}$, where for certain
values of $j$ and $j^{\prime }$, $\left\vert S_{j^{\prime }j}\left( k\right)
\right\vert \rightarrow \infty $. In the scattering state $\left\vert
q_{j,k}\right\rangle \equiv q_{j,k}^{\dag }\left\vert 0\right\rangle $,
where $\left\vert 0\right\rangle $ is the Fermi sea ground state, the
incident component of the wave function at momentum $k$ becomes negligible
relative to the scattered component. Therefore, the time-dependent wave
function on lead $j^{\prime }$ at site $n$ reads

\begin{equation}
\chi _{j,k}\left( j^{\prime },n,\bar{t}\right) \approx S_{j^{\prime
}j}\left( k\right) e^{ikn}e^{-i\epsilon _{k}\bar{t}}=S_{j^{\prime }j}\left(
k\right) e^{ikn}e^{2it\bar{t}\cos k_{1}\cosh k_{2}}e^{2t\bar{t}\sin
k_{1}\sinh k_{2}}\text{.}  \label{wfpolelead}
\end{equation}%
This expression is valid for any $j^{\prime }$ where $\left\vert
S_{j^{\prime }j}\left( k\right) \right\vert $ is divergent; for other $%
j^{\prime }$ the wave function is negligible.

We define the \textquotedblleft junction area\textquotedblright\ to include
any tight-binding site that is not part of a lead, together with the 0th
site of each lead. The total probability of the electron being inside the
junction area, $N\left( \bar{t}\right) $, obeys the probability continuity
equation

\begin{equation}
\frac{d}{d\bar{t}}N\left( \bar{t}\right) =it\sum_{j^{\prime }}\left(
c_{j^{\prime },0}^{\dag }c_{j^{\prime },1}-c_{j^{\prime },1}^{\dag
}c_{j^{\prime },0}\right) \left( \bar{t}\right) \text{,}
\end{equation}%
where the right-hand side is the current operator between site $0$ and site $%
1$ of lead $j^{\prime }$, summed over all leads. Taking the expectation
value in the state $\left\vert q_{j,k}\right\rangle $, we find

\begin{equation}
-\left( 4t\sin k_{1}\sinh k_{2}\right) C_{j}\left( k\right) e^{4t\bar{t}\sin
k_{1}\sinh k_{2}}=\sum_{j^{\prime }}\left( 2t\sin k_{1}\right)
e^{-k_{2}}\left\vert S_{j^{\prime }j}\left( k\right) \right\vert ^{2}e^{4t%
\bar{t}\sin k_{1}\sinh k_{2}}\text{.}  \label{probcons}
\end{equation}%
For the left-hand side we have used the form of the time evolution $%
e^{-i\epsilon _{k}\bar{t}}$, and $C_{j}\left( k\right) $ is a positive
time-independent constant proportional to the total probability in the
junction area; $C_{j}\left( k\right) $ is divergent whenever $\left\vert
S_{j^{\prime }j}\left( k\right) \right\vert $ is divergent.\ For the
right-hand side, we have used Eq.~(\ref{wfpolelead}) at $n=0$ and $n=1$; the
summation is over any $j^{\prime }$ where $\left\vert S_{j^{\prime }j}\left(
k\right) \right\vert $ is divergent.

Eq.~(\ref{probcons}) implies that either $\sin k_{1}=0$, in which case $%
k_{1}=0$ or $\pi $; or $\sinh k_{2}<0$, in which case $\left\vert e^{i\left(
k_{1}+ik_{2}\right) }\right\vert >1$. The poles of $S\left( k\right) $ on
the $z=e^{ik}$ plane are therefore either outside the unit circle or located
on the real axis. For the models we study in this paper, the poles of $%
S\left( k\right) $ and those of $\Gamma \left( k\right) /\left( \sin
k\right) $ coincide (see also Sec.~\ref{sec:appNI}); in other words, the
poles of $S\left( k\right) $ and $\Gamma \left( k\right) /\left( \sin
k\right) $ on the $z=e^{ik}$ plane are either outside the unit circle or on
the real axis.

We mention that similar results apply in the theory with a reduced bandwidth
and a linearized dispersion in the leads. Eqs.~(\ref{SMatminusk}) and (\ref%
{Gammaminusk}) continue to hold; on the other hand, the probability current
is proportional to $v_{F}$ instead of $2t\sin k_{1}$, and all poles of $%
S\left( k\right) $ and $\Gamma \left( k\right) /\left( \sin k\right) $\ are
located in the lower half of the $k$ plane.

\subsection{Background transmission}

This part is independent of the QD and the result should be the famous
Landauer formula:

\begin{equation}
G_{jj^{\prime },00}^{\prime D}\left( \Omega \right) =2\int_{0}^{\pi }\frac{%
dk_{1}dk_{2}}{\left( 2\pi \right) ^{2}}\frac{f\left( \epsilon
_{k_{1}}\right) -f\left( \epsilon _{k_{2}}\right) }{\epsilon
_{k_{1}}-\epsilon _{k_{2}}+\Omega ^{+}}\tr \left( \mathbb{M}_{k_{1}k_{2}}^{j}%
\mathbb{M}_{k_{2}k_{1}}^{j^{\prime }}\right) \text{.}
\end{equation}%
Inserting Eq.~(\ref{bbMmat}), taking advantage of Eq.~(\ref{SMatminusk}) and
the unitarity of the $\mathbb{U}$\ matrix, we find

\begin{align}
G_{jj^{\prime },00}^{\prime D}\left( \Omega \right) & =2\int_{-\pi }^{\pi }%
\frac{dk_{1}dk_{2}}{\left( 2\pi \right) ^{2}}\frac{f\left( \epsilon
_{k_{1}}\right) -f\left( \epsilon _{k_{2}}\right) }{\epsilon
_{k_{1}}-\epsilon _{k_{2}}+\Omega ^{+}}  \notag \\
& \times \left\{ \delta _{jj^{\prime }}\frac{1}{1-e^{i\left(
k_{1}-k_{2}+i0\right) }}\frac{1}{1-e^{i\left( k_{2}-k_{1}+i0\right) }}%
+S_{jj^{\prime }}^{\ast }\left( k_{1}\right) \delta _{jj^{\prime }}\frac{1}{%
1-e^{-i\left( k_{1}+k_{2}-i0\right) }}\frac{1}{1-e^{i\left(
k_{2}-k_{1}+i0\right) }}\right.  \notag \\
& \left. +\delta _{jj^{\prime }}S_{jj^{\prime }}\left( k_{2}\right) \frac{1}{%
1-e^{i\left( k_{1}+k_{2}+i0\right) }}\frac{1}{1-e^{i\left(
k_{2}-k_{1}+i0\right) }}+S_{jj^{\prime }}^{\ast }\left( k_{1}\right)
S_{jj^{\prime }}\left( k_{2}\right) \frac{1}{\left[ 1-e^{i\left(
k_{2}-k_{1}+i0\right) }\right] ^{2}}\right\} \text{.}
\end{align}%
By residue theorem we can perform the $k_{2}$ integral in the part
proportional to $f\left( \epsilon _{k_{1}}\right) $ and the $k_{1}$ integral
in the part proportional to $f\left( \epsilon _{k_{2}}\right) $. In the
following we assume $\Omega >0$; the case $\Omega <0$ can be dealt with
similarly.

We begin from the first term in curly brackets, which is proportional to $%
\delta _{jj^{\prime }}$. For the part proportional to $f\left( \epsilon
_{k_{1}}\right) $, making the substitution $z_{2}=e^{ik_{2}}$, and
calculating the contour integral on the counterclockwise unit circle, we find

\begin{subequations}
\begin{align}
& \int_{-\pi }^{\pi }\frac{dk_{2}}{2\pi }\frac{f\left( \epsilon
_{k_{1}}\right) }{\epsilon _{k_{1}}-\epsilon _{k_{2}}+\Omega ^{+}}\frac{1}{%
1-e^{i\left( k_{1}-k_{2}+i\eta \right) }}\frac{1}{1-e^{i\left(
k_{2}-k_{1}+i\eta \right) }}  \notag \\
& =\frac{f\left( \epsilon _{k_{1}}\right) }{2it\sin p_{1}}\frac{1}{%
1-e^{i\left( k_{1}-p_{1}\right) }}\frac{1}{1-e^{i\left( p_{1}-k_{1}\right) }}%
+\frac{f\left( \epsilon _{k_{1}}\right) }{\Omega }\frac{1}{1-e^{-2\eta }}%
\text{,}
\end{align}%
where $\eta \rightarrow 0^{+}$. ($\eta $ corresponds to the rate of
switching on the bias voltage in Kubo formalism, so the limit $\eta
\rightarrow 0$ should be taken before the DC limit $\Omega \rightarrow 0$.)
We have assumed $\epsilon _{k_{1}}+\Omega \equiv \epsilon _{p_{1}}$ where $%
0\leq p_{1}\leq \pi $ if $p_{1}$ is real; the poles of the integrand inside
the unit circle are then $z_{2}=e^{i\left( p_{1}+i0\right) }$ and $%
z_{2}=e^{i\left( k_{1}+i\eta \right) }$. At the band edges, $2t-\Omega
<\epsilon _{k_{1}}<2t$, and $p_{1}$ is purely imaginary; we can choose it to
have a positive imaginary part so the above expression remains valid.
Similarly

\begin{align}
& \int_{-\pi }^{\pi }\frac{dk_{1}}{2\pi }\frac{f\left( \epsilon
_{k_{2}}\right) }{\epsilon _{k_{1}}-\epsilon _{k_{2}}+\Omega ^{+}}\frac{1}{%
1-e^{i\left( k_{1}-k_{2}+i\eta \right) }}\frac{1}{1-e^{i\left(
k_{2}-k_{1}+i\eta \right) }}  \notag \\
& =\frac{f\left( \epsilon _{k_{2}}\right) }{2it\sin p_{2}}\frac{1}{%
1-e^{i\left( -p_{2}-k_{2}\right) }}\frac{1}{1-e^{i\left( p_{2}+k_{2}\right) }%
}+\frac{f\left( \epsilon _{k_{2}}\right) }{\Omega }\frac{1}{1-e^{-2\eta }}%
\text{,}
\end{align}%
where $\epsilon _{k_{2}}-\Omega =\epsilon _{p_{2}}$, $0\leq p_{2}\leq \pi $
if $p_{2}$ is real, or $p_{2}=-i\left\vert p_{2}\right\vert $ if $p_{2}$ is
purely imaginary. Now combine the two parts. In the $\Omega \rightarrow 0$
limit, $p_{1}\rightarrow k_{1}$ only for $p_{1}$ real and $k_{1}>0$, and $%
p_{2}\rightarrow -k_{2}$ only for $p_{2}$ real and $k_{2}<0$; the most
divergent contribution is therefore

\end{subequations}
\begin{align}
& \int_{-\pi }^{\pi }\frac{dk_{1}dk_{2}}{\left( 2\pi \right) ^{2}}\frac{%
f\left( \epsilon _{k_{1}}\right) -f\left( \epsilon _{k_{2}}\right) }{%
\epsilon _{k_{1}}-\epsilon _{k_{2}}+\Omega ^{+}}\frac{1}{1-e^{i\left(
k_{1}-k_{2}+i0\right) }}\frac{1}{1-e^{i\left( k_{2}-k_{1}+i0\right) }} 
\notag \\
& =\int_{-2t}^{2t-\Omega }\frac{d\epsilon _{k_{1}}}{2\pi i}\left[ f\left(
\epsilon _{k_{1}}\right) -f\left( \epsilon _{k_{1}}+\Omega \right) \right] 
\frac{1}{\left( 2t\sin k_{1}\right) \left( 2t\sin p_{1}\right) }\frac{1}{%
1-e^{i\left( k_{1}-p_{1}\right) }}\frac{1}{1-e^{i\left( p_{1}-k_{1}\right) }}%
+O\left( 1\right) \text{.}
\end{align}%
We have substituted the dummy variables $k_{2}\rightarrow p_{1}$, $%
p_{2}\rightarrow k_{1}$, and noted that $\epsilon _{p_{1}}=\epsilon
_{k_{1}}+\Omega $. Expanding various parts of the integrand in $\Omega
\rightarrow 0$ limit, we find

\begin{align}
& \int_{-\pi }^{\pi }\frac{dk_{1}dk_{2}}{\left( 2\pi \right) ^{2}}\frac{%
f\left( \epsilon _{k_{1}}\right) -f\left( \epsilon _{k_{2}}\right) }{%
\epsilon _{k_{1}}-\epsilon _{k_{2}}+\Omega ^{+}}\frac{1}{1-e^{i\left(
k_{1}-k_{2}+i0\right) }}\frac{1}{1-e^{i\left( k_{2}-k_{1}+i0\right) }} 
\notag \\
& =\frac{1}{2\pi i\Omega }\int_{-2t}^{2t-\Omega }d\epsilon _{k_{1}}\left[
-f^{\prime }\left( \epsilon _{k_{1}}\right) \right] +O\left( 1\right) \text{.%
}  \label{deltajjpr}
\end{align}

The two terms in $G_{jj^{\prime },00}^{\prime D}\left( \Omega \right) $
which are linear in the S-matrix do not contribute any terms of $O\left(
1/\Omega \right) $ to $G_{jj^{\prime }}^{\prime D}$: the difference of Fermi
functions is proportional to $\Omega $, but the denominators are also $%
O\left( \Omega \right) $, unlike the case for the $\delta _{jj^{\prime }}$
terms whose denominators are $O\left( \Omega ^{2}\right) $. This leaves us
with the term quadratic in the S-matrix, which can be similarly evaluated.
For the part proportional to $f\left( \epsilon _{k_{1}}\right) $,

\begin{subequations}
\begin{align}
& \int_{-\pi }^{\pi }\frac{dk_{2}}{2\pi }\frac{f\left( \epsilon
_{k_{1}}\right) }{\epsilon _{k_{1}}-\epsilon _{k_{2}}+\Omega ^{+}}%
S_{jj^{\prime }}^{\ast }\left( k_{1}\right) S_{jj^{\prime }}\left(
k_{2}\right) \frac{1}{\left[ 1-e^{i\left( k_{2}-k_{1}+i0\right) }\right] ^{2}%
}  \notag \\
& =\frac{f\left( \epsilon _{k_{1}}\right) }{2it\sin p_{1}}S_{jj^{\prime
}}^{\ast }\left( k_{1}\right) S_{jj^{\prime }}\left( p_{1}\right) \frac{1}{%
\left[ 1-e^{i\left( p_{1}-k_{1}\right) }\right] ^{2}}+\left( \text{%
contribution of poles of }S_{jj^{\prime }}\right) \text{;}
\end{align}%
the poles of $S_{jj^{\prime }}$ inside the unit circle (on the real axis)
may contribute to the contour integral, but these terms are regular in the $%
\Omega \rightarrow 0$ limit and do not contribute to the DC conductance.
Similarly

\begin{align}
& \int_{-\pi }^{\pi }\frac{dk_{1}}{2\pi }\frac{f\left( \epsilon
_{k_{2}}\right) }{\epsilon _{k_{1}}-\epsilon _{k_{2}}+\Omega ^{+}}%
S_{jj^{\prime }}^{\ast }\left( k_{1}\right) S_{jj^{\prime }}\left(
k_{2}\right) \frac{1}{\left[ 1-e^{i\left( k_{2}-k_{1}+i0\right) }\right] ^{2}%
}  \notag \\
& =\frac{f\left( \epsilon _{k_{2}}\right) }{2it\sin p_{2}}S_{jj^{\prime
}}^{\ast }\left( p_{2}\right) S_{jj^{\prime }}\left( k_{2}\right) \frac{1}{%
\left[ 1-e^{i\left( k_{2}-p_{2}\right) }\right] ^{2}}+\left( \text{%
contribution of poles of }S_{jj^{\prime }}^{\ast }\right) \text{,}
\end{align}%
Therefore

\end{subequations}
\begin{align}
& \int_{-\pi }^{\pi }\frac{dk_{1}dk_{2}}{\left( 2\pi \right) ^{2}}\frac{%
f\left( \epsilon _{k_{1}}\right) -f\left( \epsilon _{k_{2}}\right) }{%
\epsilon _{k_{1}}-\epsilon _{k_{2}}+\Omega ^{+}}S_{jj^{\prime }}^{\ast
}\left( k_{1}\right) S_{jj^{\prime }}\left( k_{2}\right) \frac{1}{\left[
1-e^{i\left( k_{2}-k_{1}+i0\right) }\right] ^{2}}  \notag \\
& =\int_{-2t}^{2t-\Omega }\frac{d\epsilon _{k_{1}}}{2\pi i}\frac{f\left(
\epsilon _{k_{1}}\right) -f\left( \epsilon _{k_{1}}+\Omega \right) }{\left(
2t\sin k_{1}\right) \left( 2t\sin p_{1}\right) }S_{jj^{\prime }}^{\ast
}\left( k_{1}\right) S_{jj^{\prime }}\left( p_{1}\right) \frac{1}{\left[
1-e^{i\left( p_{1}-k_{1}\right) }\right] ^{2}}+O\left( 1\right)  \notag \\
& =-\frac{1}{2\pi i\Omega }\int_{-2t}^{2t-\Omega }d\epsilon _{k_{1}}\left[
-f^{\prime }\left( \epsilon _{k_{1}}\right) \right] S_{jj^{\prime }}^{\ast
}\left( k_{1}\right) S_{jj^{\prime }}\left( p_{1}\right) +O\left( 1\right) 
\text{.}  \label{SstarS}
\end{align}%
From Eqs.~(\ref{deltajjpr}) and (\ref{SstarS}), we conclude that

\begin{equation}
G_{jj^{\prime},00}^{\prime D}\left( \Omega\right) =\frac{1}{\pi i\Omega}%
\int_{-2t}^{2t-\Omega}d\epsilon_{k_{1}}\left[ -f^{\prime}\left(
\epsilon_{k_{1}}\right) \right] \left[ \delta_{jj^{\prime}}-S_{jj^{%
\prime}}^{\ast}\left( k_{1}\right) S_{jj^{\prime}}\left( p_{1}\right) \right]
+O\left( 1\right) \text{;}
\end{equation}
taking the $\Omega\rightarrow0$ limit, noting that $p_{1}\rightarrow k_{1}$,
we recover the Landauer formula, Eq.~(\ref{discprobKYAbgt}).

\subsection{Terms linear in T-matrix}

We focus on $G_{jj^{\prime },0R}^{\prime D}+G_{jj^{\prime },0A}^{\prime D}$;
the calculation of $G_{jj^{\prime },R0}^{\prime D}+G_{jj^{\prime
},A0}^{\prime D}$ is analogous.

\begin{align}
& G_{jj^{\prime },0R}^{\prime D}\left( \Omega \right) +G_{jj^{\prime
},0A}^{\prime D}\left( \Omega \right)  \notag \\
& =2\int_{0}^{\pi }\frac{dk_{1}}{\left( 2\pi \right) ^{2}}\frac{dq_{1}dq_{2}%
}{\left( 2\pi \right) ^{2}}\int d\omega \frac{f\left( \omega \right)
-f\left( \epsilon _{q_{1}}\right) }{\omega -\epsilon _{q_{1}}+\Omega ^{+}}%
\tr \left\{ \mathbb{M}_{k_{1}q_{1}}^{j}\mathbb{M}_{q_{1}q_{2}}^{j^{\prime
}}\left( i\tau _{\psi }\right) \right.  \notag \\
& \times \left. \left[ g_{q_{2}}^{R}\left( \omega \right)
V_{q_{2}}G_{dd}^{R}\left( \omega \right) V_{k_{1}}g_{k_{1}}^{R}\left( \omega
\right) -g_{q_{2}}^{A}\left( \omega \right) V_{q_{2}}G_{dd}^{A}\left( \omega
\right) V_{k_{1}}g_{k_{1}}^{A}\left( \omega \right) \right] \right\} \text{.}
\end{align}%
Using Eqs.~(\ref{bbUmatrow1}), (\ref{bbMmat}), (\ref{SMatminusk}) and (\ref%
{Gammaminusk}), a huge simplification takes place:

\begin{subequations}
\begin{align}
& G_{jj^{\prime },0R}^{\prime D}\left( \Omega \right)  \notag \\
& =2\int_{-\pi }^{\pi }\frac{dk_{1}dq_{1}dq_{2}}{\left( 2\pi \right) ^{4}}%
\int d\omega \frac{f\left( \omega \right) -f\left( \epsilon _{q_{1}}\right) 
}{\omega -\epsilon _{q_{1}}+\Omega ^{+}}ig_{q_{2}}^{R}\left( \omega \right)
G_{dd}^{R}\left( \omega \right) g_{k_{1}}^{R}\left( \omega \right)
\sum_{r_{1}r_{2}}t_{r_{1}}^{\ast }t_{r_{2}}  \notag \\
& \times \Gamma _{r_{1}j}\left( k_{1}\right) \left[ \delta _{jj^{\prime }}%
\frac{1}{1-e^{i\left( k_{1}-q_{1}+i0\right) }}+S_{jj^{\prime }}\left(
q_{1}\right) \frac{1}{1-e^{i\left( k_{1}+q_{1}+i0\right) }}\right] \Gamma
_{r_{2}j^{\prime }}^{\ast }\left( q_{2}\right) \frac{1}{1-e^{i\left(
q_{1}-q_{2}+i0\right) }}\text{,}
\end{align}

\begin{align}
& G_{jj^{\prime },0A}^{\prime D}\left( \Omega \right)  \notag \\
& =2\int_{-\pi }^{\pi }\frac{dk_{1}dq_{1}dq_{2}}{\left( 2\pi \right) ^{4}}%
\int d\omega \frac{f\left( \omega \right) -f\left( \epsilon _{q_{1}}\right) 
}{\omega -\epsilon _{q_{1}}+\Omega ^{+}}\left( -i\right) g_{q_{2}}^{A}\left(
\omega \right) G_{dd}^{A}\left( \omega \right) g_{k_{1}}^{A}\left( \omega
\right) \sum_{r_{1}r_{2}}t_{r_{1}}^{\ast }t_{r_{2}}  \notag \\
& \times \Gamma _{r_{1}j}\left( k_{1}\right) \left[ \delta _{jj^{\prime }}%
\frac{1}{1-e^{i\left( k_{1}-q_{1}+i0\right) }}+S_{jj^{\prime }}\left(
q_{1}\right) \frac{1}{1-e^{i\left( k_{1}+q_{1}+i0\right) }}\right] \Gamma
_{r_{2}j^{\prime }}^{\ast }\left( q_{2}\right) \frac{1}{1-e^{i\left(
q_{1}-q_{2}+i0\right) }}\text{.}
\end{align}%
Writing $\omega =\epsilon _{k}$, where $0\leq k\leq \pi $ or $k=i\left\vert
k\right\vert $, we are now free to do the $k_{1}$ and $q_{2}$ integrals. The
poles of $\Gamma \left( k_{1}\right) $ and $\Gamma ^{\ast }\left(
q_{2}\right) $ are again not important in the DC limit:

\end{subequations}
\begin{subequations}
\begin{align}
& G_{jj^{\prime },0R}^{\prime D}\left( \Omega \right)  \notag \\
& =2\int_{-\pi }^{\pi }\frac{dq_{1}}{\left( 2\pi \right) ^{2}}\int d\epsilon
_{k}\frac{f\left( \epsilon _{k}\right) -f\left( \epsilon _{q_{1}}\right) }{%
\epsilon _{k}-\epsilon _{q_{1}}+\Omega ^{+}}i\sum_{r_{1}r_{2}}t_{r_{1}}^{%
\ast }t_{r_{2}}\frac{1}{2it\sin k}G_{dd}^{R}\left( \epsilon _{k}\right) 
\frac{1}{2it\sin k}  \notag \\
& \times \Gamma _{r_{1}j}\left( k\right) \left[ \delta _{jj^{\prime }}\frac{1%
}{1-e^{i\left( k-q_{1}+i0\right) }}+S_{jj^{\prime }}\left( q_{1}\right) 
\frac{1}{1-e^{i\left( k+q_{1}+i0\right) }}\right] \Gamma _{r_{2}j^{\prime
}}^{\ast }\left( -k\right) \frac{1}{1-e^{i\left( q_{1}+k+i0\right) }}%
+O\left( 1\right) \text{,}
\end{align}

\begin{align}
& G_{jj^{\prime },0A}^{\prime D}\left( \Omega \right)  \notag \\
& =2\int_{-\pi }^{\pi }\frac{dq_{1}}{\left( 2\pi \right) ^{2}}\int d\epsilon
_{k}\frac{f\left( \epsilon _{k}\right) -f\left( \epsilon _{q_{1}}\right) }{%
\epsilon _{k}-\epsilon _{q_{1}}+\Omega ^{+}}\left( -i\right)
\sum_{r_{1}r_{2}}t_{r_{1}}^{\ast }t_{r_{2}}\frac{1}{-2it\sin k}%
G_{dd}^{A}\left( \epsilon _{k}\right) \frac{1}{-2it\sin k}  \notag \\
& \times \Gamma _{r_{1}j}\left( -k\right) \left[ \delta _{jj^{\prime }}\frac{%
1}{1-e^{i\left( -k-q_{1}+i0\right) }}+S_{jj^{\prime }}\left( q_{1}\right) 
\frac{1}{1-e^{i\left( -k+q_{1}+i0\right) }}\right] \Gamma _{r_{2}j^{\prime
}}^{\ast }\left( k\right) \frac{1}{1-e^{i\left( q_{1}-k+i0\right) }}+O\left(
1\right) \text{.}
\end{align}%
Now do the $\epsilon _{k}$ and $q_{1}$ integrals. The $\delta _{jj^{\prime
}} $ terms are regular in the DC limit, so we only need to keep the $%
S_{jj^{\prime }}$ terms. In the $0R$ term, while the $q_{1}$ integral in the 
$f\left( \epsilon _{k}\right) $ part is straightforward, the $\epsilon _{k}$
integral in the $f\left( \epsilon _{q_{1}}\right) $ part can be simplified
by expanding around $k+q_{1}=0$:

\end{subequations}
\begin{subequations}
\begin{align}
& G_{jj^{\prime },0R}^{\prime D}\left( \Omega \right)  \notag \\
& =2\frac{1}{2\pi }\int_{-2t}^{2t-\Omega }d\epsilon _{k}\frac{f\left(
\epsilon _{k}\right) }{2it\sin p}i\sum_{r_{1}r_{2}}t_{r_{1}}^{\ast }t_{r_{2}}%
\frac{1}{2it\sin k}G_{dd}^{R}\left( \epsilon _{k}\right) \frac{\Gamma
_{r_{1}j}\left( k\right) \Gamma _{r_{2}j^{\prime }}^{\ast }\left( -k\right) 
}{2it\sin k}S_{jj^{\prime }}\left( p\right)  \notag \\
& \times \frac{1}{\left[ 1-e^{i\left( k+p+i0\right) }\right] ^{2}}%
-2\int_{-\pi }^{0}\frac{dq_{1}}{\left( 2\pi \right) ^{2}}\left( 2t\sin
q_{1}\right) S_{jj^{\prime }}\left( q_{1}\right) \int_{-\infty }^{\infty
}d\epsilon _{k}\frac{f\left( \epsilon _{q_{1}}\right) }{\left( \epsilon
_{k}-\epsilon _{q_{1}}+\Omega ^{+}\right) }  \notag \\
& \times i\sum_{r_{1}r_{2}}t_{r_{1}}^{\ast }t_{r_{2}}G_{dd}^{R}\left(
\epsilon _{k}\right) \frac{\Gamma _{r_{1}j}\left( k\right) \Gamma
_{r_{2}j^{\prime }}^{\ast }\left( -k\right) }{2t\sin k}\frac{1}{\left(
\epsilon _{k}-\epsilon _{-q_{1}}+i0\right) ^{2}}+O\left( 1\right)  \notag \\
& =O\left( 1\right) \text{.}
\end{align}%
Here, in the $f\left( \epsilon _{k}\right) $ part, we have written $\epsilon
_{k}+\Omega \equiv \epsilon _{p}$ ($0\leq p\leq \pi $) assuming $\Omega >0$,
integrated over $q_{1}$ using the complex variable $e^{iq_{1}}$, and again
neglected $O\left( 1\right) $ contributions from the poles of $S\left(
q_{1}\right) $. Because $k+p$ is always positive and never close to $0$, the
denominator for the $f\left( \epsilon _{k}\right) $ part\ is $O\left(
1\right) $; thus the $f\left( \epsilon _{k}\right) $ part is itself $O\left(
1\right) $. Meanwhile, in the $f\left( \epsilon _{q_{1}}\right) $\ part, we
have used $\left( 2t\sin k\right) \left( k+q_{1}+i0\right) \approx \left(
2t\sin q_{1}\right) \left( k+q_{1}+i0\right) \approx \epsilon _{k}-\epsilon
_{-q_{1}}+i0$ for $\left\vert k+q_{1}\right\vert \ll 1$. We then extend the $%
\epsilon _{k}$ domain of integration back to the entire real axis. Both $%
G_{dd}^{R}\left( \epsilon _{k}\right) $ and $\Gamma _{r_{1}j}\left( k\right)
\Gamma _{r_{2}j^{\prime }}^{\ast }\left( -k\right) /\left( \sin k\right) $
are analytic in the upper $\epsilon _{k}$ half plane; thus, closing the $%
\epsilon _{k}$ contour above the real axis, the $\epsilon _{k}$ integral in
the $f\left( \epsilon _{q_{1}}\right) $\ part sees no pole and vanishes.
Similarly, in the $0A$ term,

\begin{align}
& G_{jj^{\prime },0A}^{\prime D}\left( \Omega \right)  \notag \\
& =2\frac{1}{2\pi i}\int_{-2t}^{2t-\Omega }d\epsilon _{k}f\left( \epsilon
_{k}\right) \left( -i\right) \sum_{r_{1}r_{2}}t_{r_{1}}^{\ast
}t_{r_{2}}G_{dd}^{A}\left( \epsilon _{k}\right) \frac{\Gamma _{r_{1}j}\left(
-k\right) \Gamma _{r_{2}j^{\prime }}^{\ast }\left( k\right) }{2t\sin k}%
S_{jj^{\prime }}\left( p\right) \frac{1}{\Omega ^{2}}  \notag \\
& -2\int_{-2t+\Omega }^{2t}\frac{d\epsilon _{q_{1}}}{2\pi i}S_{jj^{\prime
}}\left( q_{1}\right) f\left( \epsilon _{q_{1}}\right) \left( -i\right)
\sum_{r_{1}r_{2}}t_{r_{1}}^{\ast }t_{r_{2}}G_{dd}^{A}\left( \epsilon
_{p_{1}}\right) \frac{\Gamma _{r_{1}j}\left( -p_{1}\right) \Gamma
_{r_{2}j^{\prime }}^{\ast }\left( p_{1}\right) }{2t\sin p_{1}}\frac{1}{%
\Omega ^{2}}+O\left( 1\right)  \notag \\
& =-2\frac{1}{\pi \Omega }\int_{-2t}^{2t-\Omega }d\epsilon _{k}\left[
-f^{\prime }\left( \epsilon _{k}\right) \right] \sum_{r_{1}r_{2}}t_{r_{1}}^{%
\ast }t_{r_{2}}\Gamma _{r_{1}j}\left( -k\right) \Gamma _{r_{2}j^{\prime
}}^{\ast }\left( k\right) \pi \nu _{k}G_{dd}^{A}\left( \epsilon _{k}\right)
S_{jj^{\prime }}\left( p\right) +O\left( 1\right) \text{.}
\end{align}%
We have adopted the shorthand $\epsilon _{q_{1}}-\Omega \equiv \epsilon
_{p_{1}}$ ($0\leq p_{1}\leq \pi $) and identified $k$ with $p_{1}$ and $p$
with $q_{1}$. This result, together with $G_{jj^{\prime },R0}^{\prime
D}+G_{jj^{\prime },A0}^{\prime D}$ which yields its complex conjugate, leads
to Eqs.~(\ref{discprobKYAZR}) and (\ref{discprobKYAZI}).

\subsection{Terms quadratic in T-matrix}

We focus on $G_{jj^{\prime},RR}^{\prime D}\left( \Omega\right)
+G_{jj^{\prime},RA}^{\prime D}\left( \Omega\right) $ first.

\end{subequations}
\begin{align}
& G_{jj^{\prime },RR}^{\prime D}\left( \Omega \right) +G_{jj^{\prime
},RA}^{\prime D}\left( \Omega \right)  \notag \\
& =2\int \frac{d\omega d\omega ^{\prime }}{\left( 2\pi \right) ^{2}}\frac{%
f\left( \omega \right) -f\left( \omega ^{\prime }\right) }{\omega -\omega
^{\prime }+\Omega ^{+}}\int_{0}^{\pi }\frac{dk_{1}dk_{2}}{\left( 2\pi
\right) ^{2}}\frac{dq_{1}dq_{2}}{\left( 2\pi \right) ^{2}}\tr \left\{ 
\mathbb{M}_{k_{1}k_{2}}^{j}\left( i\tau _{\psi }\right) g_{k_{2}}^{R}\left(
\omega ^{\prime }\right) V_{k_{2}}G_{dd}^{R}\left( \omega ^{\prime }\right)
V_{q_{1}}g_{q_{1}}^{R}\left( \omega ^{\prime }\right) \right.  \notag \\
& \times \left. \mathbb{M}_{q_{1}q_{2}}^{j^{\prime }}\left( i\tau _{\psi
}\right) \left[ g_{q_{2}}^{R}\left( \omega \right) V_{q_{2}}G_{dd}^{R}\left(
\omega \right) V_{k_{1}}g_{k_{1}}^{R}\left( \omega \right)
-g_{q_{2}}^{A}\left( \omega \right) V_{q_{2}}G_{dd}^{A}\left( \omega \right)
V_{k_{1}}g_{k_{1}}^{A}\left( \omega \right) \right] \right\} \text{.}
\end{align}%
Inserting Eqs.~(\ref{bbUmatrow1}), (\ref{bbMmat}) and using Eq.~(\ref%
{Gammaminusk}) again, we find

\begin{align}
& G_{jj^{\prime },RR}^{\prime D}\left( \Omega \right) +G_{jj^{\prime
},RA}^{\prime D}\left( \Omega \right)  \notag \\
& =2\int \frac{d\omega d\omega ^{\prime }}{\left( 2\pi \right) ^{2}}\frac{%
f\left( \omega \right) -f\left( \omega ^{\prime }\right) }{\omega -\omega
^{\prime }+\Omega ^{+}}\int_{-\pi }^{\pi }\frac{dk_{1}dk_{2}}{\left( 2\pi
\right) ^{2}}\frac{dq_{1}dq_{2}}{\left( 2\pi \right) ^{2}}%
\sum_{r_{1}r_{2}r_{1}^{\prime }r_{2}^{\prime }}t_{r_{1}}^{\ast
}t_{r_{2}}t_{r_{1}^{\prime }}^{\ast }t_{r_{2}^{\prime }}\Gamma
_{r_{1}j}\left( k_{1}\right)  \notag \\
& \times \Gamma _{r_{2}j}^{\ast }\left( k_{2}\right) \frac{1}{1-e^{i\left(
k_{1}-k_{2}+i0\right) }}ig_{k_{2}}^{R}\left( \omega ^{\prime }\right)
G_{dd}^{R}\left( \omega ^{\prime }\right) g_{q_{1}}^{R}\left( \omega
^{\prime }\right) \Gamma _{r_{1}^{\prime }j^{\prime }}\left( q_{1}\right)
\Gamma _{r_{2}^{\prime }j^{\prime }}^{\ast }\left( q_{2}\right)  \notag \\
& \times \frac{1}{1-e^{i\left( q_{1}-q_{2}+i0\right) }}i\left[
g_{q_{2}}^{R}\left( \omega \right) G_{dd}^{R}\left( \omega \right)
g_{k_{1}}^{R}\left( \omega \right) -g_{q_{2}}^{A}\left( \omega \right)
G_{dd}^{A}\left( \omega \right) g_{k_{1}}^{A}\left( \omega \right) \right] 
\text{.}
\end{align}%
We can integrate over all four momenta. Let $\omega =\epsilon _{k}$ where $%
0\leq k\leq \pi $ or $k=i\left\vert k\right\vert $, and $\omega ^{\prime
}=\epsilon _{k^{\prime }}$ where $0\leq k^{\prime }\leq \pi $ or $k^{\prime
}=-i\left\vert k^{\prime }\right\vert $; integrating over $k_{2}$ and $q_{1}$%
,

\begin{align}
& G_{jj^{\prime },RR}^{\prime D}\left( \Omega \right) +G_{jj^{\prime
},RA}^{\prime D}\left( \Omega \right)  \notag \\
& =2\int \frac{d\epsilon _{k}d\epsilon _{k^{\prime }}}{\left( 2\pi \right)
^{2}}\frac{f\left( \epsilon _{k}\right) -f\left( \epsilon _{k^{\prime
}}\right) }{\epsilon _{k}-\epsilon _{k^{\prime }}+\Omega ^{+}}\int_{-\pi
}^{\pi }\frac{dk_{1}dq_{2}}{\left( 2\pi \right) ^{2}}\sum_{r_{1}r_{2}r_{1}^{%
\prime }r_{2}^{\prime }}t_{r_{1}}^{\ast }t_{r_{2}}t_{r_{1}^{\prime }}^{\ast
}t_{r_{2}^{\prime }}\Gamma _{r_{1}j}\left( k_{1}\right)  \notag \\
& \times \Gamma _{r_{2}j}^{\ast }\left( -k^{\prime }\right) \frac{1}{%
1-e^{i\left( k_{1}+k^{\prime }+i0\right) }}i\frac{1}{2it\sin k^{\prime }}%
G_{dd}^{R}\left( \epsilon _{k^{\prime }}\right) \frac{1}{2it\sin k^{\prime }}%
\Gamma _{r_{1}^{\prime }j^{\prime }}\left( k^{\prime }\right) \Gamma
_{r_{2}^{\prime }j^{\prime }}^{\ast }\left( q_{2}\right)  \notag \\
& \times \frac{1}{1-e^{i\left( k^{\prime }-q_{2}+i0\right) }}i\left[
g_{q_{2}}^{R}\left( \epsilon _{k}\right) G_{dd}^{R}\left( \epsilon
_{k}\right) g_{k_{1}}^{R}\left( \epsilon _{k}\right) -g_{q_{2}}^{A}\left(
\epsilon _{k}\right) G_{dd}^{A}\left( \epsilon _{k}\right)
g_{k_{1}}^{A}\left( \epsilon _{k}\right) \right] \text{;}
\end{align}%
finally, integrating over $k_{1}$ and $q_{2}$, we find

\begin{subequations}
\begin{equation}
G_{jj^{\prime},RR}^{\prime D}\left( \Omega\right) =O\left( 1\right) \text{,}
\end{equation}

\begin{align}
& G_{jj^{\prime },RA}^{\prime D}\left( \Omega \right)  \notag \\
& =2\int \frac{d\epsilon _{k}d\epsilon _{k^{\prime }}}{\left( 2\pi \right)
^{2}}\frac{f\left( \epsilon _{k}\right) -f\left( \epsilon _{k^{\prime
}}\right) }{\epsilon _{k}-\epsilon _{k^{\prime }}+\Omega ^{+}}%
\sum_{r_{1}r_{2}r_{1}^{\prime }r_{2}^{\prime }}t_{r_{1}}^{\ast
}t_{r_{2}}t_{r_{1}^{\prime }}^{\ast }t_{r_{2}^{\prime }}\Gamma
_{r_{1}j}\left( -k\right) \Gamma _{r_{2}j}^{\ast }\left( -k^{\prime }\right)
\Gamma _{r_{1}^{\prime }j^{\prime }}\left( k^{\prime }\right)  \notag \\
& \times \Gamma _{r_{2}^{\prime }j^{\prime }}^{\ast }\left( k\right) \frac{1%
}{\left( 2t\sin k^{\prime }\right) ^{2}}\frac{1}{\left( 2t\sin k\right) ^{2}}%
\frac{1}{\left[ 1-e^{i\left( k^{\prime }-k+i0\right) }\right] ^{2}}%
G_{dd}^{R}\left( \epsilon _{k^{\prime }}\right) G_{dd}^{A}\left( \epsilon
_{k}\right) \text{.}
\end{align}%
Expanding around $k=k^{\prime }$ and integrating over $\epsilon _{k}$ and $%
\epsilon _{k^{\prime }}$, assuming $\Omega >0$, we obtain

\end{subequations}
\begin{align}
& G_{jj^{\prime },RA}^{\prime D}\left( \Omega \right)  \notag \\
& =2\int_{-2t}^{2t-\Omega }\frac{d\epsilon _{k}}{2\pi i}\Omega \left[
-f^{\prime }\left( \epsilon _{k}\right) \right] \sum_{r_{1}r_{2}r_{1}^{%
\prime }r_{2}^{\prime }}t_{r_{1}}^{\ast }t_{r_{2}}t_{r_{1}^{\prime }}^{\ast
}t_{r_{2}^{\prime }}\frac{\Gamma _{r_{1}j}\left( -k\right) \Gamma
_{r_{2}^{\prime }j^{\prime }}^{\ast }\left( k\right) }{2t\sin k}  \notag \\
& \times \frac{\Gamma _{r_{2}j}^{\ast }\left( -p\right) \Gamma
_{r_{1}^{\prime }j^{\prime }}\left( p\right) }{2t\sin p}\frac{1}{-\Omega ^{2}%
}G_{dd}^{R}\left( \epsilon _{p}\right) G_{dd}^{A}\left( \epsilon _{k}\right)
+O\left( 1\right) \text{,}  \label{GjjprDRA}
\end{align}%
where we have written $\epsilon _{k}+\Omega \equiv \epsilon _{p}$. A similar
calculation can be performed on $G_{jj^{\prime },AR}^{\prime D}\left( \Omega
\right) $ and $G_{jj^{\prime },AA}^{\prime D}\left( \Omega \right) $; both
are $O\left( 1\right) $ for the same reason that $G_{jj^{\prime
},RR}^{\prime D}\left( \Omega \right) $ is $O\left( 1\right) $. Therefore, $%
G_{jj^{\prime },RA}^{\prime D}$ is the only term quadratic in the T-matrix
which contributes to the linear DC conductance. Note that this is not the
case in Ref.~\onlinecite{PhysRevB.88.245104}, where the $RA$ term and the $%
AR $ term are complex conjugates as a result of taking the $\delta $%
-function part in Eq.~(\ref{GjjprDImIm}). It is easy to see that Eq.~(\ref%
{GjjprDRA}) reproduces Eq.~(\ref{discprobKYAZ2}).

We mention in passing that Eq.~(\ref{discprobKYA}) can be derived in the
wide band limit with essentially the same method, although the pole
structure is much simpler in that case.

\section{Non-interacting QD\label{sec:appNI}}

In this appendix we verify Eq.~(\ref{discprob}) in the case of a
non-interacting QD with $U=0$ as a consistency check on our formalism. The
connected part must vanish completely in this case; thus the disconnected
contribution to the conductance should coincide with the Landauer formula,
with an S-matrix modified by the presence of the non-interacting QD.

\subsection{Relation between $S\left( k\right) $ and $\Gamma \left( k\right) 
$}

First let us quantify the relation between the background S-matrix $S\left(
k\right) $ and the coupling site wave function matrix $\Gamma \left(
k\right) $. $S\left( k\right) $ and $\Gamma \left( k\right) $ are both
determined by the non-interacting part of the system $H_{0}$ with the QD
decoupled, and are usually intimately related: for instance they often have
the same poles on the complex momentum plane, and have resonances (sharp
changes in amplitude and phase) over the same range of real momenta. To
characterize this relation, we attach to each of the $M$ coupling sites an
additional \textquotedblleft phantom\textquotedblright\ lead. These phantom
leads have the same nearest neighbor hopping $t$ as the $N$ pre-existing
leads. The coupling site is designated as site $0$ of its phantom lead. We
let $S^{0}$ be the S-matrix of the resulting non-interacting system with $%
\left( N+M\right) $ leads; that is, for the scattering state incident from
lead $m$ with momentum $k$, the wave function on lead $m^{\prime }$ should be

\begin{equation}
\chi_{m,k}\left( m^{\prime},n\right)
=\delta_{mm^{\prime}}e^{-ikn}+S_{m^{\prime}m}^{0}\left( k\right) e^{ikn}%
\text{.}  \label{S0wf}
\end{equation}
$m$ and $m^{\prime}$ can be $j=1$, $\cdots$, $N$ or $r=1$, $\cdots$, $M$, so
that $S^{0}$ is an $\left( N+M\right) \times\left( N+M\right) $ unitary
matrix. For future convenience we partition $S^{0}$ as follows,

\begin{equation}
S^{0}\equiv \left( 
\begin{array}{cc}
S_{LL}^{0} & S_{LD}^{0} \\ 
S_{DL}^{0} & S_{DD}^{0}%
\end{array}%
\right) \text{,}  \label{S0Mat}
\end{equation}%
where $L$ and $D$ are shorthands for lead and dot respectively; $S_{LL}^{0}$
is $N\times N$ and $S_{DD}^{0}$ is $M\times M$. Note that $S_{LL}^{0}$ and $%
S_{DD}^{0}$ are not usually unitary, although all four blocks are
constrained by the overall unitarity of $S^{0}$. Also, our argument for $%
S\left( -k\right) $ in Sec.~\ref{sec:appdisc} applies to $S^{0}\left(
-k\right) $:

\begin{equation}
S^{0}\left( -k\right) =\left[ S^{0}\left( k\right) \right] ^{\dag }\text{.}
\label{S0Matminusk}
\end{equation}

Due to the non-interacting nature of the system, we can conveniently express 
$S$ and $\Gamma $ in terms of the auxiliary object $S^{0}$. With the QD
decoupled, the phantom leads should be terminated by open boundary
conditions. The wave function on phantom lead $r$ should be

\begin{equation}
\chi _{j,k}\left( r,n\right) =\frac{\Gamma _{rj}\left( k\right) }{2i\sin k}%
\left( e^{ik}e^{-ikn}-e^{-ik}e^{ikn}\right) \text{,}
\end{equation}%
so that $\chi _{j,k}\left( r,1\right) =0$ and $\chi _{j,k}\left( r,0\right)
=\Gamma _{rj}\left( k\right) $. The incident amplitude and reflected
amplitude in lead $r$ are respectively $\pm e^{\pm ik}\Gamma _{rj}\left(
k\right) /\left( 2i\sin k\right) $. We then see, from the definition of $%
S^{0}$, that $\Gamma _{rj}\left( k\right) $ and $S$ obey the following
relations:

\begin{subequations}
\begin{equation}
S_{rj}^{0}\left( k\right) +\sum_{r^{\prime }=1}^{M}S_{rr^{\prime
}}^{0}\left( k\right) \frac{\Gamma _{r^{\prime }j}\left( k\right) e^{ik}}{%
2i\sin k}=-\frac{\Gamma _{rj}\left( k\right) e^{-ik}}{2i\sin k}\text{,}
\end{equation}

\begin{equation}
S_{j^{\prime }j}\left( k\right) =S_{j^{\prime }j}^{0}\left( k\right)
+\sum_{r^{\prime }=1}^{M}S_{j^{\prime }r^{\prime }}^{0}\left( k\right) \frac{%
\Gamma _{r^{\prime }j}\left( k\right) e^{ik}}{2i\sin k}\text{.}
\end{equation}%
Therefore,

\end{subequations}
\begin{equation}
\Gamma \left( k\right) =-\left( 2i\sin k\right) \left[ e^{ik}S_{DD}^{0}%
\left( k\right) +e^{-ik}\right] ^{-1}S_{DL}^{0}\left( k\right) \text{,}
\label{Gammak}
\end{equation}%
and

\begin{equation}
S\left( k\right) =S_{LL}^{0}\left( k\right) -S_{LD}^{0}\left( k\right) \left[
e^{ik}S_{DD}^{0}\left( k\right) +e^{-ik}\right] ^{-1}S_{DL}^{0}\left(
k\right) e^{ik}\text{.}  \label{SMatk}
\end{equation}%
It is straightforward to check the unitarity of $S\left( k\right) $.

\subsection{Green's function of the QD}

We now find the retarded T-matrix of the non-interacting QD. As is standard
for the non-interacting Anderson model in Eq.~(\ref{Andersonmodel}),\cite%
{mahan2000many} the retarded Green's function for the QD is

\begin{equation}
G_{dd}^{R}\left( \omega\right) =\frac{1}{\omega^{+}-\epsilon_{d}-\Sigma
_{dd}^{R}\left( \omega\right) }\text{,}  \label{GRddNI}
\end{equation}
where $\omega^{+}\equiv\omega+i0$, and the retarded self-energy is

\begin{equation}
\Sigma_{dd}^{R}\left( \omega\right) =\int_{0}^{\pi}\frac{dq}{2\pi}\frac{%
V_{q}^{2}}{\omega^{+}-\epsilon_{q}}=\int_{0}^{\pi}\frac{dq}{2\pi}\frac{\tr %
\left( \lambda\Gamma_{q}\Gamma_{q}^{\dag}\right) }{\omega^{+}-\epsilon_{q}}%
\text{.}
\end{equation}
We have used Eq.~(\ref{Vksquared}). Inserting Eq.~(\ref{Gammak}), using the
unitarity of $S^{0}$ and Eq.~(\ref{S0Matminusk}), we obtain

\begin{align}
& \int_{0}^{\pi }\frac{dq}{2\pi }\frac{\Gamma _{q}\Gamma _{q}^{\dag }}{%
\omega ^{+}-\epsilon _{q}}  \notag \\
& =\int_{0}^{\pi }\frac{dq}{2\pi }\frac{4\sin ^{2}q}{\omega ^{+}-\epsilon
_{q}}\left[ e^{2iq}S_{DD}^{0}\left( q\right) +1\right] ^{-1}\left[
1-S_{DD}^{0}\left( q\right) \left( S_{DD}^{0}\left( q\right) \right) ^{\dag }%
\right] \left\{ \left[ e^{2iq}S_{DD}^{0}\left( q\right) +1\right]
^{-1}\right\} ^{\dag }  \notag \\
& =\int_{0}^{\pi }\frac{dq}{2\pi }\frac{4\sin ^{2}q}{\omega ^{+}-\epsilon
_{q}}\left( \left[ e^{2iq}S_{DD}^{0}\left( q\right) +1\right] ^{-1}+\left\{ %
\left[ e^{2iq}S_{DD}^{0}\left( q\right) +1\right] ^{-1}\right\} ^{\dag
}-1\right)  \notag \\
& =\int_{-\pi }^{\pi }\frac{dq}{2\pi }\frac{4\sin ^{2}q}{\omega
^{+}-\epsilon _{q}}\left\{ \left[ e^{2iq}S_{DD}^{0}\left( q\right) +1\right]
^{-1}-\frac{1}{2}\right\} \text{.}
\end{align}

With contour methods it is straightforward to show, for $0<p<\pi$ and
integer $n$, that

\begin{equation}
\int_{-\pi }^{\pi }\frac{dq}{2\pi }\frac{e^{inq}}{\epsilon _{p}-\epsilon
_{q}+i0}=\frac{e^{i\left\vert n\right\vert p}}{2it\sin p}\text{.}
\label{gencontint}
\end{equation}%
We now make the assumption that $S_{DD}^{0}\left( z=e^{iq}\right) \equiv
S_{DD}^{0}\left( q\right) $ is analytic at the origin of the complex plane.
This appears to be a reasonable assumption, because a singularity at the
origin would imply an infinite energy feature, which should decouple
completely from any properties of system probed at finite energies. Thus $%
S_{DD}^{0}\left( z\right) $\ can be expanded around the origin in a Taylor
series of $z$, and we may integrate term by term to find

\begin{equation}
\int_{-\pi}^{\pi}\frac{dq}{2\pi}\frac{e^{in_{1}q}\left[ S_{DD}^{0}\left(
q\right) \right] ^{n_{2}}}{\epsilon_{p}-\epsilon_{q}+i0}=\frac{e^{in_{1}p}%
\left[ S_{DD}^{0}\left( p\right) \right] ^{n_{2}}}{2it\sin p}\text{,}
\end{equation}
where $n_{1}$ and $n_{2}$ are non-negative integers.

Expanding $4\sin ^{2}q\left\{ \left[ e^{2iq}S_{DD}^{0}\left( q\right) +1%
\right] ^{-1}-1/2\right\} $ in powers of $e^{iq}$ and $S_{DD}^{0}\left(
q\right) $, the only negative power term in the series is $-\left(
1/2\right) e^{-2iq}$; therefore, by Eq.~(\ref{gencontint}),

\begin{equation}
\int_{0}^{\pi}\frac{dq}{2\pi}\frac{\Gamma_{q}\Gamma_{q}^{\dag}}{\epsilon
_{p}-\epsilon_{q}+i0}=\frac{1}{2it\sin p}\left( 4\sin^{2}p\left\{ \left[
e^{2ip}S_{DD}^{0}\left( p\right) +1\right] ^{-1}-\frac{1}{2}\right\} +\frac{1%
}{2}e^{-2ip}-\frac{1}{2}e^{2ip}\right) \text{,}
\end{equation}
and hence

\begin{equation}
\Sigma _{dd}^{R}\left( \epsilon _{p}\right) =-\frac{1}{t}\tr \left( \lambda
\left\{ e^{-ip}+2i\sin p\left[ e^{2ip}S_{DD}^{0}\left( p\right) +1\right]
^{-1}\right\} \right) \text{.}  \label{SigmaRddNI}
\end{equation}

Eqs.~(\ref{GRddNI}) and (\ref{SigmaRddNI}) give the retarded T-matrix T$%
_{pp}\left( \epsilon _{p}\right) =V_{p}^{2}G_{dd}^{R}\left( \epsilon
_{p}\right) $. One can easily verify that the optical theorem $\operatorname{Im}%
\left[ -\pi \nu _{p}\text{T}_{pp}\left( \epsilon _{p}\right) \right]
=\left\vert -\pi \nu _{p}\text{T}_{pp}\left( \epsilon _{p}\right)
\right\vert ^{2}$ is obeyed, which must be the case for a non-interacting
system. Inserting Eqs.~(\ref{SMatk}), (\ref{Gammak}), (\ref{GRddNI}), and (%
\ref{SigmaRddNI}) into Eq.~(\ref{discprob}), we find

\begin{equation}
\mathcal{T}_{jj^{\prime }}^{D}\left( \epsilon _{p}\right) =\delta
_{jj^{\prime }}-\left\vert S_{jj^{\prime }}^{\text{NI}}\left( p\right)
\right\vert ^{2}\text{,}
\end{equation}%
where

\begin{align}
S^{\text{NI}}\left( p\right) & =S_{LL}^{0}\left( p\right) +S_{LD}^{0}\left(
p\right) \left\{ \frac{\left( 2i\sin p\right) e^{-2ip}\left[
S_{DD}^{0}\left( p\right) +e^{-2ip}\right] ^{-1}\bar{\lambda}_{p}\left[
S_{DD}^{0}\left( p\right) +e^{-2ip}\right] ^{-1}}{1+\tr \left( \bar{\lambda}%
_{p}\left\{ e^{-ip}+2i\sin p\left[ e^{2ip}S_{DD}^{0}\left( p\right) +1\right]
^{-1}\right\} \right) }\right.  \notag \\
& \left. -\left[ S_{DD}^{0}\left( p\right) +e^{-2ip}\right] ^{-1}\right\}
S_{DL}^{0}\left( p\right) \text{.}  \label{SMatNISE}
\end{align}
Here we have defined the dimensionless coupling matrix

\begin{equation}
\bar{\lambda}_{p}\equiv\frac{\lambda}{t\left( \epsilon_{p}-\epsilon
_{d}\right) }\text{.}  \label{lambdabarp}
\end{equation}

\subsection{Landauer formula}

We turn to the alternative method of scattering state wave functions to find
the S-matrix $S^{\text{NI}}$ in the presence of the QD.

By definition, for the scattering state incident from $j$, the wave function
on lead $j^{\prime }$ is

\begin{equation}
\chi _{j,k}\left( j^{\prime },n\right) =\delta _{jj^{\prime
}}e^{-ikn}+S_{j^{\prime }j}^{\text{NI}}\left( k\right) e^{ikn}\text{.}
\end{equation}%
The coupling sites are now attached to the QD. We imagine that the QD and
the coupling sites are separated by \textquotedblleft phantom
leads\textquotedblright\ of zero length, so that the wave function on
phantom lead $r$ takes the form

\begin{equation}
\chi_{j,k}\left( r,n\right) =A_{r,j,k}e^{-ikn}+B_{r,j,k}e^{ikn}\text{,}
\end{equation}
where $n=0$ only. Since the phantom leads are also attached to the QD, the
Schroedinger equation on coupling site $r$ is

\begin{equation}
t\left( A_{r,j,k}e^{-ik}+B_{r,j,k}e^{ik}\right) =t_{r}\phi _{j}\left(
k\right) \text{,}  \label{Schdot1}
\end{equation}%
where $\phi _{j}\left( k\right) $ is the wave function on the QD. Meanwhile,
the Schroedinger equation on the QD itself is

\begin{equation}
\epsilon_{k}\phi_{j}\left( k\right) =\epsilon_{d}\phi_{j}\left( k\right)
-\sum_{r}t_{r}^{\ast}\left( A_{r,j,k}+B_{r,j,k}\right) \text{.}
\label{Schdot2}
\end{equation}
Eqs.~(\ref{Schdot1}) and (\ref{Schdot2}) allow us to express $B_{r,j,k}$ in
terms of $A_{r,j,k}$, i.e. to find the S-matrix for the non-interacting QD:

\begin{equation}
t\left( A_{r,j,k}e^{-ik}+B_{r,j,k}e^{ik}\right) =-\frac{t_{r}}{\epsilon
_{k}-\epsilon_{d}}\sum_{r^{\prime}}t_{r^{\prime}}^{\ast}\left( A_{r^{\prime
},j,k}+B_{r^{\prime},j,k}\right) \text{,}
\end{equation}
or more compactly

\begin{equation}
e^{-ik}A_{k}+e^{ik}B_{k}=-\bar{\lambda}_{k}\left( A_{k}+B_{k}\right) \text{,}
\end{equation}

\begin{equation}
B_{k}=-\left( e^{ik}+\bar{\lambda}_{k}\right) ^{-1}\left( e^{-ik}+\bar{%
\lambda}_{k}\right) A_{k}\text{.}  \label{ABNI}
\end{equation}
On the other hand, the amplitudes $A_{k}$ and $B_{k}$ are also related to
each other and to $S^{\text{NI}}$ by the phantom-lead S-matrix $S^{0}$:

\begin{equation}
S_{rj}^{0}\left( k\right) +\sum_{r^{\prime}}S_{rr^{\prime}}^{0}\left(
k\right) A_{r^{\prime},j,k}=B_{r,j,k}\text{,}  \label{SchABS0}
\end{equation}
and

\begin{equation}
S_{j^{\prime }j}^{\text{NI}}\left( k\right) =S_{j^{\prime }j}^{0}\left(
k\right) +\sum_{r^{\prime }}S_{j^{\prime }r^{\prime }}^{0}\left( k\right)
A_{r^{\prime },j,k}\text{.}  \label{SchASNIS0}
\end{equation}%
Eliminating $A_{k}$ and $B_{k}$ from Eqs.~(\ref{ABNI}), (\ref{SchABS0}) and (%
\ref{SchASNIS0}) finally yields

\begin{equation}
S^{\text{NI}}\left( k\right) =S_{LL}^{0}\left( k\right) -S_{LD}^{0}\left(
k\right) \left[ S_{DD}^{0}\left( k\right) +\left( e^{ik}+\bar{\lambda}%
_{k}\right) ^{-1}\left( e^{-ik}+\bar{\lambda}_{k}\right) \right]
^{-1}S_{DL}^{0}\left( k\right) \text{.}  \label{SMatNIcas}
\end{equation}

Using Eqs.~(\ref{couplingmatlambda}) and (\ref{lambdabarp}), it is a lengthy
but straightforward task to prove the equivalence of Eqs.~(\ref{SMatNISE})
and (\ref{SMatNIcas}). Therefore the disconnected contribution to the
conductance recovers the Landauer formula in the case of a non-interacting
QD.

\section{Fermi liquid perturbation theory\label{sec:appFermi}}

In this appendix, we discuss the perturbation theory in the Fermi liquid
regime $T\ll T_{K}\ll E_{V}$ (also assuming $E_{V}\sim E_{\text{conn}}$; see
Table~\ref{tab:table1}). We first\ present an alternative derivation of Eq.~(%
\ref{TtildeAL}), the $O\left( 1/T_{K}^{2}\right) $\ retarded\ T-matrix
obtained in Ref.~\onlinecite{PhysRevB.48.7297}. Then we perform an
additional consistency check on our formalism of eliminating the connected
contribution to the DC conductance at low temperatures: we directly compute
the\ connected contribution to $O\left( T^{2}/T_{K}^{2}\right) $, and show
that Eq.~(\ref{connelimT}) is indeed satisfied.

In momentum space, the leading irrelevant operator Eq.~(\ref{FLLIO}) takes
the form

\begin{align}
H_{int}& =\frac{2\pi v_{F}^{2}}{T_{K}}\int d\eta H\left( \eta \right) \int 
\frac{dq_{1}dq_{2}dq_{3}dq_{4}}{\left( 2\pi \right) ^{4}}e^{i\left(
q_{1}-q_{2}+q_{3}-q_{4}\right) \eta }\colon \tilde{\psi}_{q_{1}\alpha
}^{\dag }\tilde{\psi}_{q_{2}\alpha }\tilde{\psi}_{q_{3}\beta }^{\dag }\tilde{%
\psi}_{q_{4}\beta }\colon   \notag \\
& -\frac{v_{F}^{2}}{T_{K}}\int d\eta H\left( \eta \right) \int \frac{%
dq_{1}dq_{2}}{\left( 2\pi \right) ^{2}}\left( q_{1}+q_{2}\right) e^{i\left(
q_{1}-q_{2}\right) \eta }\colon \tilde{\psi}_{q_{1}\alpha }^{\dag }\tilde{%
\psi}_{q_{2}\alpha }\colon \text{.}  \label{FLLIOmom}
\end{align}%
(We measure all momenta relative to the Fermi wavevector $k_{F}$ hereafter.)
Here $\eta $ is the location of the operator; the weight function $H\left(
\eta \right) \ $is peaked at the origin and can be approximated as a $\delta 
$-function above the length scale $v_{F}/T_{K}$. To lighten notations, we
take $H\left( \eta \right) =\delta \left( \eta \right) $ whenever it is
unambiguous to do so.

At $O\left( 1/T_{K}^{2}\right) $ both terms in Eq.~(\ref{FLLIOmom})
contribute to the\ T-matrix,\ but only the first term plays a role in the
connected 4-point function.

\subsection{T-matrix}

To find the retarded T-matrix of the phase-shifted screening channel $\tilde{%
\psi}$, we begin from the imaginary time 2-point Green's function

\begin{equation}
\mathcal{\tilde{G}}_{kk^{\prime }}\left( \tau \right) \equiv -\left\langle
T_{\tau }\tilde{\psi}_{k}\left( \tau \right) \tilde{\psi}_{k^{\prime
}}^{\dag }\left( 0\right) \right\rangle \text{.}
\end{equation}%
This object is diagonal in spin indices. The three diagrams in Fig.~\ref%
{fig:FermiFD1} panel b) evaluate to

\begin{eqnarray}
\mathcal{\tilde{G}}_{kk^{\prime }}\left( \tau \right)  &=&2\pi \delta \left(
k-k^{\prime }\right) g_{k}\left( \tau \right) -\frac{v_{F}^{2}}{T_{K}}\left(
k+k^{\prime }\right) \int d\tau _{1}g_{k}\left( \tau -\tau _{1}\right)
g_{k^{\prime }}\left( \tau _{1}\right) +\left( \frac{v_{F}^{2}}{T_{K}}%
\right) ^{2}\int \frac{dq}{2\pi }\left( k+q\right)   \notag \\
&&\times \left( q+k^{\prime }\right) \int d\eta _{1}d\eta _{2}H\left( \eta
_{1}\right) H\left( \eta _{2}\right) e^{i\left( k\eta _{1}-k^{\prime }\eta
_{2}\right) }e^{iq\left( \eta _{2}-\eta _{1}\right) }\int d\tau _{1}d\tau
_{2}g_{k}\left( \tau -\tau _{1}\right) g_{q}\left( \tau _{1}-\tau
_{2}\right)   \notag \\
&&\times g_{k^{\prime }}\left( \tau _{2}\right) -4\left( \frac{2\pi v_{F}^{2}%
}{T_{K}}\right) ^{2}\int \frac{dq_{2}dq_{3}dq_{4}}{\left( 2\pi \right) ^{3}}%
\int d\eta _{1}d\eta _{2}H\left( \eta _{1}\right) H\left( \eta _{2}\right)
e^{i\left( k\eta _{1}-k^{\prime }\eta _{2}\right) }  \notag \\
&&\times e^{i\left( q_{2}-q_{3}+q_{4}\right) \left( \eta _{2}-\eta
_{1}\right) }\int d\tau _{1}d\tau _{2}g_{k}\left( \tau -\tau _{1}\right)
g_{q_{2}}\left( \tau _{1}-\tau _{2}\right) g_{q_{3}}\left( \tau _{2}-\tau
_{1}\right) g_{q_{4}}\left( \tau _{1}-\tau _{2}\right) g_{k^{\prime }}\left(
\tau _{2}\right) \text{.}
\end{eqnarray}%
Going to the Fourier space, we identify the imaginary time T-matrix as

\begin{eqnarray}
\text{\~{T}}_{kk^{\prime }}\left( i\omega _{n}\right)  &=&-\frac{v_{F}^{2}}{%
T_{K}}\left( k+k^{\prime }\right) +\left( \frac{v_{F}^{2}}{T_{K}}\right)
^{2}\int \frac{dq}{2\pi }\left( k+q\right) \left( q+k^{\prime }\right) \int
d\eta _{1}d\eta _{2}H\left( \eta _{1}\right) H\left( \eta _{2}\right)
e^{i\left( k\eta _{1}-k^{\prime }\eta _{2}\right) }  \notag \\
&&\times e^{iq\left( \eta _{2}-\eta _{1}\right) }g_{q}\left( i\omega
_{n}\right) -4\left( \frac{2\pi v_{F}^{2}}{T_{K}}\right) ^{2}\int \frac{%
dq_{2}dq_{3}dq_{4}}{\left( 2\pi \right) ^{3}}\int d\eta _{1}d\eta
_{2}H\left( \eta _{1}\right) H\left( \eta _{2}\right) e^{i\left( k\eta
_{1}-k^{\prime }\eta _{2}\right) }  \notag \\
&&\times e^{i\left( q_{2}-q_{3}+q_{4}\right) \left( \eta _{2}-\eta
_{1}\right) }\frac{1}{\beta }\sum_{\omega _{n_{2}}}\frac{1}{\beta }%
\sum_{\omega _{n_{4}}}g_{q_{2}}\left( i\omega _{n_{2}}\right)
g_{q_{3}}\left( i\omega _{n_{2}}+i\omega _{n_{4}}-i\omega _{n}\right)
g_{q_{4}}\left( i\omega _{n_{4}}\right) \text{.}
\end{eqnarray}%
where all Matsubara frequencies are fermionic, e.g. $\omega _{n}=\left(
2n+1\right) \pi /\beta $. Both frequency summations are standard,\cite%
{mahan2000many} and analytic continuation $i\omega _{n}\rightarrow \omega
^{+}$\ yields

\begin{eqnarray}
\text{\~{T}}_{kk^{\prime }}\left( \omega \right)  &=&-\frac{v_{F}^{2}}{T_{K}}%
\left( k+k^{\prime }\right) +\left( \frac{v_{F}^{2}}{T_{K}}\right) ^{2}\int 
\frac{dq}{2\pi }\left( k+q\right) \left( q+k^{\prime }\right) \int d\eta
_{1}d\eta _{2}H\left( \eta _{1}\right) H\left( \eta _{2}\right) e^{i\left(
k\eta _{1}-k^{\prime }\eta _{2}\right) }  \notag \\
&&\times \frac{e^{iq\left( \eta _{2}-\eta _{1}\right) }}{\omega
^{+}-\epsilon _{q}}-4\left( \frac{2\pi v_{F}^{2}}{T_{K}}\right) ^{2}\int
d\eta _{1}d\eta _{2}H\left( \eta _{1}\right) H\left( \eta _{2}\right)
e^{i\left( k\eta _{1}-k^{\prime }\eta _{2}\right) }\int \frac{%
dq_{2}dq_{3}dq_{4}}{\left( 2\pi \right) ^{3}}  \notag \\
&&\times e^{i\left( q_{2}-q_{3}+q_{4}\right) \left( \eta _{2}-\eta
_{1}\right) }\frac{\left[ -f_{B}\left( \epsilon _{q_{3}}-\epsilon
_{q_{4}}\right) -f\left( \epsilon _{q_{2}}\right) \right] \left[ f\left(
\epsilon _{q_{4}}\right) -f\left( \epsilon _{q_{3}}\right) \right] }{\omega
^{+}+\epsilon _{q_{3}}-\epsilon _{q_{4}}-\epsilon _{q_{2}}}\text{,}
\end{eqnarray}%
where $f_{B}\left( \omega \right) =1/\left( e^{\beta \omega }-1\right) $ is
the Bose function.

In the $q$ integral we close the contour in the upper half plane for $\eta
_{2}>\eta _{1}$, and in the lower half plane for $\eta _{2}<\eta _{1}$;\
this leads to

\begin{equation}
\int \frac{dq}{2\pi }\left( k+q\right) \left( q+k^{\prime }\right) \frac{%
e^{iq\left( \eta _{2}-\eta _{1}\right) }}{\omega ^{+}-\epsilon _{q}}=-\frac{i%
}{v_{F}}\left( k+\frac{\omega }{v_{F}}\right) \left( \frac{\omega }{v_{F}}%
+k^{\prime }\right) e^{i\frac{\omega ^{+}}{v_{F}}\left( \eta _{2}-\eta
_{1}\right) }\theta \left( \eta _{2}-\eta _{1}\right) \text{.}
\end{equation}%
For the on-shell T-matrix \~{T}$_{pp}\left( \epsilon _{p}\right) $, the
phase factors involving $\eta _{1}$ and $\eta _{2}$ cancel, and the $\eta $
integrals become $\int d\eta _{1}d\eta _{2}H\left( \eta _{1}\right) H\left(
\eta _{2}\right) \theta \left( \eta _{2}-\eta _{1}\right) =1/2$. We can
simplify the triple integral over $q_{2}$, $q_{3}$ and $q_{4}$ by the
contour method in a similar fashion, before using the following identity,

\begin{equation}
\int_{-\infty }^{\infty }d\epsilon _{q_{2}}d\epsilon _{q_{3}}d\epsilon
_{q_{4}}\left[ f_{B}\left( \epsilon _{q_{3}}-\epsilon _{q_{4}}\right)
+f\left( \epsilon _{q_{2}}\right) \right] \left[ f\left( \epsilon
_{q_{4}}\right) -f\left( \epsilon _{q_{3}}\right) \right] \delta \left(
\omega +\epsilon _{q_{3}}-\epsilon _{q_{4}}-\epsilon _{q_{2}}\right) =\frac{1%
}{2}\left( \pi ^{2}T^{2}+\omega ^{2}\right) \text{,}  \label{tripFermi}
\end{equation}%
which has been given in Ref.~\onlinecite{JLowTempPhys.17.31} in the context
of an inelastic scattering collision integral. Collecting all three terms,
we recover Eq.~(\ref{TtildeAL}).

\subsection{Connected contribution to the conductance}

Inserting Eqs.~(\ref{GCkkqqimag}) and (\ref{psitildescatt}) into the 4-point
function Eq.~(\ref{GprCimag1}), and performing the $k_{1}$, $k_{2}$, $q_{1}$
and $q_{2}$ integrals, we obtain

\begin{align}
\mathcal{G}_{jj^{\prime }}^{\prime C}\left( i\omega _{p}\right) & =\int 
\frac{dp_{1}dp_{2}dp_{3}dp_{4}}{\left( 2\pi \right) ^{4}}\mathcal{\tilde{G}}%
_{p_{1}p_{2}p_{3}p_{4}}^{C}\left( i\omega _{p}\right)
\sum_{j_{1}j_{2}}\sum_{j_{1}^{\prime }j_{2}^{\prime }}\mathbb{U}_{1,j_{1}}%
\mathbb{U}_{1,j_{2}}^{\ast }\mathbb{U}_{1,j_{1}^{\prime }}\mathbb{U}%
_{1,j_{2}^{\prime }}^{\ast }  \notag \\
& \times \left( \delta _{jj_{1}}\delta _{jj_{2}}\frac{i}{p_{1}-p_{2}+i0}%
+S_{jj_{1}}^{\ast }S_{jj_{2}}\frac{i}{p_{2}-p_{1}+i0}\right)   \notag \\
& \times \left( \delta _{j^{\prime }j_{1}^{\prime }}\delta _{j^{\prime
}j_{2}^{\prime }}\frac{i}{p_{3}-p_{4}+i0}+S_{j^{\prime }j_{1}^{\prime
}}^{\ast }S_{j^{\prime }j_{2}^{\prime }}\frac{i}{p_{4}-p_{3}+i0}\right) 
\text{;}  \label{GprCimagFL}
\end{align}%
we have ignored the momentum dependence of $\mathbb{U}$ and $S$ in the Fermi
liquid regime (which is justified at $T_{K}\ll E_{\text{conn}}$). Here the $%
\delta _{P}$-independent connected four-point correlation function for $%
\tilde{\psi}$ is defined as

\begin{equation}
\mathcal{\tilde{G}}_{p_{1}p_{2}p_{3}p_{4}}^{C}\left( i\omega _{p}\right)
\equiv -\int_{0}^{\beta }d\tau e^{i\omega _{p}\tau }\sum_{\sigma \sigma
^{\prime }}\left\langle T_{\tau }\tilde{\psi}_{p_{1}\sigma }^{\dag }\left(
\tau \right) \tilde{\psi}_{p_{2}\sigma }\left( \tau \right) \tilde{\psi}%
_{p_{3}\sigma ^{\prime }}^{\dag }\left( 0\right) \tilde{\psi}_{p_{4}\sigma
^{\prime }}\left( 0\right) \right\rangle _{C}\text{.}
\label{GtildeCppppimag}
\end{equation}%
We observe that $\delta _{\psi \psi }$ drops out of $\mathcal{G}_{jj^{\prime
}}^{\prime C}$ completely, which reflects the inelastic nature of the
connected contribution.

\begin{figure}[ptb]
\includegraphics[width=0.4\textwidth]{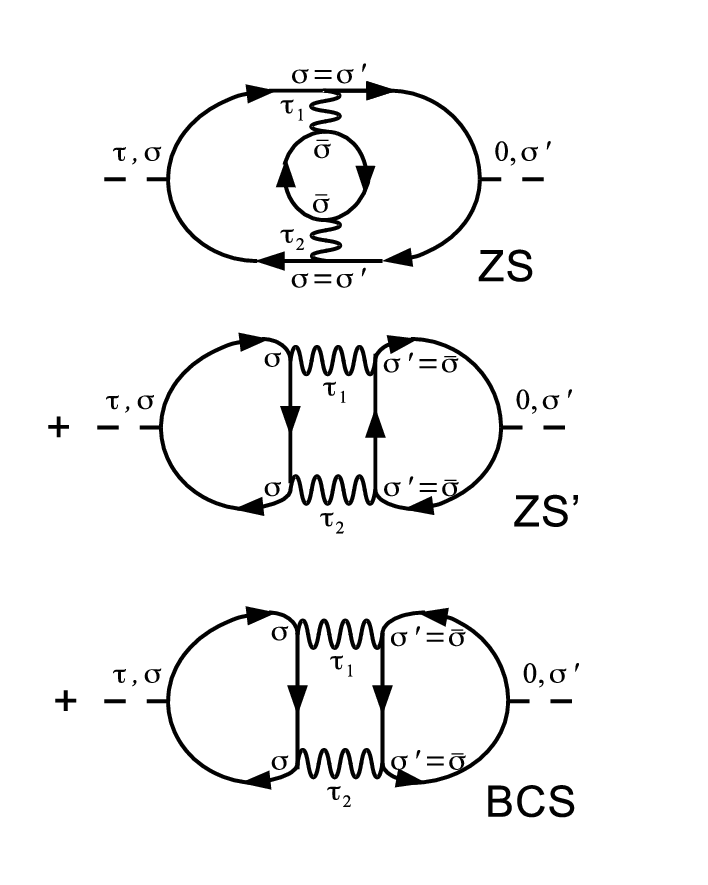}%
\caption{The three connected diagrams at $O\left(  T^{2}/T_{K}^{2} \right)  $
contributing to the conductance. ZS, ZS' and BCS label only the topology of
the diagrams and not necessarily the physics.\label{fig:FermiFD2}}
\end{figure}

To $O\left( 1/T_{K}^{2}\right) $, there are three diagrams resulting in
nonzero connected contributions to the linear DC conductance, depicted in
Fig.~\ref{fig:FermiFD2}. The corresponding 4-point functions read

\begin{subequations}
\begin{align}
\mathcal{\tilde{G}}_{p_{1}p_{2}p_{3}p_{4}}^{C\text{,BCS}}\left( i\omega
_{p}\right) & =-4\left( \frac{2\pi v_{F}^{2}}{T_{K}}\right)
^{2}\int_{0}^{\beta }d\tau e^{i\omega _{p}\tau }\int_{0}^{\beta }d\tau
_{1}d\tau _{2}\sum_{\sigma \sigma ^{\prime }}\delta _{\sigma \bar{\sigma}%
^{\prime }}\int \frac{dq_{1}dq_{3}}{\left( 2\pi \right) ^{2}}  \notag \\
& \times g_{p_{1}}\left( \tau _{1}-\tau \right) g_{p_{2}}\left( \tau -\tau
_{2}\right) g_{p_{3}}\left( \tau _{1}\right) g_{p_{4}}\left( -\tau
_{2}\right) g_{q_{1}}\left( \tau _{2}-\tau _{1}\right) g_{q_{3}}\left( \tau
_{2}-\tau _{1}\right) \text{,}
\end{align}

\begin{align}
\mathcal{\tilde{G}}_{p_{1}p_{2}p_{3}p_{4}}^{C\text{,ZS}}\left( i\omega
_{p}\right) & =-4\left( \frac{2\pi v_{F}^{2}}{T_{K}}\right)
^{2}\int_{0}^{\beta }d\tau e^{i\omega _{p}\tau }\int_{0}^{\beta }d\tau
_{1}d\tau _{2}\sum_{\sigma \sigma ^{\prime }}\delta _{\sigma \sigma ^{\prime
}}\int \frac{dq_{3}dq_{4}}{\left( 2\pi \right) ^{2}}  \notag \\
& \times g_{p_{1}}\left( \tau _{1}-\tau \right) g_{p_{2}}\left( \tau -\tau
_{2}\right) g_{p_{3}}\left( \tau _{2}\right) g_{p_{4}}\left( -\tau
_{1}\right) g_{q_{3}}\left( \tau _{2}-\tau _{1}\right) g_{q_{4}}\left( \tau
_{1}-\tau _{2}\right) \text{,}
\end{align}

\begin{align}
\mathcal{\tilde{G}}_{p_{1}p_{2}p_{3}p_{4}}^{C\text{,ZS'}}\left( i\omega
_{p}\right) & =-4\left( \frac{2\pi v_{F}^{2}}{T_{K}}\right)
^{2}\int_{0}^{\beta }d\tau e^{i\omega _{p}\tau }\int_{0}^{\beta }d\tau
_{1}d\tau _{2}\sum_{\sigma \sigma ^{\prime }}\delta _{\sigma \bar{\sigma}%
^{\prime }}\int \frac{dq_{1}dq_{4}}{\left( 2\pi \right) ^{2}}  \notag \\
& \times g_{p_{1}}\left( \tau _{1}-\tau \right) g_{p_{2}}\left( \tau -\tau
_{2}\right) g_{p_{3}}\left( \tau _{2}\right) g_{p_{4}}\left( -\tau
_{1}\right) g_{q_{1}}\left( \tau _{2}-\tau _{1}\right) g_{q_{4}}\left( \tau
_{1}-\tau _{2}\right) \text{.}
\end{align}%
Here the terminology of BCS, ZS and ZS$^{\prime }$ is borrowed from Ref.~%
\onlinecite{RevModPhys.66.129} and refers only to the topology of the
diagrams.

We illustrate the calculation with the BCS diagram; ZS and ZS' again turn
out to be completely analogous. Going to the Fourier space,

\end{subequations}
\begin{align}
& \mathcal{\tilde{G}}_{p_{1}p_{2}p_{3}p_{4}}^{C\text{,BCS}}\left( i\omega
_{p}\right)   \notag \\
& =-\left( \frac{2\pi v_{F}^{2}}{T_{K}}\right) ^{2}8\int \frac{dq_{1}dq_{3}}{%
\left( 2\pi \right) ^{2}}\frac{1}{\beta }\sum_{\omega _{n_{1}}}\frac{1}{%
\beta }\sum_{\omega _{n_{3}}}\frac{1}{\beta }\sum_{\omega
_{n_{5}}}g_{p_{1}}\left( i\omega _{n_{1}}\right) g_{p_{2}}\left( i\omega
_{n_{1}}+i\omega _{p}\right)   \notag \\
& \times g_{p_{3}}\left( i\omega _{n_{3}}\right) g_{p_{4}}\left( i\omega
_{n_{3}}-i\omega _{p}\right) g_{q_{1}}\left( i\omega _{n_{5}}\right)
g_{q_{3}}\left( i\omega _{n_{1}}+i\omega _{n_{3}}-i\omega _{n_{5}}\right) 
\text{;}
\end{align}%
the $\omega _{n_{5}}$ summation is standard, whereas the $\omega _{n_{1}}$
and $\omega _{n_{3}}$ summations require the following identities:

\begin{subequations}
\begin{align}
& \frac{1}{\beta }\sum_{\omega _{n_{3}}}\frac{1}{i\omega _{n_{3}}-\epsilon
_{p_{3}}}\frac{1}{i\omega _{n_{3}}-i\omega _{p}-\epsilon _{p_{4}}}\frac{1}{%
i\omega _{n_{1}}+i\omega _{n_{3}}-\epsilon _{q_{1}}-\epsilon _{q_{3}}} 
\notag \\
& =f\left( \epsilon _{p_{3}}\right) \frac{1}{\epsilon _{p_{3}}-i\omega
_{p}-\epsilon _{p_{4}}}\frac{1}{i\omega _{n_{1}}+\epsilon _{p_{3}}-\epsilon
_{q_{1}}-\epsilon _{q_{3}}}+f\left( \epsilon _{p_{4}}\right) \frac{1}{%
i\omega _{p}+\epsilon _{p_{4}}-\epsilon _{p_{3}}}\frac{1}{i\omega
_{n_{1}}+i\omega _{p}+\epsilon _{p_{4}}-\epsilon _{q_{1}}-\epsilon _{q_{3}}}
\notag \\
& -f_{B}\left( \epsilon _{q_{1}}+\epsilon _{q_{3}}\right) \frac{1}{\epsilon
_{q_{1}}+\epsilon _{q_{3}}-i\omega _{n_{1}}-\epsilon _{p_{3}}}\frac{1}{%
\epsilon _{q_{1}}+\epsilon _{q_{3}}-i\omega _{n_{1}}-i\omega _{p}-\epsilon
_{p_{4}}}\text{,}
\end{align}%
and

\begin{align}
& -\frac{1}{\beta }\sum_{\omega _{n_{1}}}\frac{1}{i\omega _{n_{1}}-\epsilon
_{p_{1}}}\frac{1}{i\omega _{n_{1}}+i\omega _{p}-\epsilon _{p_{2}}}\frac{1}{%
i\omega _{n_{1}}+i\omega _{p}+\epsilon _{p_{4}}-\epsilon _{q_{1}}-\epsilon
_{q_{3}}}  \notag \\
& =\int_{-\infty }^{\infty }\frac{d\epsilon }{2\pi i}f\left( \epsilon
\right) \frac{1}{\epsilon -i\omega _{p}-\epsilon _{p_{1}}}\left( \frac{1}{%
\epsilon ^{+}-\epsilon _{p_{2}}}\frac{1}{\epsilon ^{+}+\epsilon
_{p_{4}}-\epsilon _{q_{1}}-\epsilon _{q_{3}}}\right.   \notag \\
& \left. -\frac{1}{\epsilon ^{-}-\epsilon _{p_{2}}}\frac{1}{\epsilon
^{-}+\epsilon _{p_{4}}-\epsilon _{q_{1}}-\epsilon _{q_{3}}}\right) -f\left(
\epsilon _{p_{1}}\right) \frac{1}{\epsilon _{p_{1}}+i\omega _{p}-\epsilon
_{p_{2}}}\frac{1}{\epsilon _{p_{1}}+i\omega _{p}+\epsilon _{p_{4}}-\epsilon
_{q_{1}}-\epsilon _{q_{3}}}\text{,}
\end{align}%
where $\epsilon ^{\pm }\equiv \epsilon \pm i0^{+}$. The second identity can
be derived by allowing the complex plane contour to wrap around the line $%
\operatorname{Im}z=\omega _{p}$.\cite{mahan2000many}

After applying the identities above, performing analytic continuation $%
i\omega _{p}\rightarrow \Omega ^{+}$, and performing all $p$ integrals that
are approachable by the contour method in Eq.~(\ref{GprCimagFL}), we find

\end{subequations}
\begin{align}
& G_{jj^{\prime }}^{\prime C\text{,BCS}}\left( \Omega \right)   \notag \\
& =-\left( \frac{2\pi v_{F}^{2}}{T_{K}}\right) ^{2}8\sum_{j_{1}j_{2}}\mathbb{%
U}_{1,j_{1}}\mathbb{U}_{1,j_{2}}^{\ast }\mathbb{U}_{1,j^{\prime }}\mathbb{U}%
_{1,j^{\prime }}^{\ast }S_{jj_{1}}^{\ast }S_{jj_{2}}\frac{1}{\Omega }\int 
\frac{dq_{1}dq_{3}}{\left( 2\pi \right) ^{2}}\left[ f\left( -\epsilon
_{q_{1}}\right) -f\left( \epsilon _{q_{3}}\right) \right]   \notag \\
& \times \left\{ \int \frac{dp_{1}dp_{4}}{\left( 2\pi \right) ^{2}}\left[
f\left( \epsilon _{p_{4}}\right) +f_{B}\left( \epsilon _{q_{1}}+\epsilon
_{q_{3}}\right) \right] \frac{f\left( \epsilon _{p_{1}}+\Omega \right)
-f\left( \epsilon _{p_{1}}\right) }{\Omega }\frac{1}{\epsilon
_{p_{1}}+\Omega ^{+}+\epsilon _{p_{4}}-\epsilon _{q_{1}}-\epsilon _{q_{3}}}%
\right.   \notag \\
& \left. -\int \frac{dp_{2}dp_{3}}{\left( 2\pi \right) ^{2}}\left[ f\left(
\epsilon _{p_{3}}\right) +f_{B}\left( \epsilon _{q_{1}}+\epsilon
_{q_{3}}\right) \right] \frac{f\left( \epsilon _{p_{2}}-\Omega \right)
-f\left( \epsilon _{p_{2}}\right) }{-\Omega }\frac{1}{\epsilon
_{p_{2}}-\Omega ^{+}+\epsilon _{p_{3}}-\epsilon _{q_{1}}-\epsilon _{q_{3}}}%
\right\} \text{.}
\end{align}%
In the DC limit, the principal value parts of the integrands cancel to $%
O\left( 1/\Omega \right) $, while the $\delta $-function parts remain:

\begin{align}
& G_{jj^{\prime }}^{\prime C\text{,BCS}}\left( \Omega \right) =-\frac{\left(
2\pi \right) ^{2}}{T_{K}^{2}}8\sum_{j_{1}j_{2}}\mathbb{U}_{1,j_{1}}\mathbb{U}%
_{1,j_{2}}^{\ast }\mathbb{U}_{1,j^{\prime }}\mathbb{U}_{1,j^{\prime }}^{\ast
}S_{jj_{1}}^{\ast }S_{jj_{2}}\frac{1}{\Omega }\frac{2i\pi }{\left( 2\pi
\right) ^{4}}\int d\epsilon _{p_{1}}\left[ -f^{\prime }\left( \epsilon
_{p_{1}}\right) \right]   \notag \\
& \times \int d\epsilon _{p_{4}}d\epsilon _{q_{1}}d\epsilon _{q_{3}}\left[
f\left( \epsilon _{p_{4}}\right) +f_{B}\left( \epsilon _{q_{1}}+\epsilon
_{q_{3}}\right) \right] \left[ f\left( -\epsilon _{q_{1}}\right) -f\left(
\epsilon _{q_{3}}\right) \right] \delta \left( \epsilon _{p_{1}}+\epsilon
_{p_{4}}-\epsilon _{q_{1}}-\epsilon _{q_{3}}\right) +O\left( 1\right)  
\notag \\
& =-\frac{4}{T_{K}^{2}}\sum_{j_{1}j_{2}}\mathbb{U}_{1,j_{1}}\mathbb{U}%
_{1,j_{2}}^{\ast }\mathbb{U}_{1,j^{\prime }}\mathbb{U}_{1,j^{\prime }}^{\ast
}S_{jj_{1}}^{\ast }S_{jj_{2}}\frac{1}{\Omega }\frac{i}{2\pi }\int d\epsilon
_{p_{1}}\left[ -f^{\prime }\left( \epsilon _{p_{1}}\right) \right] \left(
\pi ^{2}T^{2}+\epsilon _{p_{1}}^{2}\right) +O\left( 1\right) \text{.}
\label{GprCFLBCS}
\end{align}%
In the second step we have again invoked Eq.~(\ref{tripFermi}).

Each of the ZS and ZS' contributions ends up being the opposite of the BCS
contribution,

\begin{equation}
G_{jj^{\prime }}^{\prime C\text{,ZS}}\left( \Omega \right) =G_{jj^{\prime
}}^{\prime C\text{,ZS'}}\left( \Omega \right) =-G_{jj^{\prime }}^{\prime C%
\text{,BCS}}\left( \Omega \right) \text{.}
\end{equation}%
therefore, using Eq.~(\ref{bbUmatrow1}) for the $\mathbb{U}$ matrix
elements, we can express the total connected contribution to the conductance
to $O\left( 1/T_{K}^{2}\right) $ as

\begin{equation}
G_{jj^{\prime }}^{C}=-\frac{e^{2}}{h}\lim_{\Omega \rightarrow 0}\left( 2\pi
i\Omega \right) G_{jj^{\prime }}^{\prime C\text{,BCS}}\left( \Omega \right) =%
\frac{2e^{2}}{h}\int d\omega \left[ -f^{\prime }\left( \omega \right) \right]
\mathcal{T}_{jj^{\prime }}^{C}\left( \omega \right) \text{,}
\end{equation}%
where

\begin{equation}
\mathcal{T}_{jj^{\prime }}^{C}\left( \omega \right) =-\frac{2}{V_{k_{F}}^{4}}%
\left[ S\left( k_{F}\right) \Gamma ^{\dag }\left( k_{F}\right) \lambda
\Gamma \left( k_{F}\right) S^{\dag }\left( k_{F}\right) \right] _{jj}\left[
\Gamma ^{\dag }\left( k_{F}\right) \lambda \Gamma \left( k_{F}\right) \right]
_{j^{\prime }j^{\prime }}\frac{\pi ^{2}T^{2}+\omega ^{2}}{T_{K}^{2}}\text{.}
\label{TCFL}
\end{equation}%
We have reintroduced the coupling matrix Eq.~(\ref{couplingmatlambda}). The $%
\omega $ integral can be done explicitly:

\begin{equation}
G_{jj^{\prime }}^{C}=-\frac{2e^{2}}{h}\frac{8}{3V_{k_{F}}^{4}}\left\vert %
\left[ S\left( k_{F}\right) \Gamma ^{\dag }\left( k_{F}\right) \lambda
\Gamma \left( k_{F}\right) \right] _{jj^{\prime }}\right\vert ^{2}\left( 
\frac{\pi T}{T_{K}}\right) ^{2}\text{,}  \label{GCFL}
\end{equation}%
i.e. the lowest order connected contribution to the conductance is $O\left(
T^{2}/T_{K}^{2}\right) $, characteristic of a Fermi liquid.

Eq.~(\ref{TCFL}) is in explicit agreement with Eq.~(\ref{connelimT}). We can
also check its consistency with the Eq.~(\ref{TDrowsum}) and single-particle
T-matrix inelasticity. Recall that, by virtue of Eq.~(\ref{connelimG}), we
should have the following approximate identity for $\omega \approx 0$ in the
Fermi liquid regime:

\begin{equation}
\mathcal{T}_{jj^{\prime }}^{C}\left( \omega \right) =-\frac{1}{V_{k_{F}}^{2}}%
\left[ S\left( k_{F}\right) \Gamma ^{\dag }\left( k_{F}\right) \lambda
\Gamma \left( k_{F}\right) S^{\dag }\left( k_{F}\right) \right]
_{jj}\sum_{j^{\prime \prime }}\mathcal{T}_{j^{\prime \prime }j^{\prime
}}^{D}\left( \omega \right) \text{.}  \label{TCFL1}
\end{equation}%
On the other hand, Eqs.~(\ref{TvsTtilde}) and (\ref{TtildeAL}) yield for the
on-shell T-matrix

\begin{equation}
\operatorname{Im}\left[ -\pi \nu \text{T}\left( \omega \right) \right] -\left\vert
-\pi \nu \text{T}\left( \omega \right) \right\vert ^{2}=\frac{\pi
^{2}T^{2}+\omega ^{2}}{2T_{K}^{2}}\text{;}  \label{Tinel}
\end{equation}%
therefore, plugging Eq.~(\ref{Tinel}) into Eq.~(\ref{TDrowsum}), we find
that for $\omega \approx 0$,

\begin{equation}
\sum_{j^{\prime \prime }}\mathcal{T}_{j^{\prime \prime }j^{\prime
}}^{D}\left( \omega \right) =\frac{\pi ^{2}T^{2}+\omega ^{2}}{2T_{K}^{2}}%
\frac{4}{V_{k_{F}}^{2}}\left[ \Gamma ^{\dag }\left( k_{F}\right) \lambda
\Gamma \left( k_{F}\right) \right] _{j^{\prime }j^{\prime }}\text{.}
\label{TDrowsum1}
\end{equation}%
Eqs.~(\ref{TCFL1}) and (\ref{TDrowsum1}) are fully consistent with Eq.~(\ref%
{TCFL}).

\section{Non-interacting Schroedinger equations for the open long ring\label%
{sec:appopen}}

In this appendix we sketch how to obtain Eq.~(\ref{longringBGamma}) which
expresses, in terms of incident amplitudes $A$, the scattered amplitudes in
the main leads $B_{1}$, $B_{2}$ and the coupling site wave functions $\Gamma
_{1}$, $\Gamma _{2}$. Because the Schroedinger equation is linear and all
incident amplitudes are independent, we can let all but one of the incident
amplitudes be zero at a time, and obtain the full solution by means of
linear superposition.

When the incident amplitudes from the side leads are all zero $A_{n}^{\left(
\alpha \right) }=0$, according to Eq.~(\ref{lossteeth}) the wave function at
wave vector $k$ takes the form

\begin{equation}
\left\{ 
\begin{array}{c}
A_{j}e^{-ikn}+B_{j}e^{ikn}\left( \text{main lead }j=1,2\text{, }%
n=0,1,2,\cdots \right) \\ 
D_{L}^{\left( L\right) }\eta _{1}^{n}+D_{R}^{\left( L\right) }\eta
_{1}^{-n}\left( \text{left QD arm, }n=0,1,\cdots ,d_{L}-1\right) \\ 
D_{L}^{\left( ref\right) }\eta _{1}^{n}+D_{R}^{\left( ref\right) }\eta
_{1}^{-n}\left( \text{reference arm, }n=1,2,\cdots ,d_{ref}\right) \\ 
D_{L}^{\left( R\right) }\eta _{1}^{n}+D_{R}^{\left( R\right) }\eta
_{1}^{-n}\left( \text{right QD arm, }n=0,1,\cdots ,d_{R}-1\right)%
\end{array}%
\right. \text{.}  \label{longringmain}
\end{equation}%
To characterize the two Y-junctions in the AB ring, it is convenient to
introduce the auxiliary objects $S_{L}^{\prime }$ and $S_{R}^{\prime }$:

\begin{equation}
\left( 
\begin{array}{c}
B_{1} \\ 
D_{R}^{\left( L\right) }\eta _{1}^{-d_{L}+1} \\ 
D_{L}^{\left( ref\right) }\eta _{1}%
\end{array}%
\right) =S_{L}^{\prime }\left( 
\begin{array}{c}
A_{1} \\ 
D_{L}^{\left( L\right) }\eta _{1}^{d_{L}-1} \\ 
D_{R}^{\left( ref\right) }\eta _{1}^{-1}%
\end{array}%
\right) \text{,}  \label{longringSprL}
\end{equation}

\begin{equation}
\left( 
\begin{array}{c}
B_{2} \\ 
D_{R}^{\left( R\right) }\eta _{1}^{-d_{R}+1} \\ 
D_{R}^{\left( ref\right) }\eta _{1}^{-d_{ref}}%
\end{array}%
\right) =S_{R}^{\prime }\left( 
\begin{array}{c}
A_{2} \\ 
D_{L}^{\left( R\right) }\eta _{1}^{d_{R}-1} \\ 
D_{L}^{\left( ref\right) }\eta _{1}^{d_{ref}}%
\end{array}%
\right) \text{.}  \label{longringSprR}
\end{equation}%
The physical meaning of $S_{L}^{\prime }$ and $S_{R}^{\prime }$ is discussed
below Eq.~(\ref{longringBGamma}). For our model we can find $S_{L}^{\prime }$
explicitly by solving the Schroedinger equations,

\begin{subequations}
\label{longringSprL1}
\begin{equation}
t_{JL}^{L}\phi _{L}=t\left( A_{1}e^{ik}+B_{1}e^{-ik}\right) \text{,}
\end{equation}

\begin{equation}
t_{JQ}^{L}\phi _{L}=t\left( D_{L}^{\left( L\right) }\eta
_{1}^{d_{L}}+D_{R}^{\left( L\right) }\eta _{1}^{-d_{L}}\right) \text{,}
\end{equation}

\begin{equation}
t_{JR}^{L}\phi _{L}=t\left( D_{L}^{\left( ref\right) }+D_{R}^{\left(
ref\right) }\right) \text{,}
\end{equation}

\begin{align}
\left( -2t\cos k\right) \phi _{L}& =-t_{JL}^{L}\left( A_{1}+B_{1}\right)
-t_{JQ}^{L}\left( D_{L}^{\left( L\right) }\eta _{1}^{d_{L}-1}+D_{R}^{\left(
L\right) }\eta _{1}^{-d_{L}+1}\right)   \notag \\
& -t_{JR}^{L}\left( D_{L}^{\left( ref\right) }\eta _{1}+D_{R}^{\left(
ref\right) }\eta _{1}^{-1}\right) \text{,}
\end{align}%
where $\phi _{L}$ is the wave function on the central site of the left
Y-junction. We can solve for $S_{R}^{\prime }$ in a similar fashion.

For given $A_{1}$ and $A_{2}$, Eq.~(\ref{longringmain}) has $8$ unknowns.
Now that $S_{L}^{\prime }$ and $S_{R}^{\prime }$ are known, Eqs.~(\ref%
{longringSprL}) and (\ref{longringSprR}) provide us with $6$ equations. The
remaining two equations are the open boundary conditions at the ends of the
two QD arms, appropriate when the QD is decoupled:

\end{subequations}
\begin{equation}
D_{L}^{\left( L\right) }\eta _{1}^{-1}+D_{R}^{\left( L\right) }\eta _{1}=0%
\text{,}  \label{longringdissopenL}
\end{equation}

\begin{equation}
D_{L}^{\left( R\right) }\eta _{1}^{-1}+D_{R}^{\left( R\right) }\eta _{1}=0%
\text{.}  \label{longringdissopenR}
\end{equation}%
It is straightforward to solve the closed set of equations.

On the other hand, when we let one of the incident amplitudes in the side
leads be nonzero, there are two additional amplitudes in the wave function.
For instance, if $A_{m}^{\left( L\right) }\neq 0$ for a given $m$, we need
to effectuate the following changes to the wave function on the left QD arm
in Eq.~(\ref{longringmain}):

\begin{equation}
\left\{ 
\begin{array}{c}
D_{L}^{\left( L1\right) }\eta _{1}^{n}+D_{R}^{\left( L1\right) }\eta
_{1}^{-n}\left( \text{left QD arm, }n=0,1,\cdots ,m\right) \\ 
D_{L}^{\left( L2\right) }\eta _{1}^{n}+D_{R}^{\left( L2\right) }\eta
_{1}^{-n}\left( \text{left QD arm, }n=m,m+1,\cdots ,d_{L}-1\right)%
\end{array}%
\right. \text{;}
\end{equation}%
$D_{L,R}^{\left( L\right) }$ should be replaced by $D_{L,R}^{\left(
L1\right) }$ in Eq.~(\ref{longringdissopenL}) and by $D_{L,R}^{\left(
L2\right) }$ in Eq.~(\ref{longringSprL}). Furthermore, we should have two
boundary conditions at site $m$:

\begin{equation}
D_{L}^{\left( L1\right) }\eta _{1}^{m}+D_{R}^{\left( L1\right) }\eta
_{1}^{-m}=D_{L}^{\left( L2\right) }\eta _{1}^{m}+D_{R}^{\left( L2\right)
}\eta _{1}^{-m}\text{,}
\end{equation}

\begin{align}
& -\left( \eta _{1}+\frac{1}{\eta _{1}}\right) \left( D_{L}^{\left(
L1\right) }\eta _{1}^{m}+D_{R}^{\left( L1\right) }\eta _{1}^{-m}\right)  
\notag \\
& =-\left( D_{L}^{\left( L1\right) }\eta _{1}^{m-1}+D_{R}^{\left( L1\right)
}\eta _{1}^{-m+1}\right) -\left( D_{L}^{\left( L2\right) }\eta
_{1}^{m+1}+D_{R}^{\left( L2\right) }\eta _{1}^{-\left( m+1\right) }\right)
+e^{ik}\left( 2i\sin k\right) \frac{t_{x}}{t}A_{m}^{\left( L\right) }\text{,}
\end{align}%
thus closing the set of equations. The last equation is none other than Eq.~(%
\ref{teethScheq}).

Solving all four different sets of equations in the limit of $d_{L}\sim
d_{R}\sim d_{ref}/2\gg 1$ and $\left\vert \eta _{1}\right\vert ^{d_{L}}\ll 1$%
\ and combining the solutions, we promptly arrive at Eq.~(\ref%
{longringBGamma}).

\bibliography{LongABK}

\providecommand{\noopsort}[1]{}\providecommand{\singleletter}[1]{#1}%
\begin{thebibliography}{66}%
\makeatletter
\providecommand \@ifxundefined [1]{%
 \@ifx{#1\undefined}
}%
\providecommand \@ifnum [1]{%
 \ifnum #1\expandafter \@firstoftwo
 \else \expandafter \@secondoftwo
 \fi
}%
\providecommand \@ifx [1]{%
 \ifx #1\expandafter \@firstoftwo
 \else \expandafter \@secondoftwo
 \fi
}%
\providecommand \natexlab [1]{#1}%
\providecommand \enquote  [1]{``#1''}%
\providecommand \bibnamefont  [1]{#1}%
\providecommand \bibfnamefont [1]{#1}%
\providecommand \citenamefont [1]{#1}%
\providecommand \href@noop [0]{\@secondoftwo}%
\providecommand \href [0]{\begingroup \@sanitize@url \@href}%
\providecommand \@href[1]{\@@startlink{#1}\@@href}%
\providecommand \@@href[1]{\endgroup#1\@@endlink}%
\providecommand \@sanitize@url [0]{\catcode `\\12\catcode `\$12\catcode
  `\&12\catcode `\#12\catcode `\^12\catcode `\_12\catcode `\%12\relax}%
\providecommand \@@startlink[1]{}%
\providecommand \@@endlink[0]{}%
\providecommand \url  [0]{\begingroup\@sanitize@url \@url }%
\providecommand \@url [1]{\endgroup\@href {#1}{\urlprefix }}%
\providecommand \urlprefix  [0]{URL }%
\providecommand \Eprint [0]{\href }%
\providecommand \doibase [0]{http://dx.doi.org/}%
\providecommand \selectlanguage [0]{\@gobble}%
\providecommand \bibinfo  [0]{\@secondoftwo}%
\providecommand \bibfield  [0]{\@secondoftwo}%
\providecommand \translation [1]{[#1]}%
\providecommand \BibitemOpen [0]{}%
\providecommand \bibitemStop [0]{}%
\providecommand \bibitemNoStop [0]{.\EOS\space}%
\providecommand \EOS [0]{\spacefactor3000\relax}%
\providecommand \BibitemShut  [1]{\csname bibitem#1\endcsname}%
\let\auto@bib@innerbib\@empty
\bibitem [{\citenamefont {Kondo}(1964)}]{ProgTheorPhys.32.37}%
  \BibitemOpen
  \bibfield  {author} {\bibinfo {author} {\bibfnamefont {J.}~\bibnamefont
  {Kondo}},\ }\href {\doibase 10.1143/PTP.32.37} {\bibfield  {journal}
  {\bibinfo  {journal} {Prog. Theor. Phys.}\ }\textbf {\bibinfo {volume}
  {32}},\ \bibinfo {pages} {37} (\bibinfo {year} {1964})}\BibitemShut {NoStop}%
\bibitem [{\citenamefont {Hewson}(1997)}]{hewson1997kondo}%
  \BibitemOpen
  \bibfield  {author} {\bibinfo {author} {\bibfnamefont {A.}~\bibnamefont
  {Hewson}},\ }\href {https://books.google.ca/books?id=fPzgHneNFDAC} {\emph
  {\bibinfo {title} {The Kondo Problem to Heavy Fermions}}},\ Cambridge Studies
  in Magnetism\ (\bibinfo  {publisher} {Cambridge University Press},\ \bibinfo
  {year} {1997})\BibitemShut {NoStop}%
\bibitem [{\citenamefont {Anderson}(1970)}]{JPhysC.3.2346}%
  \BibitemOpen
  \bibfield  {author} {\bibinfo {author} {\bibfnamefont {P.~W.}\ \bibnamefont
  {Anderson}},\ }\href {http://stacks.iop.org/0022-3719/3/i=12/a=008}
  {\bibfield  {journal} {\bibinfo  {journal} {J. Phys. C}\ }\textbf {\bibinfo
  {volume} {3}},\ \bibinfo {pages} {2436} (\bibinfo {year} {1970})}\BibitemShut
  {NoStop}%
\bibitem [{\citenamefont {Wilson}(1975)}]{RevModPhys.47.773}%
  \BibitemOpen
  \bibfield  {author} {\bibinfo {author} {\bibfnamefont {K.~G.}\ \bibnamefont
  {Wilson}},\ }\href {\doibase 10.1103/RevModPhys.47.773} {\bibfield  {journal}
  {\bibinfo  {journal} {Rev. Mod. Phys.}\ }\textbf {\bibinfo {volume} {47}},\
  \bibinfo {pages} {773} (\bibinfo {year} {1975})}\BibitemShut {NoStop}%
\bibitem [{\citenamefont {Nozi{\`e}res}(1974)}]{JLowTempPhys.17.31}%
  \BibitemOpen
  \bibfield  {author} {\bibinfo {author} {\bibfnamefont {P.}~\bibnamefont
  {Nozi{\`e}res}},\ }\href {\doibase 10.1007/BF00654541} {\bibfield  {journal}
  {\bibinfo  {journal} {J. Low Temp. Phys.}\ }\textbf {\bibinfo {volume}
  {17}},\ \bibinfo {pages} {31} (\bibinfo {year} {1974})}\BibitemShut {NoStop}%
\bibitem [{\citenamefont {Zar\'and}\ \emph {et~al.}(2004)\citenamefont
  {Zar\'and}, \citenamefont {Borda}, \citenamefont {von Delft},\ and\
  \citenamefont {Andrei}}]{PhysRevLett.93.107204}%
  \BibitemOpen
  \bibfield  {author} {\bibinfo {author} {\bibfnamefont {G.}~\bibnamefont
  {Zar\'and}}, \bibinfo {author} {\bibfnamefont {L.}~\bibnamefont {Borda}},
  \bibinfo {author} {\bibfnamefont {J.}~\bibnamefont {von Delft}}, \ and\
  \bibinfo {author} {\bibfnamefont {N.}~\bibnamefont {Andrei}},\ }\href
  {\doibase 10.1103/PhysRevLett.93.107204} {\bibfield  {journal} {\bibinfo
  {journal} {Phys. Rev. Lett.}\ }\textbf {\bibinfo {volume} {93}},\ \bibinfo
  {pages} {107204} (\bibinfo {year} {2004})}\BibitemShut {NoStop}%
\bibitem [{\citenamefont {Borda}\ \emph {et~al.}(2007)\citenamefont {Borda},
  \citenamefont {Fritz}, \citenamefont {Andrei},\ and\ \citenamefont
  {Zar\'and}}]{PhysRevB.75.235112}%
  \BibitemOpen
  \bibfield  {author} {\bibinfo {author} {\bibfnamefont {L.}~\bibnamefont
  {Borda}}, \bibinfo {author} {\bibfnamefont {L.}~\bibnamefont {Fritz}},
  \bibinfo {author} {\bibfnamefont {N.}~\bibnamefont {Andrei}}, \ and\ \bibinfo
  {author} {\bibfnamefont {G.}~\bibnamefont {Zar\'and}},\ }\href {\doibase
  10.1103/PhysRevB.75.235112} {\bibfield  {journal} {\bibinfo  {journal} {Phys.
  Rev. B}\ }\textbf {\bibinfo {volume} {75}},\ \bibinfo {pages} {235112}
  (\bibinfo {year} {2007})}\BibitemShut {NoStop}%
\bibitem [{\citenamefont {Micklitz}\ \emph {et~al.}(2006)\citenamefont
  {Micklitz}, \citenamefont {Altland}, \citenamefont {Costi},\ and\
  \citenamefont {Rosch}}]{PhysRevLett.96.226601}%
  \BibitemOpen
  \bibfield  {author} {\bibinfo {author} {\bibfnamefont {T.}~\bibnamefont
  {Micklitz}}, \bibinfo {author} {\bibfnamefont {A.}~\bibnamefont {Altland}},
  \bibinfo {author} {\bibfnamefont {T.~A.}\ \bibnamefont {Costi}}, \ and\
  \bibinfo {author} {\bibfnamefont {A.}~\bibnamefont {Rosch}},\ }\href
  {\doibase 10.1103/PhysRevLett.96.226601} {\bibfield  {journal} {\bibinfo
  {journal} {Phys. Rev. Lett.}\ }\textbf {\bibinfo {volume} {96}},\ \bibinfo
  {pages} {226601} (\bibinfo {year} {2006})}\BibitemShut {NoStop}%
\bibitem [{\citenamefont {Pierre}\ and\ \citenamefont
  {Birge}(2002)}]{PhysRevLett.89.206804}%
  \BibitemOpen
  \bibfield  {author} {\bibinfo {author} {\bibfnamefont {F.}~\bibnamefont
  {Pierre}}\ and\ \bibinfo {author} {\bibfnamefont {N.~O.}\ \bibnamefont
  {Birge}},\ }\href {\doibase 10.1103/PhysRevLett.89.206804} {\bibfield
  {journal} {\bibinfo  {journal} {Phys. Rev. Lett.}\ }\textbf {\bibinfo
  {volume} {89}},\ \bibinfo {pages} {206804} (\bibinfo {year}
  {2002})}\BibitemShut {NoStop}%
\bibitem [{\citenamefont {Goldhaber-Gordon}\ \emph {et~al.}(1998)\citenamefont
  {Goldhaber-Gordon}, \citenamefont {Shtrikman}, \citenamefont {Mahalu},
  \citenamefont {Abusch-Magder}, \citenamefont {Meirav},\ and\ \citenamefont
  {Kastner}}]{Nature.391.156}%
  \BibitemOpen
  \bibfield  {author} {\bibinfo {author} {\bibfnamefont {D.}~\bibnamefont
  {Goldhaber-Gordon}}, \bibinfo {author} {\bibfnamefont {H.}~\bibnamefont
  {Shtrikman}}, \bibinfo {author} {\bibfnamefont {D.}~\bibnamefont {Mahalu}},
  \bibinfo {author} {\bibfnamefont {D.}~\bibnamefont {Abusch-Magder}}, \bibinfo
  {author} {\bibfnamefont {U.}~\bibnamefont {Meirav}}, \ and\ \bibinfo {author}
  {\bibfnamefont {M.~A.}\ \bibnamefont {Kastner}},\ }\href {\doibase
  10.1038/34373} {\bibfield  {journal} {\bibinfo  {journal} {Nature}\ }\textbf
  {\bibinfo {volume} {391}},\ \bibinfo {pages} {156} (\bibinfo {year}
  {1998})}\BibitemShut {NoStop}%
\bibitem [{\citenamefont {Cronenwett}\ \emph {et~al.}(1998)\citenamefont
  {Cronenwett}, \citenamefont {Oosterkamp},\ and\ \citenamefont
  {Kouwenhoven}}]{Science.281.540}%
  \BibitemOpen
  \bibfield  {author} {\bibinfo {author} {\bibfnamefont {S.~M.}\ \bibnamefont
  {Cronenwett}}, \bibinfo {author} {\bibfnamefont {T.~H.}\ \bibnamefont
  {Oosterkamp}}, \ and\ \bibinfo {author} {\bibfnamefont {L.~P.}\ \bibnamefont
  {Kouwenhoven}},\ }\href {\doibase 10.1126/science.281.5376.540} {\bibfield
  {journal} {\bibinfo  {journal} {Science}\ }\textbf {\bibinfo {volume}
  {281}},\ \bibinfo {pages} {540} (\bibinfo {year} {1998})}\BibitemShut
  {NoStop}%
\bibitem [{\citenamefont {Simmel}\ \emph {et~al.}(1999)\citenamefont {Simmel},
  \citenamefont {Blick}, \citenamefont {Kotthaus}, \citenamefont
  {Wegscheider},\ and\ \citenamefont {Bichler}}]{PhysRevLett.83.804}%
  \BibitemOpen
  \bibfield  {author} {\bibinfo {author} {\bibfnamefont {F.}~\bibnamefont
  {Simmel}}, \bibinfo {author} {\bibfnamefont {R.~H.}\ \bibnamefont {Blick}},
  \bibinfo {author} {\bibfnamefont {J.~P.}\ \bibnamefont {Kotthaus}}, \bibinfo
  {author} {\bibfnamefont {W.}~\bibnamefont {Wegscheider}}, \ and\ \bibinfo
  {author} {\bibfnamefont {M.}~\bibnamefont {Bichler}},\ }\href {\doibase
  10.1103/PhysRevLett.83.804} {\bibfield  {journal} {\bibinfo  {journal} {Phys.
  Rev. Lett.}\ }\textbf {\bibinfo {volume} {83}},\ \bibinfo {pages} {804}
  (\bibinfo {year} {1999})}\BibitemShut {NoStop}%
\bibitem [{\citenamefont {van~der Wiel}\ \emph {et~al.}(2000)\citenamefont
  {van~der Wiel}, \citenamefont {Franceschi}, \citenamefont {Fujisawa},
  \citenamefont {Elzerman}, \citenamefont {Tarucha},\ and\ \citenamefont
  {Kouwenhoven}}]{Science.289.2105}%
  \BibitemOpen
  \bibfield  {author} {\bibinfo {author} {\bibfnamefont {W.~G.}\ \bibnamefont
  {van~der Wiel}}, \bibinfo {author} {\bibfnamefont {S.~D.}\ \bibnamefont
  {Franceschi}}, \bibinfo {author} {\bibfnamefont {T.}~\bibnamefont
  {Fujisawa}}, \bibinfo {author} {\bibfnamefont {J.~M.}\ \bibnamefont
  {Elzerman}}, \bibinfo {author} {\bibfnamefont {S.}~\bibnamefont {Tarucha}}, \
  and\ \bibinfo {author} {\bibfnamefont {L.~P.}\ \bibnamefont {Kouwenhoven}},\
  }\href {\doibase 10.1126/science.289.5487.2105} {\bibfield  {journal}
  {\bibinfo  {journal} {Science}\ }\textbf {\bibinfo {volume} {289}},\ \bibinfo
  {pages} {2105} (\bibinfo {year} {2000})}\BibitemShut {NoStop}%
\bibitem [{\citenamefont {Pustilnik}\ and\ \citenamefont
  {Glazman}(2004)}]{JPhysCondensMatter.16.R513}%
  \BibitemOpen
  \bibfield  {author} {\bibinfo {author} {\bibfnamefont {M.}~\bibnamefont
  {Pustilnik}}\ and\ \bibinfo {author} {\bibfnamefont {L.}~\bibnamefont
  {Glazman}},\ }\href {http://stacks.iop.org/0953-8984/16/i=16/a=R01}
  {\bibfield  {journal} {\bibinfo  {journal} {J. Phys. Condens. Matter}\
  }\textbf {\bibinfo {volume} {16}},\ \bibinfo {pages} {R513} (\bibinfo {year}
  {2004})}\BibitemShut {NoStop}%
\bibitem [{\citenamefont {Glazman}\ and\ \citenamefont
  {Pustilnik}(2005)}]{condmat0501007}%
  \BibitemOpen
  \bibfield  {author} {\bibinfo {author} {\bibfnamefont {L.~I.}\ \bibnamefont
  {Glazman}}\ and\ \bibinfo {author} {\bibfnamefont {M.}~\bibnamefont
  {Pustilnik}},\ }\enquote {\bibinfo {title} {Nanophysics: Coherence and
  transport},}\ \ (\bibinfo  {publisher} {Elsevier},\ \bibinfo {year} {2005})\
  pp.\ \bibinfo {pages} {427--478}\BibitemShut {NoStop}%
\bibitem [{\citenamefont {Pereira}\ \emph {et~al.}(2008)\citenamefont
  {Pereira}, \citenamefont {Laflorencie}, \citenamefont {Affleck},\ and\
  \citenamefont {Halperin}}]{PhysRevB.77.125327}%
  \BibitemOpen
  \bibfield  {author} {\bibinfo {author} {\bibfnamefont {R.~G.}\ \bibnamefont
  {Pereira}}, \bibinfo {author} {\bibfnamefont {N.}~\bibnamefont
  {Laflorencie}}, \bibinfo {author} {\bibfnamefont {I.}~\bibnamefont
  {Affleck}}, \ and\ \bibinfo {author} {\bibfnamefont {B.~I.}\ \bibnamefont
  {Halperin}},\ }\href {\doibase 10.1103/PhysRevB.77.125327} {\bibfield
  {journal} {\bibinfo  {journal} {Phys. Rev. B}\ }\textbf {\bibinfo {volume}
  {77}},\ \bibinfo {pages} {125327} (\bibinfo {year} {2008})}\BibitemShut
  {NoStop}%
\bibitem [{\citenamefont {Simon}\ and\ \citenamefont
  {Affleck}(2002)}]{PhysRevLett.89.206602}%
  \BibitemOpen
  \bibfield  {author} {\bibinfo {author} {\bibfnamefont {P.}~\bibnamefont
  {Simon}}\ and\ \bibinfo {author} {\bibfnamefont {I.}~\bibnamefont
  {Affleck}},\ }\href {\doibase 10.1103/PhysRevLett.89.206602} {\bibfield
  {journal} {\bibinfo  {journal} {Phys. Rev. Lett.}\ }\textbf {\bibinfo
  {volume} {89}},\ \bibinfo {pages} {206602} (\bibinfo {year}
  {2002})}\BibitemShut {NoStop}%
\bibitem [{\citenamefont {Simon}\ and\ \citenamefont
  {Affleck}(2003)}]{PhysRevB.68.115304}%
  \BibitemOpen
  \bibfield  {author} {\bibinfo {author} {\bibfnamefont {P.}~\bibnamefont
  {Simon}}\ and\ \bibinfo {author} {\bibfnamefont {I.}~\bibnamefont
  {Affleck}},\ }\href {\doibase 10.1103/PhysRevB.68.115304} {\bibfield
  {journal} {\bibinfo  {journal} {Phys. Rev. B}\ }\textbf {\bibinfo {volume}
  {68}},\ \bibinfo {pages} {115304} (\bibinfo {year} {2003})}\BibitemShut
  {NoStop}%
\bibitem [{\citenamefont {Cornaglia}\ and\ \citenamefont
  {Balseiro}(2003)}]{PhysRevLett.90.216801}%
  \BibitemOpen
  \bibfield  {author} {\bibinfo {author} {\bibfnamefont {P.~S.}\ \bibnamefont
  {Cornaglia}}\ and\ \bibinfo {author} {\bibfnamefont {C.~A.}\ \bibnamefont
  {Balseiro}},\ }\href {\doibase 10.1103/PhysRevLett.90.216801} {\bibfield
  {journal} {\bibinfo  {journal} {Phys. Rev. Lett.}\ }\textbf {\bibinfo
  {volume} {90}},\ \bibinfo {pages} {216801} (\bibinfo {year}
  {2003})}\BibitemShut {NoStop}%
\bibitem [{\citenamefont {Park}\ \emph {et~al.}(2013)\citenamefont {Park},
  \citenamefont {Lee}, \citenamefont {Oreg},\ and\ \citenamefont
  {Sim}}]{PhysRevLett.110.246603}%
  \BibitemOpen
  \bibfield  {author} {\bibinfo {author} {\bibfnamefont {J.}~\bibnamefont
  {Park}}, \bibinfo {author} {\bibfnamefont {S.-S.~B.}\ \bibnamefont {Lee}},
  \bibinfo {author} {\bibfnamefont {Y.}~\bibnamefont {Oreg}}, \ and\ \bibinfo
  {author} {\bibfnamefont {H.-S.}\ \bibnamefont {Sim}},\ }\href {\doibase
  10.1103/PhysRevLett.110.246603} {\bibfield  {journal} {\bibinfo  {journal}
  {Phys. Rev. Lett.}\ }\textbf {\bibinfo {volume} {110}},\ \bibinfo {pages}
  {246603} (\bibinfo {year} {2013})}\BibitemShut {NoStop}%
\bibitem [{\citenamefont {Thimm}\ \emph {et~al.}(1999)\citenamefont {Thimm},
  \citenamefont {Kroha},\ and\ \citenamefont {von
  Delft}}]{PhysRevLett.82.2143}%
  \BibitemOpen
  \bibfield  {author} {\bibinfo {author} {\bibfnamefont {W.~B.}\ \bibnamefont
  {Thimm}}, \bibinfo {author} {\bibfnamefont {J.}~\bibnamefont {Kroha}}, \ and\
  \bibinfo {author} {\bibfnamefont {J.}~\bibnamefont {von Delft}},\ }\href
  {\doibase 10.1103/PhysRevLett.82.2143} {\bibfield  {journal} {\bibinfo
  {journal} {Phys. Rev. Lett.}\ }\textbf {\bibinfo {volume} {82}},\ \bibinfo
  {pages} {2143} (\bibinfo {year} {1999})}\BibitemShut {NoStop}%
\bibitem [{\citenamefont {Cornaglia}\ and\ \citenamefont
  {Balseiro}(2002)}]{PhysRevB.66.115303}%
  \BibitemOpen
  \bibfield  {author} {\bibinfo {author} {\bibfnamefont {P.~S.}\ \bibnamefont
  {Cornaglia}}\ and\ \bibinfo {author} {\bibfnamefont {C.~A.}\ \bibnamefont
  {Balseiro}},\ }\href {\doibase 10.1103/PhysRevB.66.115303} {\bibfield
  {journal} {\bibinfo  {journal} {Phys. Rev. B}\ }\textbf {\bibinfo {volume}
  {66}},\ \bibinfo {pages} {115303} (\bibinfo {year} {2002})}\BibitemShut
  {NoStop}%
\bibitem [{\citenamefont {Kaul}\ \emph {et~al.}(2005)\citenamefont {Kaul},
  \citenamefont {Ullmo}, \citenamefont {Chandrasekharan},\ and\ \citenamefont
  {Baranger}}]{EurophysLett.71.973}%
  \BibitemOpen
  \bibfield  {author} {\bibinfo {author} {\bibfnamefont {R.~K.}\ \bibnamefont
  {Kaul}}, \bibinfo {author} {\bibfnamefont {D.}~\bibnamefont {Ullmo}},
  \bibinfo {author} {\bibfnamefont {S.}~\bibnamefont {Chandrasekharan}}, \ and\
  \bibinfo {author} {\bibfnamefont {H.~U.}\ \bibnamefont {Baranger}},\ }\href
  {http://stacks.iop.org/0295-5075/71/i=6/a=973} {\bibfield  {journal}
  {\bibinfo  {journal} {Europhys. Lett.}\ }\textbf {\bibinfo {volume} {71}},\
  \bibinfo {pages} {973} (\bibinfo {year} {2005})}\BibitemShut {NoStop}%
\bibitem [{\citenamefont {Simon}\ \emph {et~al.}(2006)\citenamefont {Simon},
  \citenamefont {Salomez},\ and\ \citenamefont
  {Feinberg}}]{PhysRevB.73.205325}%
  \BibitemOpen
  \bibfield  {author} {\bibinfo {author} {\bibfnamefont {P.}~\bibnamefont
  {Simon}}, \bibinfo {author} {\bibfnamefont {J.}~\bibnamefont {Salomez}}, \
  and\ \bibinfo {author} {\bibfnamefont {D.}~\bibnamefont {Feinberg}},\ }\href
  {\doibase 10.1103/PhysRevB.73.205325} {\bibfield  {journal} {\bibinfo
  {journal} {Phys. Rev. B}\ }\textbf {\bibinfo {volume} {73}},\ \bibinfo
  {pages} {205325} (\bibinfo {year} {2006})}\BibitemShut {NoStop}%
\bibitem [{\citenamefont {Kaul}\ \emph {et~al.}(2006)\citenamefont {Kaul},
  \citenamefont {Zar\'and}, \citenamefont {Chandrasekharan}, \citenamefont
  {Ullmo},\ and\ \citenamefont {Baranger}}]{PhysRevLett.96.176802}%
  \BibitemOpen
  \bibfield  {author} {\bibinfo {author} {\bibfnamefont {R.~K.}\ \bibnamefont
  {Kaul}}, \bibinfo {author} {\bibfnamefont {G.}~\bibnamefont {Zar\'and}},
  \bibinfo {author} {\bibfnamefont {S.}~\bibnamefont {Chandrasekharan}},
  \bibinfo {author} {\bibfnamefont {D.}~\bibnamefont {Ullmo}}, \ and\ \bibinfo
  {author} {\bibfnamefont {H.~U.}\ \bibnamefont {Baranger}},\ }\href {\doibase
  10.1103/PhysRevLett.96.176802} {\bibfield  {journal} {\bibinfo  {journal}
  {Phys. Rev. Lett.}\ }\textbf {\bibinfo {volume} {96}},\ \bibinfo {pages}
  {176802} (\bibinfo {year} {2006})}\BibitemShut {NoStop}%
\bibitem [{\citenamefont {Liu}\ \emph {et~al.}(2012{\natexlab{a}})\citenamefont
  {Liu}, \citenamefont {Burdin}, \citenamefont {Baranger},\ and\ \citenamefont
  {Ullmo}}]{EurophysLett.97.17006}%
  \BibitemOpen
  \bibfield  {author} {\bibinfo {author} {\bibfnamefont {D.~E.}\ \bibnamefont
  {Liu}}, \bibinfo {author} {\bibfnamefont {S.}~\bibnamefont {Burdin}},
  \bibinfo {author} {\bibfnamefont {H.~U.}\ \bibnamefont {Baranger}}, \ and\
  \bibinfo {author} {\bibfnamefont {D.}~\bibnamefont {Ullmo}},\ }\href
  {http://stacks.iop.org/0295-5075/97/i=1/a=17006} {\bibfield  {journal}
  {\bibinfo  {journal} {Europhys. Lett.}\ }\textbf {\bibinfo {volume} {97}},\
  \bibinfo {pages} {17006} (\bibinfo {year} {2012}{\natexlab{a}})}\BibitemShut
  {NoStop}%
\bibitem [{\citenamefont {Liu}\ \emph {et~al.}(2012{\natexlab{b}})\citenamefont
  {Liu}, \citenamefont {Burdin}, \citenamefont {Baranger},\ and\ \citenamefont
  {Ullmo}}]{PhysRevB.85.155455}%
  \BibitemOpen
  \bibfield  {author} {\bibinfo {author} {\bibfnamefont {D.~E.}\ \bibnamefont
  {Liu}}, \bibinfo {author} {\bibfnamefont {S.}~\bibnamefont {Burdin}},
  \bibinfo {author} {\bibfnamefont {H.~U.}\ \bibnamefont {Baranger}}, \ and\
  \bibinfo {author} {\bibfnamefont {D.}~\bibnamefont {Ullmo}},\ }\href
  {\doibase 10.1103/PhysRevB.85.155455} {\bibfield  {journal} {\bibinfo
  {journal} {Phys. Rev. B}\ }\textbf {\bibinfo {volume} {85}},\ \bibinfo
  {pages} {155455} (\bibinfo {year} {2012}{\natexlab{b}})}\BibitemShut
  {NoStop}%
\bibitem [{\citenamefont {Simon}\ \emph {et~al.}(2005)\citenamefont {Simon},
  \citenamefont {Entin-Wohlman},\ and\ \citenamefont
  {Aharony}}]{PhysRevB.72.245313}%
  \BibitemOpen
  \bibfield  {author} {\bibinfo {author} {\bibfnamefont {P.}~\bibnamefont
  {Simon}}, \bibinfo {author} {\bibfnamefont {O.}~\bibnamefont
  {Entin-Wohlman}}, \ and\ \bibinfo {author} {\bibfnamefont {A.}~\bibnamefont
  {Aharony}},\ }\href {\doibase 10.1103/PhysRevB.72.245313} {\bibfield
  {journal} {\bibinfo  {journal} {Phys. Rev. B}\ }\textbf {\bibinfo {volume}
  {72}},\ \bibinfo {pages} {245313} (\bibinfo {year} {2005})}\BibitemShut
  {NoStop}%
\bibitem [{\citenamefont {Yoshii}\ and\ \citenamefont
  {Eto}(2011)}]{PhysRevB.83.165310}%
  \BibitemOpen
  \bibfield  {author} {\bibinfo {author} {\bibfnamefont {R.}~\bibnamefont
  {Yoshii}}\ and\ \bibinfo {author} {\bibfnamefont {M.}~\bibnamefont {Eto}},\
  }\href {\doibase 10.1103/PhysRevB.83.165310} {\bibfield  {journal} {\bibinfo
  {journal} {Phys. Rev. B}\ }\textbf {\bibinfo {volume} {83}},\ \bibinfo
  {pages} {165310} (\bibinfo {year} {2011})}\BibitemShut {NoStop}%
\bibitem [{\citenamefont {Affleck}\ and\ \citenamefont
  {Simon}(2001)}]{PhysRevLett.86.2854}%
  \BibitemOpen
  \bibfield  {author} {\bibinfo {author} {\bibfnamefont {I.}~\bibnamefont
  {Affleck}}\ and\ \bibinfo {author} {\bibfnamefont {P.}~\bibnamefont
  {Simon}},\ }\href {\doibase 10.1103/PhysRevLett.86.2854} {\bibfield
  {journal} {\bibinfo  {journal} {Phys. Rev. Lett.}\ }\textbf {\bibinfo
  {volume} {86}},\ \bibinfo {pages} {2854} (\bibinfo {year}
  {2001})}\BibitemShut {NoStop}%
\bibitem [{\citenamefont {Simon}\ and\ \citenamefont
  {Affleck}(2001)}]{PhysRevB.64.085308}%
  \BibitemOpen
  \bibfield  {author} {\bibinfo {author} {\bibfnamefont {P.}~\bibnamefont
  {Simon}}\ and\ \bibinfo {author} {\bibfnamefont {I.}~\bibnamefont
  {Affleck}},\ }\href {\doibase 10.1103/PhysRevB.64.085308} {\bibfield
  {journal} {\bibinfo  {journal} {Phys. Rev. B}\ }\textbf {\bibinfo {volume}
  {64}},\ \bibinfo {pages} {085308} (\bibinfo {year} {2001})}\BibitemShut
  {NoStop}%
\bibitem [{\citenamefont {Komijani}\ \emph {et~al.}(2013)\citenamefont
  {Komijani}, \citenamefont {Yoshii},\ and\ \citenamefont
  {Affleck}}]{PhysRevB.88.245104}%
  \BibitemOpen
  \bibfield  {author} {\bibinfo {author} {\bibfnamefont {Y.}~\bibnamefont
  {Komijani}}, \bibinfo {author} {\bibfnamefont {R.}~\bibnamefont {Yoshii}}, \
  and\ \bibinfo {author} {\bibfnamefont {I.}~\bibnamefont {Affleck}},\ }\href
  {\doibase 10.1103/PhysRevB.88.245104} {\bibfield  {journal} {\bibinfo
  {journal} {Phys. Rev. B}\ }\textbf {\bibinfo {volume} {88}},\ \bibinfo
  {pages} {245104} (\bibinfo {year} {2013})}\BibitemShut {NoStop}%
\bibitem [{\citenamefont {Glazman}\ and\ \citenamefont
  {Raikh}(1988)}]{JETPLett.47.452}%
  \BibitemOpen
  \bibfield  {author} {\bibinfo {author} {\bibfnamefont {L.}~\bibnamefont
  {Glazman}}\ and\ \bibinfo {author} {\bibfnamefont {M.}~\bibnamefont
  {Raikh}},\ }\href {http://www.jetpletters.ac.ru/ps/1095/article_16538.shtml}
  {\bibfield  {journal} {\bibinfo  {journal} {JETP Lett.}\ }\textbf {\bibinfo
  {volume} {47}},\ \bibinfo {pages} {452} (\bibinfo {year} {1988})}\BibitemShut
  {NoStop}%
\bibitem [{\citenamefont {Bruder}\ \emph {et~al.}(1996)\citenamefont {Bruder},
  \citenamefont {Fazio},\ and\ \citenamefont {Schoeller}}]{PhysRevLett.76.114}%
  \BibitemOpen
  \bibfield  {author} {\bibinfo {author} {\bibfnamefont {C.}~\bibnamefont
  {Bruder}}, \bibinfo {author} {\bibfnamefont {R.}~\bibnamefont {Fazio}}, \
  and\ \bibinfo {author} {\bibfnamefont {H.}~\bibnamefont {Schoeller}},\ }\href
  {\doibase 10.1103/PhysRevLett.76.114} {\bibfield  {journal} {\bibinfo
  {journal} {Phys. Rev. Lett.}\ }\textbf {\bibinfo {volume} {76}},\ \bibinfo
  {pages} {114} (\bibinfo {year} {1996})}\BibitemShut {NoStop}%
\bibitem [{\citenamefont {Pustilnik}\ and\ \citenamefont
  {Glazman}(2001)}]{PhysRevB.64.045328}%
  \BibitemOpen
  \bibfield  {author} {\bibinfo {author} {\bibfnamefont {M.}~\bibnamefont
  {Pustilnik}}\ and\ \bibinfo {author} {\bibfnamefont {L.~I.}\ \bibnamefont
  {Glazman}},\ }\href {\doibase 10.1103/PhysRevB.64.045328} {\bibfield
  {journal} {\bibinfo  {journal} {Phys. Rev. B}\ }\textbf {\bibinfo {volume}
  {64}},\ \bibinfo {pages} {045328} (\bibinfo {year} {2001})}\BibitemShut
  {NoStop}%
\bibitem [{\citenamefont {Malecki}\ and\ \citenamefont
  {Affleck}(2010)}]{PhysRevB.82.165426}%
  \BibitemOpen
  \bibfield  {author} {\bibinfo {author} {\bibfnamefont {J.}~\bibnamefont
  {Malecki}}\ and\ \bibinfo {author} {\bibfnamefont {I.}~\bibnamefont
  {Affleck}},\ }\href {\doibase 10.1103/PhysRevB.82.165426} {\bibfield
  {journal} {\bibinfo  {journal} {Phys. Rev. B}\ }\textbf {\bibinfo {volume}
  {82}},\ \bibinfo {pages} {165426} (\bibinfo {year} {2010})}\BibitemShut
  {NoStop}%
\bibitem [{\citenamefont {Meir}\ and\ \citenamefont
  {Wingreen}(1992)}]{PhysRevLett.68.2512}%
  \BibitemOpen
  \bibfield  {author} {\bibinfo {author} {\bibfnamefont {Y.}~\bibnamefont
  {Meir}}\ and\ \bibinfo {author} {\bibfnamefont {N.~S.}\ \bibnamefont
  {Wingreen}},\ }\href {\doibase 10.1103/PhysRevLett.68.2512} {\bibfield
  {journal} {\bibinfo  {journal} {Phys. Rev. Lett.}\ }\textbf {\bibinfo
  {volume} {68}},\ \bibinfo {pages} {2512} (\bibinfo {year}
  {1992})}\BibitemShut {NoStop}%
\bibitem [{\citenamefont {Jauho}\ \emph {et~al.}(1994)\citenamefont {Jauho},
  \citenamefont {Wingreen},\ and\ \citenamefont {Meir}}]{PhysRevB.50.5528}%
  \BibitemOpen
  \bibfield  {author} {\bibinfo {author} {\bibfnamefont {A.-P.}\ \bibnamefont
  {Jauho}}, \bibinfo {author} {\bibfnamefont {N.~S.}\ \bibnamefont {Wingreen}},
  \ and\ \bibinfo {author} {\bibfnamefont {Y.}~\bibnamefont {Meir}},\ }\href
  {\doibase 10.1103/PhysRevB.50.5528} {\bibfield  {journal} {\bibinfo
  {journal} {Phys. Rev. B}\ }\textbf {\bibinfo {volume} {50}},\ \bibinfo
  {pages} {5528} (\bibinfo {year} {1994})}\BibitemShut {NoStop}%
\bibitem [{\citenamefont {Dinu}\ \emph {et~al.}(2007)\citenamefont {Dinu},
  \citenamefont {\ifmmode~\mbox{\c{T}}\else \c{T}\fi{}olea},\ and\
  \citenamefont {Aldea}}]{PhysRevB.76.113302}%
  \BibitemOpen
  \bibfield  {author} {\bibinfo {author} {\bibfnamefont {I.~V.}\ \bibnamefont
  {Dinu}}, \bibinfo {author} {\bibfnamefont {M.}~\bibnamefont
  {\ifmmode~\mbox{\c{T}}\else \c{T}\fi{}olea}}, \ and\ \bibinfo {author}
  {\bibfnamefont {A.}~\bibnamefont {Aldea}},\ }\href {\doibase
  10.1103/PhysRevB.76.113302} {\bibfield  {journal} {\bibinfo  {journal} {Phys.
  Rev. B}\ }\textbf {\bibinfo {volume} {76}},\ \bibinfo {pages} {113302}
  (\bibinfo {year} {2007})}\BibitemShut {NoStop}%
\bibitem [{\citenamefont {Zaffalon}\ \emph {et~al.}(2008)\citenamefont
  {Zaffalon}, \citenamefont {Bid}, \citenamefont {Heiblum}, \citenamefont
  {Mahalu},\ and\ \citenamefont {Umansky}}]{PhysRevLett.100.226601}%
  \BibitemOpen
  \bibfield  {author} {\bibinfo {author} {\bibfnamefont {M.}~\bibnamefont
  {Zaffalon}}, \bibinfo {author} {\bibfnamefont {A.}~\bibnamefont {Bid}},
  \bibinfo {author} {\bibfnamefont {M.}~\bibnamefont {Heiblum}}, \bibinfo
  {author} {\bibfnamefont {D.}~\bibnamefont {Mahalu}}, \ and\ \bibinfo {author}
  {\bibfnamefont {V.}~\bibnamefont {Umansky}},\ }\href {\doibase
  10.1103/PhysRevLett.100.226601} {\bibfield  {journal} {\bibinfo  {journal}
  {Phys. Rev. Lett.}\ }\textbf {\bibinfo {volume} {100}},\ \bibinfo {pages}
  {226601} (\bibinfo {year} {2008})}\BibitemShut {NoStop}%
\bibitem [{\citenamefont {Schuster}\ \emph {et~al.}(1997)\citenamefont
  {Schuster}, \citenamefont {Buks}, \citenamefont {Heiblum}, \citenamefont
  {Mahalu}, \citenamefont {Umansky},\ and\ \citenamefont
  {Shtrikman}}]{Nature.385.417}%
  \BibitemOpen
  \bibfield  {author} {\bibinfo {author} {\bibfnamefont {R.}~\bibnamefont
  {Schuster}}, \bibinfo {author} {\bibfnamefont {E.}~\bibnamefont {Buks}},
  \bibinfo {author} {\bibfnamefont {M.}~\bibnamefont {Heiblum}}, \bibinfo
  {author} {\bibfnamefont {D.}~\bibnamefont {Mahalu}}, \bibinfo {author}
  {\bibfnamefont {V.}~\bibnamefont {Umansky}}, \ and\ \bibinfo {author}
  {\bibfnamefont {H.}~\bibnamefont {Shtrikman}},\ }\href {\doibase
  10.1038/385417a0} {\bibfield  {journal} {\bibinfo  {journal} {Nature}\
  }\textbf {\bibinfo {volume} {385}},\ \bibinfo {pages} {417} (\bibinfo {year}
  {1997})}\BibitemShut {NoStop}%
\bibitem [{\citenamefont {Ji}\ \emph {et~al.}(2000)\citenamefont {Ji},
  \citenamefont {Heiblum}, \citenamefont {Sprinzak}, \citenamefont {Mahalu},\
  and\ \citenamefont {Shtrikman}}]{Science.290.779}%
  \BibitemOpen
  \bibfield  {author} {\bibinfo {author} {\bibfnamefont {Y.}~\bibnamefont
  {Ji}}, \bibinfo {author} {\bibfnamefont {M.}~\bibnamefont {Heiblum}},
  \bibinfo {author} {\bibfnamefont {D.}~\bibnamefont {Sprinzak}}, \bibinfo
  {author} {\bibfnamefont {D.}~\bibnamefont {Mahalu}}, \ and\ \bibinfo {author}
  {\bibfnamefont {H.}~\bibnamefont {Shtrikman}},\ }\href {\doibase
  10.1126/science.290.5492.779} {\bibfield  {journal} {\bibinfo  {journal}
  {Science}\ }\textbf {\bibinfo {volume} {290}},\ \bibinfo {pages} {779}
  (\bibinfo {year} {2000})}\BibitemShut {NoStop}%
\bibitem [{\citenamefont {Ji}\ \emph {et~al.}(2002)\citenamefont {Ji},
  \citenamefont {Heiblum},\ and\ \citenamefont
  {Shtrikman}}]{PhysRevLett.88.076601}%
  \BibitemOpen
  \bibfield  {author} {\bibinfo {author} {\bibfnamefont {Y.}~\bibnamefont
  {Ji}}, \bibinfo {author} {\bibfnamefont {M.}~\bibnamefont {Heiblum}}, \ and\
  \bibinfo {author} {\bibfnamefont {H.}~\bibnamefont {Shtrikman}},\ }\href
  {\doibase 10.1103/PhysRevLett.88.076601} {\bibfield  {journal} {\bibinfo
  {journal} {Phys. Rev. Lett.}\ }\textbf {\bibinfo {volume} {88}},\ \bibinfo
  {pages} {076601} (\bibinfo {year} {2002})}\BibitemShut {NoStop}%
\bibitem [{\citenamefont {Avinun-Kalish}\ \emph {et~al.}(2005)\citenamefont
  {Avinun-Kalish}, \citenamefont {Heiblum}, \citenamefont {Zarchin},
  \citenamefont {Mahalu},\ and\ \citenamefont {Umansky}}]{Nature.436.529}%
  \BibitemOpen
  \bibfield  {author} {\bibinfo {author} {\bibfnamefont {M.}~\bibnamefont
  {Avinun-Kalish}}, \bibinfo {author} {\bibfnamefont {M.}~\bibnamefont
  {Heiblum}}, \bibinfo {author} {\bibfnamefont {O.}~\bibnamefont {Zarchin}},
  \bibinfo {author} {\bibfnamefont {D.}~\bibnamefont {Mahalu}}, \ and\ \bibinfo
  {author} {\bibfnamefont {V.}~\bibnamefont {Umansky}},\ }\href {\doibase
  10.1038/nature03899} {\bibfield  {journal} {\bibinfo  {journal} {Nature}\
  }\textbf {\bibinfo {volume} {436}},\ \bibinfo {pages} {529} (\bibinfo {year}
  {2005})}\BibitemShut {NoStop}%
\bibitem [{\citenamefont {Takada}\ \emph {et~al.}(2014)\citenamefont {Takada},
  \citenamefont {B\"auerle}, \citenamefont {Yamamoto}, \citenamefont
  {Watanabe}, \citenamefont {Hermelin}, \citenamefont {Meunier}, \citenamefont
  {Alex}, \citenamefont {Weichselbaum}, \citenamefont {von Delft},
  \citenamefont {Ludwig}, \citenamefont {Wieck},\ and\ \citenamefont
  {Tarucha}}]{PhysRevLett.113.126601}%
  \BibitemOpen
  \bibfield  {author} {\bibinfo {author} {\bibfnamefont {S.}~\bibnamefont
  {Takada}}, \bibinfo {author} {\bibfnamefont {C.}~\bibnamefont {B\"auerle}},
  \bibinfo {author} {\bibfnamefont {M.}~\bibnamefont {Yamamoto}}, \bibinfo
  {author} {\bibfnamefont {K.}~\bibnamefont {Watanabe}}, \bibinfo {author}
  {\bibfnamefont {S.}~\bibnamefont {Hermelin}}, \bibinfo {author}
  {\bibfnamefont {T.}~\bibnamefont {Meunier}}, \bibinfo {author} {\bibfnamefont
  {A.}~\bibnamefont {Alex}}, \bibinfo {author} {\bibfnamefont {A.}~\bibnamefont
  {Weichselbaum}}, \bibinfo {author} {\bibfnamefont {J.}~\bibnamefont {von
  Delft}}, \bibinfo {author} {\bibfnamefont {A.}~\bibnamefont {Ludwig}},
  \bibinfo {author} {\bibfnamefont {A.~D.}\ \bibnamefont {Wieck}}, \ and\
  \bibinfo {author} {\bibfnamefont {S.}~\bibnamefont {Tarucha}},\ }\href
  {\doibase 10.1103/PhysRevLett.113.126601} {\bibfield  {journal} {\bibinfo
  {journal} {Phys. Rev. Lett.}\ }\textbf {\bibinfo {volume} {113}},\ \bibinfo
  {pages} {126601} (\bibinfo {year} {2014})}\BibitemShut {NoStop}%
\bibitem [{\citenamefont {Carmi}\ \emph {et~al.}(2012)\citenamefont {Carmi},
  \citenamefont {Oreg}, \citenamefont {Berkooz},\ and\ \citenamefont
  {Goldhaber-Gordon}}]{PhysRevB.86.115129}%
  \BibitemOpen
  \bibfield  {author} {\bibinfo {author} {\bibfnamefont {A.}~\bibnamefont
  {Carmi}}, \bibinfo {author} {\bibfnamefont {Y.}~\bibnamefont {Oreg}},
  \bibinfo {author} {\bibfnamefont {M.}~\bibnamefont {Berkooz}}, \ and\
  \bibinfo {author} {\bibfnamefont {D.}~\bibnamefont {Goldhaber-Gordon}},\
  }\href {\doibase 10.1103/PhysRevB.86.115129} {\bibfield  {journal} {\bibinfo
  {journal} {Phys. Rev. B}\ }\textbf {\bibinfo {volume} {86}},\ \bibinfo
  {pages} {115129} (\bibinfo {year} {2012})}\BibitemShut {NoStop}%
\bibitem [{\citenamefont {Aharony}\ \emph {et~al.}(2002)\citenamefont
  {Aharony}, \citenamefont {Entin-Wohlman}, \citenamefont {Halperin},\ and\
  \citenamefont {Imry}}]{PhysRevB.66.115311}%
  \BibitemOpen
  \bibfield  {author} {\bibinfo {author} {\bibfnamefont {A.}~\bibnamefont
  {Aharony}}, \bibinfo {author} {\bibfnamefont {O.}~\bibnamefont
  {Entin-Wohlman}}, \bibinfo {author} {\bibfnamefont {B.~I.}\ \bibnamefont
  {Halperin}}, \ and\ \bibinfo {author} {\bibfnamefont {Y.}~\bibnamefont
  {Imry}},\ }\href {\doibase 10.1103/PhysRevB.66.115311} {\bibfield  {journal}
  {\bibinfo  {journal} {Phys. Rev. B}\ }\textbf {\bibinfo {volume} {66}},\
  \bibinfo {pages} {115311} (\bibinfo {year} {2002})}\BibitemShut {NoStop}%
\bibitem [{\citenamefont {Aharony}\ and\ \citenamefont
  {Entin-Wohlman}(2005)}]{PhysRevB.72.073311}%
  \BibitemOpen
  \bibfield  {author} {\bibinfo {author} {\bibfnamefont {A.}~\bibnamefont
  {Aharony}}\ and\ \bibinfo {author} {\bibfnamefont {O.}~\bibnamefont
  {Entin-Wohlman}},\ }\href {\doibase 10.1103/PhysRevB.72.073311} {\bibfield
  {journal} {\bibinfo  {journal} {Phys. Rev. B}\ }\textbf {\bibinfo {volume}
  {72}},\ \bibinfo {pages} {073311} (\bibinfo {year} {2005})}\BibitemShut
  {NoStop}%
\bibitem [{\citenamefont {Entin-Wohlman}\ \emph {et~al.}(2005)\citenamefont
  {Entin-Wohlman}, \citenamefont {Aharony},\ and\ \citenamefont
  {Meir}}]{PhysRevB.71.035333}%
  \BibitemOpen
  \bibfield  {author} {\bibinfo {author} {\bibfnamefont {O.}~\bibnamefont
  {Entin-Wohlman}}, \bibinfo {author} {\bibfnamefont {A.}~\bibnamefont
  {Aharony}}, \ and\ \bibinfo {author} {\bibfnamefont {Y.}~\bibnamefont
  {Meir}},\ }\href {\doibase 10.1103/PhysRevB.71.035333} {\bibfield  {journal}
  {\bibinfo  {journal} {Phys. Rev. B}\ }\textbf {\bibinfo {volume} {71}},\
  \bibinfo {pages} {035333} (\bibinfo {year} {2005})}\BibitemShut {NoStop}%
\bibitem [{\citenamefont {Kashcheyevs}\ \emph {et~al.}(2006)\citenamefont
  {Kashcheyevs}, \citenamefont {Aharony},\ and\ \citenamefont
  {Entin-Wohlman}}]{PhysRevB.73.125338}%
  \BibitemOpen
  \bibfield  {author} {\bibinfo {author} {\bibfnamefont {V.}~\bibnamefont
  {Kashcheyevs}}, \bibinfo {author} {\bibfnamefont {A.}~\bibnamefont
  {Aharony}}, \ and\ \bibinfo {author} {\bibfnamefont {O.}~\bibnamefont
  {Entin-Wohlman}},\ }\href {\doibase 10.1103/PhysRevB.73.125338} {\bibfield
  {journal} {\bibinfo  {journal} {Phys. Rev. B}\ }\textbf {\bibinfo {volume}
  {73}},\ \bibinfo {pages} {125338} (\bibinfo {year} {2006})}\BibitemShut
  {NoStop}%
\bibitem [{\citenamefont {Schrieffer}\ and\ \citenamefont
  {Wolff}(1966)}]{PhysRev.149.491}%
  \BibitemOpen
  \bibfield  {author} {\bibinfo {author} {\bibfnamefont {J.~R.}\ \bibnamefont
  {Schrieffer}}\ and\ \bibinfo {author} {\bibfnamefont {P.~A.}\ \bibnamefont
  {Wolff}},\ }\href {\doibase 10.1103/PhysRev.149.491} {\bibfield  {journal}
  {\bibinfo  {journal} {Phys. Rev.}\ }\textbf {\bibinfo {volume} {149}},\
  \bibinfo {pages} {491} (\bibinfo {year} {1966})}\BibitemShut {NoStop}%
\bibitem [{\citenamefont {Affleck}\ and\ \citenamefont
  {Ludwig}(1993)}]{PhysRevB.48.7297}%
  \BibitemOpen
  \bibfield  {author} {\bibinfo {author} {\bibfnamefont {I.}~\bibnamefont
  {Affleck}}\ and\ \bibinfo {author} {\bibfnamefont {A.~W.~W.}\ \bibnamefont
  {Ludwig}},\ }\href {\doibase 10.1103/PhysRevB.48.7297} {\bibfield  {journal}
  {\bibinfo  {journal} {Phys. Rev. B}\ }\textbf {\bibinfo {volume} {48}},\
  \bibinfo {pages} {7297} (\bibinfo {year} {1993})}\BibitemShut {NoStop}%
\bibitem [{\citenamefont {Ng}\ and\ \citenamefont
  {Lee}(1988)}]{PhysRevLett.61.1768}%
  \BibitemOpen
  \bibfield  {author} {\bibinfo {author} {\bibfnamefont {T.~K.}\ \bibnamefont
  {Ng}}\ and\ \bibinfo {author} {\bibfnamefont {P.~A.}\ \bibnamefont {Lee}},\
  }\href {\doibase 10.1103/PhysRevLett.61.1768} {\bibfield  {journal} {\bibinfo
   {journal} {Phys. Rev. Lett.}\ }\textbf {\bibinfo {volume} {61}},\ \bibinfo
  {pages} {1768} (\bibinfo {year} {1988})}\BibitemShut {NoStop}%
\bibitem [{\citenamefont {Mahan}(2000)}]{mahan2000many}%
  \BibitemOpen
  \bibfield  {author} {\bibinfo {author} {\bibfnamefont {G.}~\bibnamefont
  {Mahan}},\ }\href {https://books.google.ca/books?id=xzSgZ4-yyMEC} {\emph
  {\bibinfo {title} {Many-Particle Physics}}},\ Physics of Solids and Liquids\
  (\bibinfo  {publisher} {Springer},\ \bibinfo {year} {2000})\BibitemShut
  {NoStop}%
\bibitem [{\citenamefont {Affleck}\ \emph {et~al.}(2008)\citenamefont
  {Affleck}, \citenamefont {Borda},\ and\ \citenamefont
  {Saleur}}]{PhysRevB.77.180404}%
  \BibitemOpen
  \bibfield  {author} {\bibinfo {author} {\bibfnamefont {I.}~\bibnamefont
  {Affleck}}, \bibinfo {author} {\bibfnamefont {L.}~\bibnamefont {Borda}}, \
  and\ \bibinfo {author} {\bibfnamefont {H.}~\bibnamefont {Saleur}},\ }\href
  {\doibase 10.1103/PhysRevB.77.180404} {\bibfield  {journal} {\bibinfo
  {journal} {Phys. Rev. B}\ }\textbf {\bibinfo {volume} {77}},\ \bibinfo
  {pages} {180404} (\bibinfo {year} {2008})}\BibitemShut {NoStop}%
\bibitem [{\citenamefont {Miroshnichenko}\ \emph {et~al.}(2010)\citenamefont
  {Miroshnichenko}, \citenamefont {Flach},\ and\ \citenamefont
  {Kivshar}}]{RevModPhys.82.2257}%
  \BibitemOpen
  \bibfield  {author} {\bibinfo {author} {\bibfnamefont {A.~E.}\ \bibnamefont
  {Miroshnichenko}}, \bibinfo {author} {\bibfnamefont {S.}~\bibnamefont
  {Flach}}, \ and\ \bibinfo {author} {\bibfnamefont {Y.~S.}\ \bibnamefont
  {Kivshar}},\ }\href {\doibase 10.1103/RevModPhys.82.2257} {\bibfield
  {journal} {\bibinfo  {journal} {Rev. Mod. Phys.}\ }\textbf {\bibinfo {volume}
  {82}},\ \bibinfo {pages} {2257} (\bibinfo {year} {2010})}\BibitemShut
  {NoStop}%
\bibitem [{\citenamefont {Vojta}\ and\ \citenamefont
  {Fritz}(2004)}]{PhysRevB.70.094502}%
  \BibitemOpen
  \bibfield  {author} {\bibinfo {author} {\bibfnamefont {M.}~\bibnamefont
  {Vojta}}\ and\ \bibinfo {author} {\bibfnamefont {L.}~\bibnamefont {Fritz}},\
  }\href {\doibase 10.1103/PhysRevB.70.094502} {\bibfield  {journal} {\bibinfo
  {journal} {Phys. Rev. B}\ }\textbf {\bibinfo {volume} {70}},\ \bibinfo
  {pages} {094502} (\bibinfo {year} {2004})}\BibitemShut {NoStop}%
\bibitem [{\citenamefont {Fritz}\ and\ \citenamefont
  {Vojta}(2004)}]{PhysRevB.70.214427}%
  \BibitemOpen
  \bibfield  {author} {\bibinfo {author} {\bibfnamefont {L.}~\bibnamefont
  {Fritz}}\ and\ \bibinfo {author} {\bibfnamefont {M.}~\bibnamefont {Vojta}},\
  }\href {\doibase 10.1103/PhysRevB.70.214427} {\bibfield  {journal} {\bibinfo
  {journal} {Phys. Rev. B}\ }\textbf {\bibinfo {volume} {70}},\ \bibinfo
  {pages} {214427} (\bibinfo {year} {2004})}\BibitemShut {NoStop}%
\bibitem [{\citenamefont {Gerland}\ \emph {et~al.}(2000)\citenamefont
  {Gerland}, \citenamefont {von Delft}, \citenamefont {Costi},\ and\
  \citenamefont {Oreg}}]{PhysRevLett.84.3710}%
  \BibitemOpen
  \bibfield  {author} {\bibinfo {author} {\bibfnamefont {U.}~\bibnamefont
  {Gerland}}, \bibinfo {author} {\bibfnamefont {J.}~\bibnamefont {von Delft}},
  \bibinfo {author} {\bibfnamefont {T.~A.}\ \bibnamefont {Costi}}, \ and\
  \bibinfo {author} {\bibfnamefont {Y.}~\bibnamefont {Oreg}},\ }\href {\doibase
  10.1103/PhysRevLett.84.3710} {\bibfield  {journal} {\bibinfo  {journal}
  {Phys. Rev. Lett.}\ }\textbf {\bibinfo {volume} {84}},\ \bibinfo {pages}
  {3710} (\bibinfo {year} {2000})}\BibitemShut {NoStop}%
\bibitem [{\citenamefont {Hofstetter}\ \emph {et~al.}(2001)\citenamefont
  {Hofstetter}, \citenamefont {K\"onig},\ and\ \citenamefont
  {Schoeller}}]{PhysRevLett.87.156803}%
  \BibitemOpen
  \bibfield  {author} {\bibinfo {author} {\bibfnamefont {W.}~\bibnamefont
  {Hofstetter}}, \bibinfo {author} {\bibfnamefont {J.}~\bibnamefont {K\"onig}},
  \ and\ \bibinfo {author} {\bibfnamefont {H.}~\bibnamefont {Schoeller}},\
  }\href {\doibase 10.1103/PhysRevLett.87.156803} {\bibfield  {journal}
  {\bibinfo  {journal} {Phys. Rev. Lett.}\ }\textbf {\bibinfo {volume} {87}},\
  \bibinfo {pages} {156803} (\bibinfo {year} {2001})}\BibitemShut {NoStop}%
\bibitem [{\citenamefont {Silvestrov}\ and\ \citenamefont
  {Imry}(2003)}]{PhysRevLett.90.106602}%
  \BibitemOpen
  \bibfield  {author} {\bibinfo {author} {\bibfnamefont {P.~G.}\ \bibnamefont
  {Silvestrov}}\ and\ \bibinfo {author} {\bibfnamefont {Y.}~\bibnamefont
  {Imry}},\ }\href {\doibase 10.1103/PhysRevLett.90.106602} {\bibfield
  {journal} {\bibinfo  {journal} {Phys. Rev. Lett.}\ }\textbf {\bibinfo
  {volume} {90}},\ \bibinfo {pages} {106602} (\bibinfo {year}
  {2003})}\BibitemShut {NoStop}%
\bibitem [{\citenamefont {Oreg}\ and\ \citenamefont
  {Goldhaber-Gordon}(2003)}]{PhysRevLett.90.136602}%
  \BibitemOpen
  \bibfield  {author} {\bibinfo {author} {\bibfnamefont {Y.}~\bibnamefont
  {Oreg}}\ and\ \bibinfo {author} {\bibfnamefont {D.}~\bibnamefont
  {Goldhaber-Gordon}},\ }\href {\doibase 10.1103/PhysRevLett.90.136602}
  {\bibfield  {journal} {\bibinfo  {journal} {Phys. Rev. Lett.}\ }\textbf
  {\bibinfo {volume} {90}},\ \bibinfo {pages} {136602} (\bibinfo {year}
  {2003})}\BibitemShut {NoStop}%
\bibitem [{\citenamefont {Potok}\ \emph {et~al.}(2007)\citenamefont {Potok},
  \citenamefont {Rau}, \citenamefont {Shtrikman}, \citenamefont {Oreg},\ and\
  \citenamefont {Goldhaber-Gordon}}]{Nature.446.167}%
  \BibitemOpen
  \bibfield  {author} {\bibinfo {author} {\bibfnamefont {R.~M.}\ \bibnamefont
  {Potok}}, \bibinfo {author} {\bibfnamefont {I.~G.}\ \bibnamefont {Rau}},
  \bibinfo {author} {\bibfnamefont {H.}~\bibnamefont {Shtrikman}}, \bibinfo
  {author} {\bibfnamefont {Y.}~\bibnamefont {Oreg}}, \ and\ \bibinfo {author}
  {\bibfnamefont {D.}~\bibnamefont {Goldhaber-Gordon}},\ }\href {\doibase
  10.1038/nature05556} {\bibfield  {journal} {\bibinfo  {journal} {Nature}\
  }\textbf {\bibinfo {volume} {446}},\ \bibinfo {pages} {167} (\bibinfo {year}
  {2007})}\BibitemShut {NoStop}%
\bibitem [{\citenamefont {Bellazzini}\ \emph {et~al.}(2007)\citenamefont
  {Bellazzini}, \citenamefont {Mintchev},\ and\ \citenamefont
  {Sorba}}]{JPhysA.40.2485}%
  \BibitemOpen
  \bibfield  {author} {\bibinfo {author} {\bibfnamefont {B.}~\bibnamefont
  {Bellazzini}}, \bibinfo {author} {\bibfnamefont {M.}~\bibnamefont
  {Mintchev}}, \ and\ \bibinfo {author} {\bibfnamefont {P.}~\bibnamefont
  {Sorba}},\ }\href {http://stacks.iop.org/1751-8121/40/i=10/a=017} {\bibfield
  {journal} {\bibinfo  {journal} {J. Phys. A}\ }\textbf {\bibinfo {volume}
  {40}},\ \bibinfo {pages} {2485} (\bibinfo {year} {2007})}\BibitemShut
  {NoStop}%
\bibitem [{\citenamefont {Perelomov}\ \emph {et~al.}(1998)\citenamefont
  {Perelomov}, \citenamefont {Zel'dovi{\v{c}}},\ and\ \citenamefont
  {Zelʹdovich}}]{perelomov1998quantum}%
  \BibitemOpen
  \bibfield  {author} {\bibinfo {author} {\bibfnamefont {A.}~\bibnamefont
  {Perelomov}}, \bibinfo {author} {\bibfnamefont {{\^A}.}~\bibnamefont
  {Zel'dovi{\v{c}}}}, \ and\ \bibinfo {author} {\bibfnamefont {I.}~\bibnamefont
  {Zelʹdovich}},\ }\href {https://books.google.ca/books?id=cHR2gk5lDzIC}
  {\emph {\bibinfo {title} {Quantum Mechanics: Selected Topics}}},\ Selected
  Topics Series\ (\bibinfo  {publisher} {World Scientific},\ \bibinfo {year}
  {1998})\BibitemShut {NoStop}%
\bibitem [{\citenamefont {Shankar}(1994)}]{RevModPhys.66.129}%
  \BibitemOpen
  \bibfield  {author} {\bibinfo {author} {\bibfnamefont {R.}~\bibnamefont
  {Shankar}},\ }\href {\doibase 10.1103/RevModPhys.66.129} {\bibfield
  {journal} {\bibinfo  {journal} {Rev. Mod. Phys.}\ }\textbf {\bibinfo {volume}
  {66}},\ \bibinfo {pages} {129} (\bibinfo {year} {1994})}\BibitemShut
  {NoStop}%
\end{thebibliography}%

\end{document}